\documentclass[a4paper,fleqn,usenatbib]{mnras}

\usepackage{newtxtext,newtxmath}
\usepackage[T1]{fontenc}
\usepackage{ae,aecompl}
\usepackage[flushleft]{threeparttable}
\usepackage{booktabs,caption}

\usepackage{graphicx}	
\usepackage{amssymb}	

\title[A Synthetical Approach to Stellar Classification]{The Average Physical Properties of A-G Stars Derived from \MakeLowercase{\textit{uvby}}-H$\beta$ Str\"omgren-Crawford Photometry as the Basis for a  Spectral-Classification Synthetical Approach}

\author[Dalle Mese, L\'opez-Cruz, Schuster, et al.]{
G. Dalle Mese$^{1,2}$\thanks{E-mail: giannina@inaoep.mx (GDM)},O. L\'opez-Cruz$^{1}$,W. J. Schuster$^{3}$, C. Chavarr\'ia-K$^{3}$, \and H.~J. Ibarra-Medel$^{4}$
\\
$^{1}$Instituto Nacional de Astrof\'isica, \'Optica y Electr\'onica (INAOE). Luis Enrique Erro 1 Santa Mar\'ia Tonantzintla, Puebla C.P. 72840,  M\'exico\\
$^{2}$Facultad de Ciencias de la Tierra y el Espacio, Universidad Aut\'onoma de Sinaloa (FACITE-UAS). Josefa Ortiz de Dom\'inguez S/N, C.U. Culiac\'an, Sin.  C.P. 80040, M\'exico\\
$^{3}$Instituto de Astronom\'ia, Universidad Nacional Aut\'onoma de M\'exico (IA-UNAM). Campus Ensenada, AP 106, 22860. Ensenada B.C., M\'exico\\
$^{4}$University of Illinois Urbana-Champaign, Department of Astronomy. 103 Astronomy Bulding, 1002 W Green St, Urbana IL 61801, U.S.A.
}

\date{Accepted XXX. Received YYY; in original form \today}

\pubyear{2018}
\hypersetup{draft}

\begin{document}
\label{firstpage}
\pagerange{\pageref{firstpage}--\pageref{lastpage}}
\maketitle

\begin{abstract}

We have revisited and updated the $uvby$ Str\"omgren colour and colour-index distributions of A, F and early G-type main sequence stars. For this aim, we selected 7054 dwarf stars along with 65 MK standard stars within the same spectral range but covering all luminosity classes. The standard stars were selected following the MK mandate strictly, using spectra taken at classification resolution recorded on photographic plates. We used the colours  of these stars to determine  the effective temperature and surface gravity. After correcting for systematic offsets using fundamental parameters and  considering a few exceptions, we find  an one-to-one correspondence, among MK spectral types, Str\"omgren photometry, and their associated physical properties. 

We have applied a principal component analysis to the mean Str\"omgren indices for dwarf stars complemented by MK standards for higher luminosity classes. We have used the projections to introduce three new photometric metaindices, namely SM1, SM2, and SM3. We have defined a 3D-box, which allowed us to segregate dwarf stars from bright giants and supergiant stars, with the aid of the metaindices. Two of the planes show that the projections of dwarfs and supergiants are ordered by temperature; however, the temperature dependence for the supergiants is not as well defined as for the dwarfs. Following the MK Process, we were able to form an automatic classifier. We present some applications and assigned synthetical spectral types. We suggest that our metaindices formalism allows the extension of Str\"omgren photometric outside its original mandate (i.e., later types), without requiring the introduction of additional photometric filters.  
 
\end{abstract}




\section{Introduction}\label{errors}

The Morgan-Keenan system (MK, hereafter) is an autonomous and self-consistent approach for stellar spectral classification. It relies on a well-defined set of standard stars, chosen on the basis of the phenomenology of spectral lines, blends and bands, according to a general progression of spectral type (abscissa) and luminosity (ordinate) \citep[e.g.,][]{morgan, Mi84, Mo84, Ga85,grayc}; hence, the standard stars define a two-dimensional grid of well separated spectral types and luminosity classes. To classify a given star, the classifier compares and interpolates its spectrum across the grid, considering the particular features to a class, but paying attention to the whole spectrum. The system was defined for Population I stars using the blue-violet region of the stellar spectrum. On the other hand, the MK Process \citep[e.g.,][]{Mo84} as a methodology for classification has been successfully extended to other regions of the electromagnetic spectrum \citep{grayc}. As any properly defined classification system, one of the main uses of the MK system is to separate rare or peculiar objects from the broad population of ``normal stars" \citep[e.g.,][]{Mo84,Ga85}; notwithstanding, one of the most creative and powerful applications of the MK  system was implemented by \citet{morgan52,morgan53}. \citeauthor{morgan53} carried out a comprehensive search for galactic OB stars to  map their distribution.  Aided with spectroscopic parallaxes and correcting for extinction,  Morgan and his collaborators were able to outline the spiral structure of the Galaxy, for the first time.  

There have been several attempts to apply the MK system automatically. \citet{manteiga} presented a list of spectral line-strengths, band fluxes and their ratios of MK standards; however, a closer advance to the emulate a classifier using the  MK system was developed by \citet{graymk}, who generated a computer program to classify stellar spectra by direct comparison with a set of MK standards.

\subsection{Intermediate band photometry and stellar classification}

\begin{figure}
\centering
\includegraphics[width=\columnwidth]{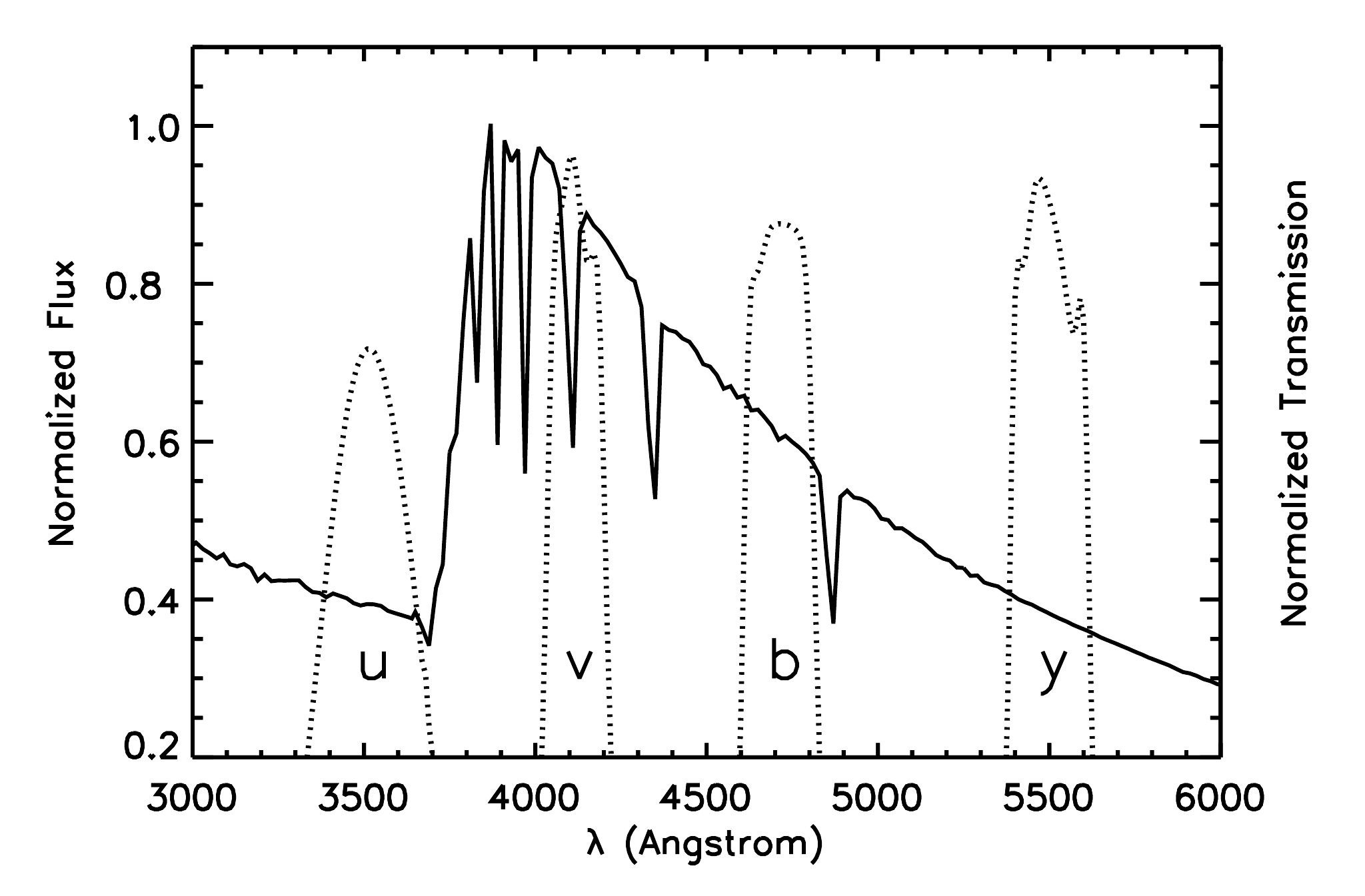}
\caption{The $u \sim 3500$ \AA, $v \sim 4100$ \AA, $b\sim4700$ \AA $\ $ and $y \sim 5500$ \AA $\ $ filter passbands overpplotted in a A0 V star spectrum generated with \textsc{CHORIZOS} \citep{maiz_2004} from a Kurucz model \citep{kurucz}.}
\label{vega}
\end{figure}

Early on, it was realized that stellar photometry provided a complementary quantitative approach to spectral classification \citep[e.g.,][]{MK53,strom_1966,C84, SDB90,C94}. The original complementary found between the MK types and stellar photometry \citep{MK53,strom_1966} has become clearer as we have gained access larger data bases  of stellar measurements and spectral classifications, and more realistic stellar atmosphere models.

Figure \ref{vega} shows the transmission curves (in dashed lines) for each of the filters that define \citep[e.g.,][]{S92,B05} the photometric system introduced by Bengt Str\"omgren during the 1950s \citep[See review by][]{strom_1966}. In general, intermediate-band photometric systems are able to reproduce the results derived from low resolution spectroscopy. \citeauthor{strom_1966} carefully configure the set of  filters to define this photometric system, considering the broad stellar spectral properties, his aim was to closely match the MK classification system.  \citeauthor{strom_1966} introduced colors and two indices resulting after  the  linear combination of colors. The color $(b-y)$ is sensitive to temperature but insensitive to  metallicity (because of small and comparable line blanketing in $b$ and $y$), the $(v-b)$ color is very sensitive to metallicity particularly for F and G stars due to the large blanketing in $v$ and much smaller blanketing in $b$,  while  the $y$ band reproduces Johnson's $V$ band. The indices  $c_1=(u-v)-(v-b)$ and $m_1=(v-b)-(b-y)$ measure physical properties which are unaccessible  employing  broad band filters: the $c_1$  measures the Balmer discontinuity, while the $m_1$ estimates line blanketing of A-G type stars, respectively. The Str\"omgren photometric system was originally designed to approximate spectral  classifications for A2 to G2 stars. Successful extensions to the range of applicability to earlier or later types out of the aforemetioned range, have been achieved by introducing extra filters. For example the $H_\beta$ index introduced by \citet{1958ApJ...128..185C} to include stars earlier than A2. This extended system is called the {\it uvby}-${\rm H}_\beta$ Str\"omgren-Crawford Photometric System \citep{1966VA......8..149C}. \citet{1991AJ....101.1902A} introduced the $hk$ index  design to cover the Ca II $H$ and $K$ lines, this extension allowed the consideration of metal poor dwarfs and red giants.

In Figure \ref{vega} the $u \sim 3500$ \AA, $v \sim 4100$ \AA, $b\sim4700$ \AA $\ $ and $y \sim 5500$ \AA $\ $ filter passbands are overpplotted onto a A0 V stellar spectrum generated with \textsc{CHORIZOS} \citep[][see \S\ref{physicalparameters}, below]{maiz_2004} from a Kurucz model \citep{kurucz} while the \textit{uvby} filter profiles are from The Isaac Newton Group filter database. Only the $v$ filter contains a strong absorption line, the H$\delta$ Balmer. The other filters do not contain very strong lines up to spectral type G0. Str\"omgren photometry is more sensitive to subtle variations in spectral energy distribution than broadband photometry, but less time consuming than spectroscopy.

Recently, \citet{monguio} introduced a complex classification scheme based on MK standards and developed an automatic method for deriving the stellar physical parameters from Str\"omgren photometry. This attempt responded, at least partially, to the concerns raised by sticklers who hold that spectral types assigned from photometric are improper, because they are generated outside the MK mandate \citep[e.g.,][]{Ga85}, and, therefore, not accurate enough. Nevertheless,  below, we show that the stellar physical properties derived photometrically approximate the physical basis of the MK sequence \citep[See \S2.4 in][]{grayc}; therefore, we can extend the photometric approach to spectral classification based on the MK process, and provide complementary classifications to large numbers of  faint stars, for which proper spectral inspection would be too time consuming. 

 Also we introduce a new photometry-based classification scheme, which is able to approximate MK types for normal stars. Our scheme is based on a principal component analysis (PCA) of the Str\"omgren indices \citep[cf.,][]{stein}. With the aid of our classification scheme we are able to segregate dwarf from supergiants, and assign spectral types automatically. \citep{terranegra}  presented a simpler photometrically approach to spectral classification using Str\"omgren indexes. 

This paper is organized as follows: \S\ref{errors} we give specific details on the MK system, the MK Process and the $uvby$-H$\beta$ Str\"omgren-Crawford Photometry. In \S\ref{ss} we describe the selection of the sample and the reddening corrections; the properties of the observed photometry are presented in \S \ref{distri}; the derivation of physical parameters is given in \S\ref{physicalparameters}; \S\ref{fundamental} describes how we derived a transformation to match our results with fundamental effective temperatures effective temperature and also is presented a comparison with standard stars; results are listed in \S \ref{results}; in \S \ref{pcasec}, we show  Str\"omgren photometric indices space allows a reduction in dimension; hence, the main sequence can be reduced to straight line using a projection generated by principal component analysis (PCA). We also present applications of the proposed  classification scheme. In last section (\S \ref{summ}) we present our conclusions.

\section{Sample selection}
\label{ss}

 A recent compilation of 298 639 $uvby$-H$\beta$ photometric measurements of 60 668 stars by \citet{paunzen}, almos includes the fully earlier catalogue by  \citet{hauck_1998}, which contains 63 313 stars. Although, \citeauthor{paunzen}'s compilation has reprocessed all the previous measurements and is much larger than \citeauthor{hauck_1998}'s, due to the inclusion of CCD-based photometry, we found that analysis using either compilation were indistinguishable; therefore, we opted for the original compilation by \citet{hauck_1998}. MK Spectral classifications were taken from the ongoing compilation by \citet{skiff}, and parallaxes from HIPPARCOS \citep{perry}. 

A preselection was done, considering that  stars earlier than A0 V are not well represented in  \citet{paunzen} compilation, see also Figure 2 in \citet{NH2}. The apparent absence of these stars can be explained by their short lifetimes and by their position within the Galaxy: stars earlier than A0 are highly massive and  are often found in high extinction regions. On the other hand, we considered stars three types later than G2 V just for completeness; hence, we explored spectral classes outside the original mandate of the Str\"omgren system.

 We have  given  preference  to the classifications published  in  the  University of Michigan Catalogue of Two-Dimensional Spectral Types for the HD Stars, Volumes 1–5  {\citep[]{houkc1975, houkf1978, houk82, houksm, houksw}}. The first reason is that the classifications  were done following the MK system and mandate strictly (objective prism spectra on photographic plates). The second, most important,  reason is  that  96\% of the classification provided in the Michigan Catalogues were assigned by a single expert classifier, Nancy \citeauthor{houk78}  \citep{NH1,NH2}. Hence, 86\% MK of the classifications included  in our study  were taken from the Michigan Catalogues. Spectra of bright stars were saturated in the plates used by \citeauthor{houk78}, classifications to those stars were provided independently Robert F. Garrison, a list of 348 classifications assigned  by R. F. Garrison using slit spectra on photographic plates are presented in  Table 8 in \citet{NH2}.  
 
 In the MK system peculiar stars are marked by the letter "p" or the abbreviation "pec", or by  the lack of convergence to a single type and luminosity class. Peculiarities are   indicated by the presence of strong metallic lines, or by different types using hydrogen lines and other discrepant spectral features; hence, we have  excluded peculiar stars including Am, Ap or any chemically peculiar (CP) star. Moreover,  we checked that none of the CP  candidate stars published by \citet{RM09} were included. 
 
 For the rest of the stars included in this paper, we revised the sources to avoided MK-like classifications assigned by the modeling of the stellar spectral energy distribution \citep[e.g.,][]{AP12}. We have made an effort to include only  stars with the highest quality types assigned according to MK system. Our  final sample contains 7054 normal star with full spectral classifications (single spectral type and luminosity class) covering from A0 V to G5 V, having accurate $uvby$-H$\beta$ Str\"omgren-Crawford photometry, and a magnitude limit of $V = 11.0$ mag.

\begin{figure}
\centering
  \includegraphics[width=\columnwidth]{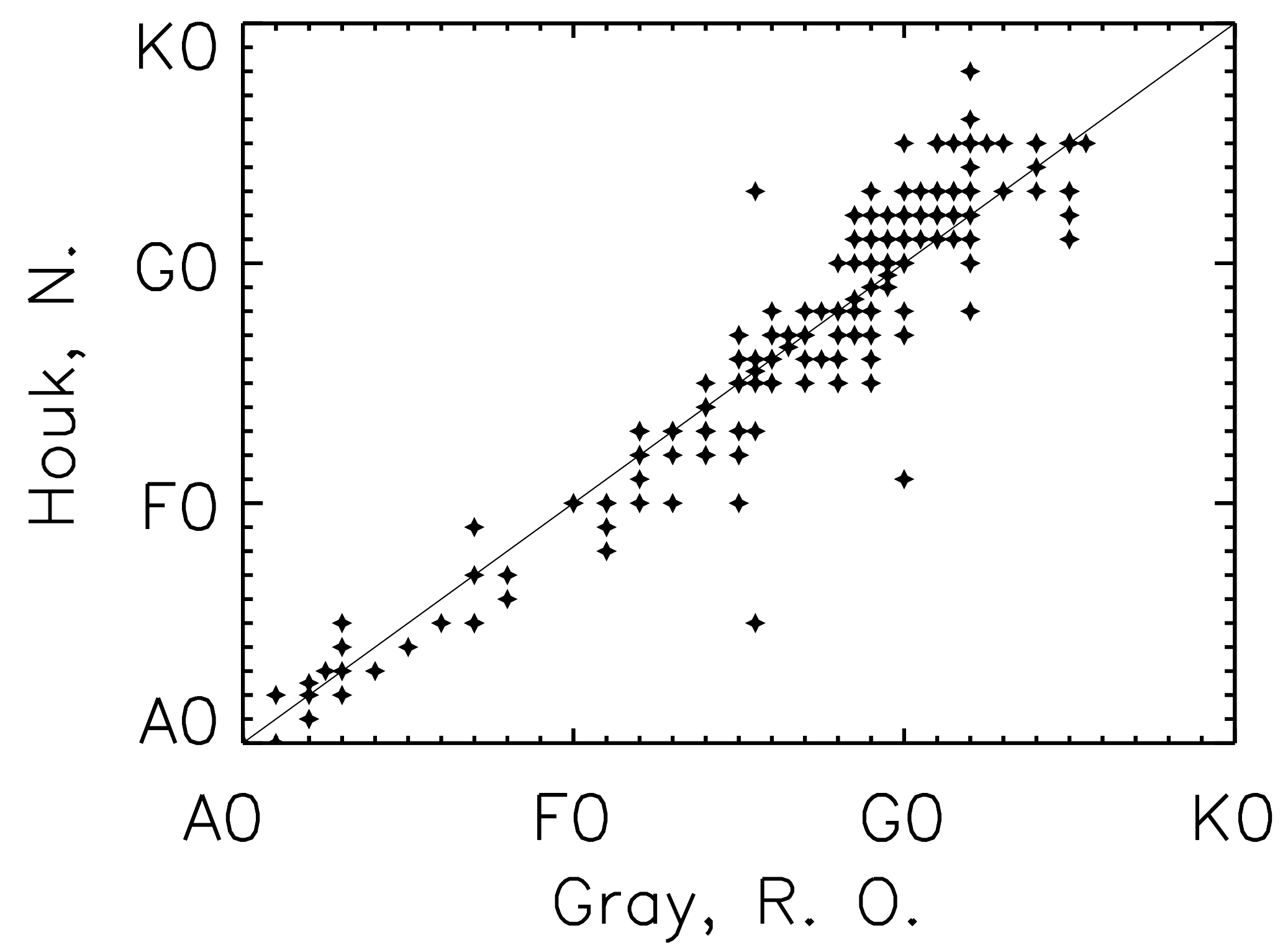}
  \caption{Comparison of the spectral types assigned in \citet[]{houk78,houk82} and those assigned by \citet{graygar}. A one-to-one line is plotted to guide the eye. A slight systematic difference is seen for stars later than G0, in which the types assigned by \citeauthor{graygar} are ealier than \citeauthor{houk78}'s. This systematic deviation, may result from differences in grid of standards employed in each study. For example, \citeauthor{houk78} used the star HD 114710 as G2 V standard, which has been revised to G0 V \citep{gray2,frasca}.}
  \label{nancycomparison}
\end{figure}

We compared \citeauthor{houk78} with those derived from digital spectra by \citet[]{gray2, gray2003, graygar}. Figure \ref{nancycomparison} shows the comparison between these classifiers. A slight trend is noted where early-type stars classified by \citeauthor{gray2} appear to be slightly later than the types by \citeauthor{houk78}. On the other hand, stars classified by \citeauthor{houk78} around G0 are systematically later than those of \citet[]{gray2, gray2003, graygar}. E. Mamajek (Personal communication) noted that \citeauthor{houk82} chose the star HD 114710 as the standard for the G2 V class; nevertheless, this star was later revised to G0 V  \citep[e.g.,][]{gray2,frasca}; hence, the main source of discrepancy seen in Figure \ref{nancycomparison} is generated by selected  grid of standard employed by each study. However, we will see below a clear progression of the spectral properties and the photometric and physical properties of the stars. A note regarding the definition of G2 V class and the Sun as a standard star is provided by E. Mamajeck,\footnote{\url{http://www.pas.rochester.edu/~emamajek/memo_G2V.html}}: a  more philosophical, discussion on the usage and abusage of MK mandate is provided by \citet{Ga85}. 

\begin{table*}\scriptsize
\caption{Grid of MK Standards stars selected by OL-C \citep{LC91}.}\label{lctab}
 \begin{center}
 \begin{tabular}{llclll}
\hline
\hline
  \multicolumn{1}{c}{ Identifier} &
  \multicolumn{1}{c}{ Spectral type} &
  \multicolumn{1}{c}{Source} &
  \multicolumn{1}{c}{ $(b-y)_0$}&
  \multicolumn{1}{c}{ $m_0$} &
  \multicolumn{1}{c}{ $c_0$}\\
\hline
HD 196379  &  A9~II &MK73&   0.236   &     0.091   &    1.4810 \\
HD 23585 &F0 V	   &MK78 &0.190  $\pm$0.005 &0.178  $\pm$0.010 &0.717  $\pm$0.008 \\
HD 220392  &  F0~IV & GG89&  0.147  $\pm$0.003&  0.187  $\pm$0.002&  0.967$\pm$  0.001\\
57 Tau     &  F0~IV &MK78&   0.171 $\pm$ 0.002 & 0.194 $\pm$ 0.005&  0.770 $\pm$ 0.003\\
$\eta$  CMi    &   F0~IIIb & GG89&  0.132 $\pm$ 0.005 & 0.193 $\pm$ 0.002& 1.007$\pm$  0.007\\
38 Cnc    &   F0~IIIb &GG89 &  0.153    &  0.180   &   0.995\\
$\zeta$   Leo    &   F0~III$^{*}$ &MK78&   0.196  &     0.169  &    0.986\\
HD 6130   &   F0~II  &MK73&  0.362 $\pm$ 0.002&  0.069 $\pm$ 0.001 &  1.179 $\pm$ 0.013\\
$\alpha$    Lep    &   F0~Ib$^{*}$ &MK78&   0.139 $\pm$ 0.001&  0.149$\pm$  0.002 &1.503 $\pm$ 0.004\\
$\phi$ Cas    &   F0~Ia &MK78&   0.492 $\pm$ 0.002&  -0.002$\pm$  0.011 &  1.438 $\pm$ 0.012\\
 HD 113139&F2 V$^{*}$ &MK78&0.244  $\pm$ 0.002 &0.170  $\pm$ 0.003 &0.576  $\pm$ 0.003 \\
32 Tau &   F2~IV &GG89&   0.225  &    0.152    &  0.660\\
$\beta$ Cas   &    F2~III &GG89&   0.216  & 0.177&    0.785  \\
$\nu$ Aql    &   F2~Ib &GG89&   0.420 $\pm$ 0.005& 0.069$\pm$  0.006& 1.449$\pm$  0.012\\
$\iota^1$ Sco   &    F2~Ia &MK78&   0.337   &   0.107   &   1.459\\
 HD 26015&F3 V &MK78 &0.270     &0.164     &0.488   \\ 
20 CVn   &    F3~III &MK78&   0.176 $\pm$ 0.003& 0.232$\pm$  0.005 &0.915  $\pm$0.003\\
 HD 27524 &F5 V & MK73&0.285  $\pm$ 0.002 &0.161  $\pm$ 0.002 &0.461  $\pm$ 0.004\\
$\alpha$    CMi   &    F5~IV-V &MK78&   0.272 $\pm$ 0.000  & 0.167 $\pm$ 0.000 & 0.532 $\pm$ 0.000  \\
48    Gem    &   F5~III-IV &MK78&   0.246    &  0.170   &   0.786\\
HD     186155 &   F5~II-III &MK78&   0.264$\pm$  0.003& 0.203 $\pm$ 0.002  & 0.718 $\pm$ 0.005\\
$\alpha$ Per    &   F5~Ib$^{*}$ &MK78&   0.301  $\pm$0.004  & 0.196$\pm$  0.004&1.073 $\pm$ 0.001\\
HD 10494  &   F5~Ia &MK78&   0.875    & -0.026    &  1.458 \\
HD 30652 & F6 V$^{*}$ & MK73&0.298  $\pm$ 0.002 &0.163  $\pm$ 0.002 &0.415  $\pm$ 0.004\\
$\alpha$     Tri    &   F6~IV &MK73&   0.316   &   0.156  &    0.501\\
HD     160365 &   F6~III &MK73&   0.374$\pm$  0.002  & 0.162  $\pm$0.003 &0.555 $\pm$ 0.003\\
HD 222368 & F7 V& MK53&0.330  $\pm$ 0.002 &0.163  $\pm$ 0.004 &0.399  $\pm$ 0.003\\
HD 27808 &F8 V & MK78& 0.338  $\pm$ 0.002 &0.172  $\pm$ 0.005 &0.382  $\pm$ 0.007\\
HD 217096 &   F8~III-IV &MK73&   0.366 $\pm$ 0.006  & 0.188$\pm$  0.002  & 0.489 $\pm$ 0.013\\
$\upsilon$   Peg   &    F8~III &MK78&   0.390  & 0.187$\pm$  0.002  & 0.461$\pm$  0.001\\
$\gamma$   Cyg   &    F8~Ib &MK78&   0.393 $\pm$ 0.004  & 0.296  $\pm$0.005 &0.884 $\pm$ 0.002\\
$\delta$    CMa  &    F8~Ia$^{*}$  &MK78&   0.375   &   0.322   &   0.929     \\
HD 27383 & F9 V& MK73 & 0.352  $\pm$ 0.003 &0.192  $\pm$ 0.003 &0.370  $\pm$ 0.005\\
HD 109358 &G0 V$^{*}$& K76/K85&0.385     &0.182  $\pm$ 0.001 &0.298  $\pm$ 0.006 \\
$\eta$   Boo   &    G0~IV$^{*}$  &MK78/K85&  0.374  $\pm$0.003 &0.203$\pm$  0.003 &0.488$\pm$  0.020\\
$\psi^3$    Psc   &    G0III &MK73&   0.440$\pm$  0.003& 0.221 $\pm$ 0.010 &0.439 $\pm$ 0.013\\
$\beta$   Cam   &    G0~Ib-IIa &MK53/K85&   0.559 $\pm$ 0.010 & 0.328  $\pm$0.006  & 0.485  $\pm$0.011\\
$\mu$    Per   &    G0~Ib &MK53/K85&   0.616 $\pm$ 0.003& 0.273 $\pm$ 0.007  & 0.545 $\pm$ 0.015\\
$\beta$   Aqr   &    G0~Ib &MK53/K85&   0.512$\pm$  0.001  & 0.325$\pm$  0.001  & 0.566 $\pm$ 0.016\\
HD     91629 &    G0~Ia &MK73&   0.481  & 0.256$\pm$  0.004  & 0.609$\pm$  0.008\\
HD     18391 &    G0~Ia FeI &MK53&   1.386   &   0.333  &    0.411 \\
HD     101947 &   G0~0-Ia FeI &K85&   0.468 $\pm$ 0.015& 0.279 $\pm$ 0.011  & 0.599 $\pm$ 0.053\\
HD 27836 &G1 V& MK73&0.384  $\pm$ 0.003 &0.200  $\pm$ 0.002 &0.330  $\pm$ 0.006\\
$\zeta$   Her   &    G1~IV$^{*}$ &MK73/K85&   0.413 $\pm$ 0.003& 0.207 $\pm$ 0.004  & 0.408 $\pm$ 0.005\\
$\epsilon$   Leo   &    G1~II  &K76/K85&  0.507$\pm$  0.005 &0.276 $\pm$ 0.004  & 0.452 $\pm$ 0.009\\
$\phi$   Vir    &   G2~IV &K76/K85&   0.439$\pm$  0.003& 0.207 $\pm$ 0.003& 0.383 $\pm$ 0.003\\
84    Her   &    G2~IIIb &K76/K85&   0.424 $\pm$ 0.005& 0.219 $\pm$ 0.005 &0.452 $\pm$ 0.004\\
$\beta$  Dra   &    G2~Ib-IIa &K85&   0.608$\pm$  0.004  & 0.323$\pm$  0.003& 0.423 $\pm$ 0.006\\
$\alpha$  Aqr   &    G2~Ib &K53/K85&   0.588 $\pm$ 0.007  & 0.384  $\pm$0.007  & 0.447 $\pm$ 0.007\\
HD     96746 &    G2~Iab &K76&   0.574  &    0.306&      0.534  \\
R Pup &    G2~0-Ia&K76/K85&    0.701$\pm$  0.003 &0.253$\pm$  0.009& 0.868$\pm$  0.023\\
HD 10307 & G2 V& MK53& 0.389     &0.198  $\pm$ 0.005 &0.348  $\pm$ 0.011 \\
HD     3421  &    G2.5~IIa &K85&   0.555   &   0.322  &    0.375\\
37    LMi   &    G2.5~IIa  &K85&  0.512$\pm$  0.002  & 0.294 $\pm$ 0.005& 0.478 $\pm$ 0.003\\
HD 186427 &G3 V& MK85& 0.416     &0.226     &0.354  \\
22    Vul   &    G3~Ib-IIp &MK73&   0.652   &   0.363    &  0.351  \\
$\beta$  Sct   &    G4~IIa  &K76/K85 & 0.709$\pm$  0.006  & 0.352$\pm$  0.005& 0.430$\pm$  0.013\\
HD 20630 &G5 V$^{*}$& MK53/K85&0.419  $\pm$ 0.003 &0.235  $\pm$ 0.005 &0.307  $\pm$ 0.003 \\
$\mu$   Her   &    G5~IV  &K53/K85 &  0.468 $\pm$ 0.005 &0.271$\pm$  0.010 &0.407 $\pm$ 0.007\\
HD     27022 &    G5~IIb &K76/K85 &   0.511 $\pm$ 0.001  & 0.286 $\pm$ 0.004  & 0.405$\pm$  0.003\\
$\omega$   Gem   &    G5~Ib-IIa &K76&   0.575    &  0.361  &    0.372 \\
25    Gem   &    G5~Ib &MK53&   0.661    &  0.430    &  0.330   \\
9     Peg   &    G5~Ib &MK53/K85 &   0.706$\pm$  0.002  & 0.480$\pm$  0.007  & 0.349 $\pm$ 0.008\\
HD 49396  &   G5~Iab &K76&   0.647$\pm$  0.002  & 0.457$\pm$  0.007  & 0.422$\pm$  0.008\\
HD     217476  &  G5 0-Ia &K76&   0.803  & 0.224  & 1.039 \\
\hline
\end{tabular}
\end{center}
     \raggedright Grid of MK standards set up by \citet{LC91} with photometric measurements taken  from \citet{hauck_1998}. These stars were selected by \citeauthor{LC91} using photographic plates taken at classification resolution, with strict observance of the MK mandate. They conform a smooth progression of types and luminosity classes. In column 3, the sources are indicated as follows: MK53 \citep{MK53}, MK73 \citep{morgan}, MK78 \citep{MK78}, K76 \citep{K76}, K85 \citep{K85}, and GG89 \citep{GG89}. The starts marked by $^{*}$ in column 2, were defined as anchor points by \citet{garrison}.
\end{table*}

\subsubsection{MK Standards}

In order to extend our analysis to higher luminosity classes and observe consistency with the MK Process, we have included in this study a grid of 65 MK standard stars selected  by OL-C \citep[][see Table \ref{lctab} and \S \ref{stat} below]{LC91}. These stars were observed with a dispersion of 65\AA/mm on photographic plates. The plates were taken with similar spectrographs (in Chile or Mexico) and were processed in the same  manner. They are, now, part of the Garrison Archive at the University of Toronto. The original idea was to provide a densely populated MK grid for visual spectral classification. Each spectrum was examined and checked to follow a smooth progression both in spectral type and luminosity class. Table \ref{lctab}, also includes 10 "anchor points" as defined by \citet{garrison}. These
anchor points are the most stable standard MK standards, as their spectral types have not changed throughout the historical revisions of the MK classification system. Those 10 anchor points are marked by an asterisk on column 2. Nevertheless, owing to the powerful interpolating ability of the eye-brain neurological complex, the trained classifier is able to assigned accurate types, even in the presence of large gaps in the grid of standard stars. The grid defined by the stars in  Table \ref{lctab} was used to search for F supergiant stars at high galactic latitude \citep{LC91, LC93, LC93a}. Due to the intrinsic properties of evolved stars, there are not enough  MK classified stars to generate an statistical analysis, the situation is complicated by spectral-type variations. The situation can be grasp in Figure 2 \citet{NH2}. Hence, we resorted to  the MK standard stars in Table \ref{lctab} to extend our analysis to  evolved star, this is just a first order approximation. Appendix A in \citet{grayc} provides updated  MK standard stars that can complement our Table.

\subsection{Reddening correction}\label{red}

Understanding the local interstellar reddening is very important in the derivation of the intrinsic color $(b-y)_0$ calibration and hence the other indices $m_0$ and $c_0$ as well (hereafter the subscript 0 means that the color or index re reddening free or has been derredened). The nearest interstellar dust patches are at about 70 pc from the Sun; this region devoid of dust is commonly identified with the \textit{Local Hot Bubble}. Indeed, several investigations indicate that the effects of interstellar reddening are negligible for stars closer than 70 pc from the Sun \citep[e.g.,][]{tim, leroy, holm,luck}. On the other hand, the study of \citet{reis} uses the same classical methods for estimating color excess employed in this paper. \citeauthor{reis} results  agree with those of \citet{perryj}, suggesting  that  the effects of interstellar reddening may be ignored for stars as far as  80 pc. We have been more stringent imposing a cut at  70 pc to delimit the \textit{Local Hot Bubble}.

\begin{figure}
\centering
  \includegraphics[scale=0.4,angle=0]{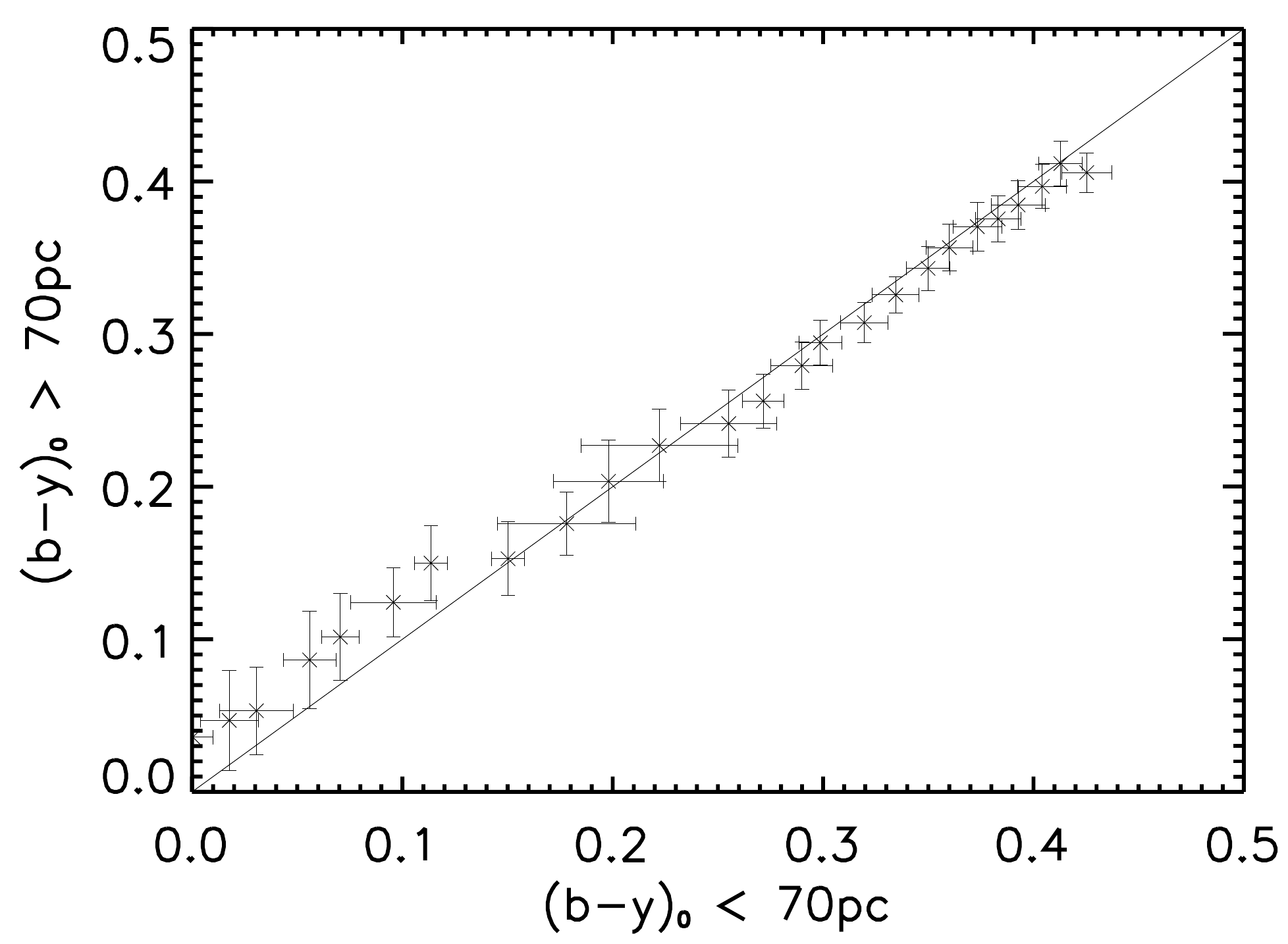}
  \caption{Comparison of the mean $(b-y)_0$ color for stars closer than 70 pc and the sample that needed a reddening correction.}
  \label{7070}
\end{figure}

Table \ref{tab70} represent the sample within 70 pc from the Sun, in which no reddenings correction were required; in \S \ref{physicalparameters} we explain how we obtained the  values for the effective temperature ($T_{eff}$) and surface gravity ($\log g$); while the absolute magnitude ($M_V$) was computed using HIPPARCOS data.

We separated the sample into two groups: stars that are closer than 70 pc from the Sun are assumed reddening free, this subsample consists of 1496 stars. For the rest of the stars we derived the derreddening correction in four different ways, (according to the spectral types): stars with A0-A2 spectral types were corrected according to \citet{hild}, A3-F4 by the method of \citet{crawford66}, F5-G2 using the calibration by \citet{schuster_1989}, and G2-G5 with the equation from \citet{olsen} using the routine published by \citet{moon}.  4322 out of the 5630 stars farther than 70 pc from the Sun have reported H$\beta$ measurements.  

In Figure \ref{7070} we plot in the abscissa the color $(b-y)_0$ for stars closer than 70 pc, where each point represents a given spectral type from A0 to G5 V stars, where values around 0 corresponds to A0 stars; values greater than 0.4 for late-G type stars. In the ordinate the is the same scale for the same color but for stars farther than 70 pc corrected by reddening as described in above. Although both axis seems consistent, there is a slight systematic difference  for stars earlier than A6, in which stars closer that 70 pc appear hotter by $(b-y)=0.02$ mag.  For our final analysis  we merged  both samples, this discrepancy will slightly affect the mean values; however, its most important effect is the increasing of the dispersion in the color and index distributions.


\section{The Distribution of Observed Properties}
\label{distri}

\begin{figure}
\centering
\includegraphics[width=\columnwidth, scale=0.6]{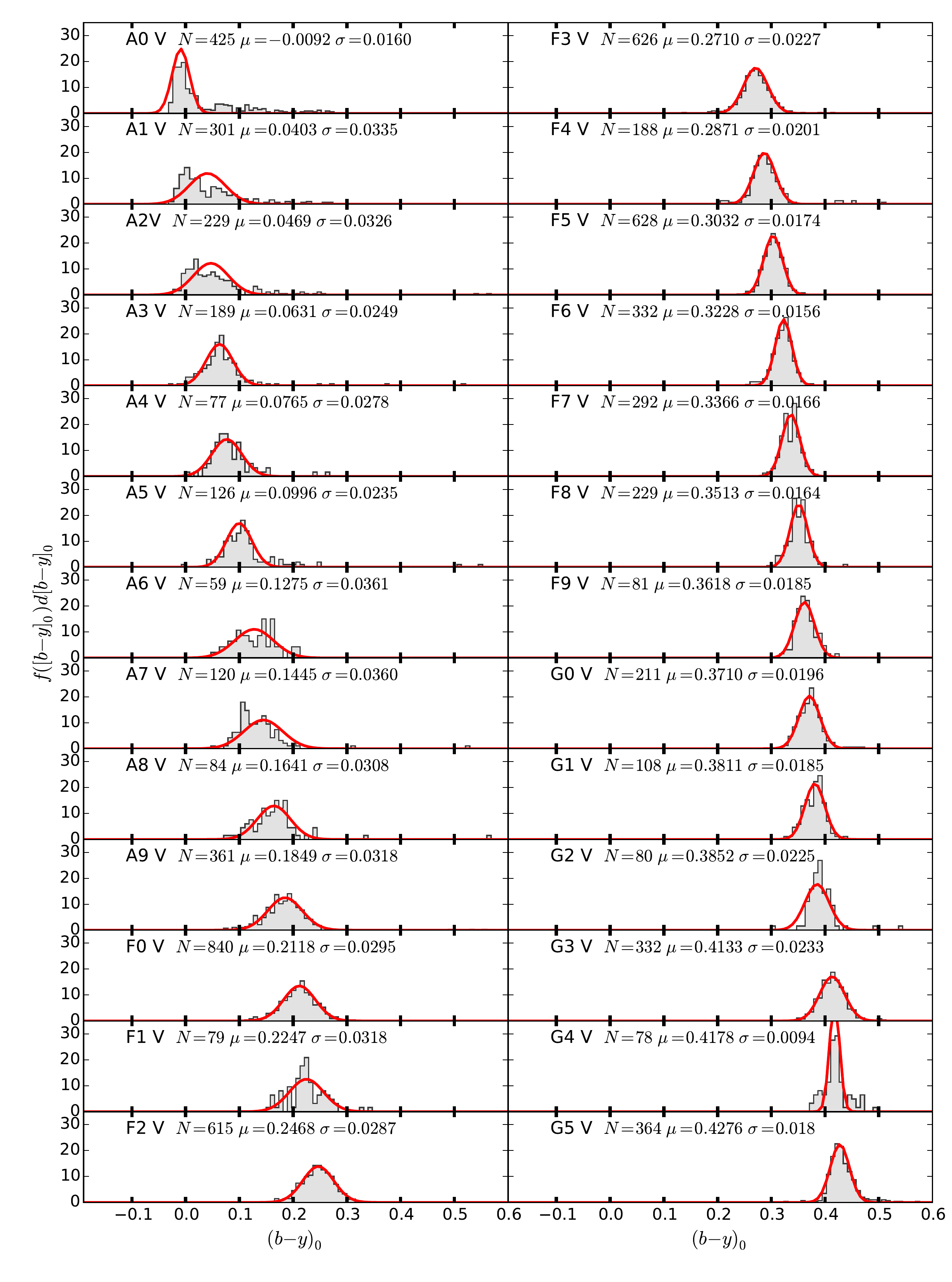}
  \caption{The  statistical distribution of the Str\"omgren photometric color $(b-y)_0$ for spectral types A0 V to G5 V of our sample of 7054 stars.}
  \label{f5}
\end{figure}

\begin{figure}
\centering
\includegraphics[width=\columnwidth, scale=0.6]{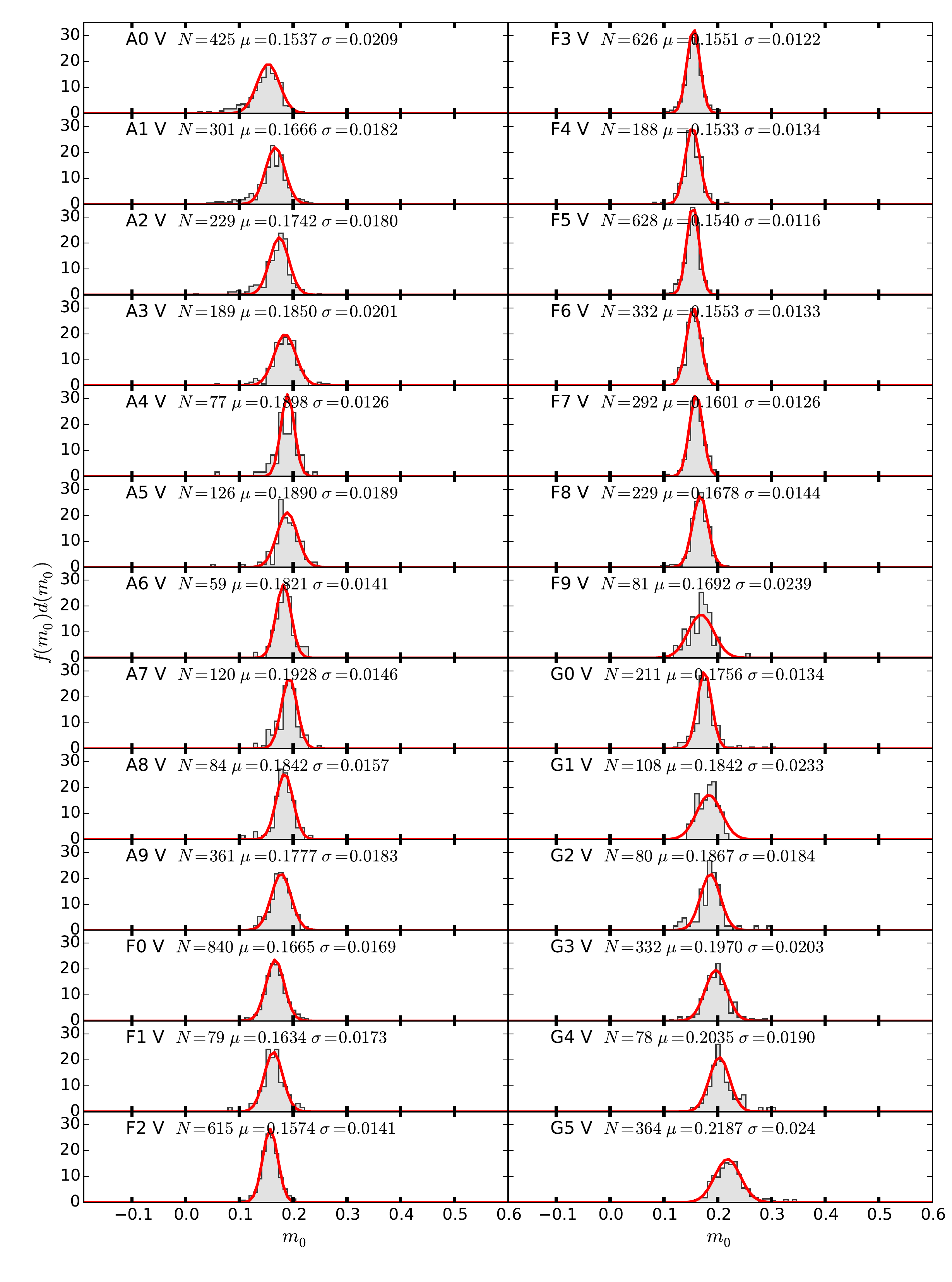}
  \caption{The  statistical distribution of the Str\"omgren photometric index $m_0$ for spectral types A0 V to G5 V of our sample of 7054 stars.}
  \label{fm1}
\end{figure}

\begin{figure}
\centering
\includegraphics[width=\columnwidth, scale=0.6]{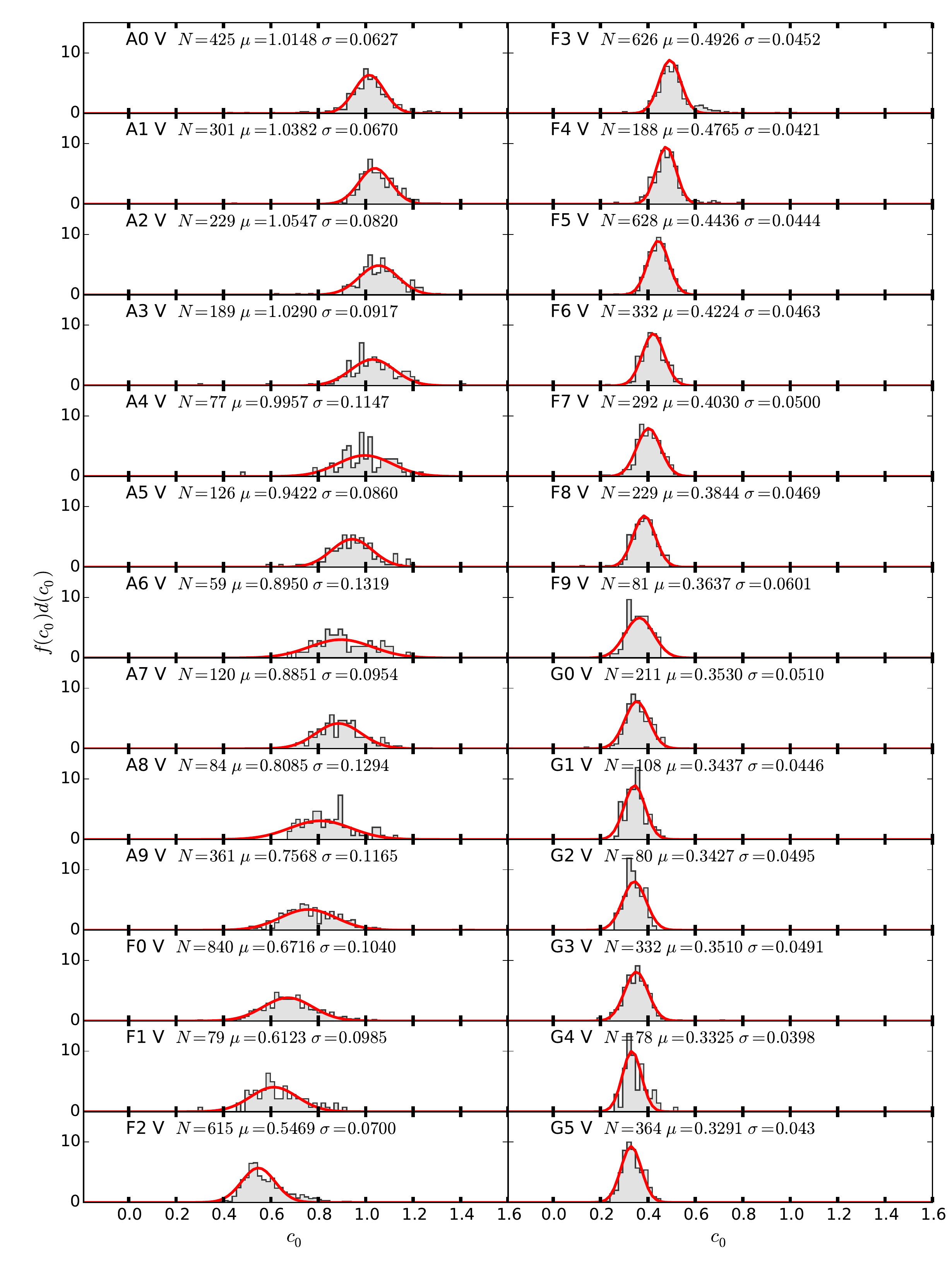}
  \caption{The statistical distribution of the Str\"omgren photometric index $c_0$ for spectral types A0 V to G5 V of our sample of 7054 stars.}
  \label{fc1}
\end{figure}

We generated the distributions of colors and color indices by spectral type for the 7054 main sequence stars from A0 to G5 of the sample. In Figure \ref{f5} we show the sequence of the $(b-y)_0$ index separated by spectral type. 

From the Central Limit Theorem, considering that we are comparing independent measurements (realizations)  subject to variety of  errors, we would expect the values of the indices per spectral type should follow a Normal distribution. We observed that highly discordant values occupied the tails of the distribution, while most of measurements cluster around their, respective, mean values. Indeed, from a Shapiro-Wilk test applied to the individual distributions, we obtained mean values of the test of $\overline{W}= 0.942$, $\overline{W}=0.965$, and $\overline{W}= 0.956$, for the indices  $(b-y)_0$, $m_0$ and  $c_0$, respectively. Although, the significance for the global distribution was low in certain cases; restricting the analysis around the mode yielded the non-rejection of the hypothesis of normality. Therefore, we suggest that the indices by spectral type are normally distributed. Therefore, we used Gaussian fits to model the distribution of indices by spectral type (red solid line). The distributions of $m_0$ and $c_0$ indices of our sample are shown in Figures \ref{fm1} and \ref{fc1}, respectively. 

 Figure \ref{f5} shows  a clear progression of the modes for each spectral type as a function of $(b-y)_0$.  This  progression also confirms the one-to-one correspondence between stellar spectral type and  effective temperature for main sequence stars (see \S\ref{results}). If we assign spectral types based on Str\"omgren photometry, our uncertainty is $\pm$2 types. This is due to the broadness of the color distribution. This  is comparable to the results of \citet{SDB90}, who used the seven-filter Vilnius photometric system. 

As seen in Figure \ref{fm1}, the metallicity index $m_0$ doesn't follow any overall sequential pattern with spectral type. This fact could be explained by the selection itself, for we have included only non-peculiar stars within  the solar-neighbourhood. Nevertheless, a slight progression $m_0 \rightarrow0.22$ is seen for stars later than F9, but its significance is weak with respect to the average $\langle m_0 \rangle =0.18$ for all the stars considered in this paper, which closely corresponds to the mean $m_0$  for G2 V stars. 

On the other hand, a progression of the indicator of the Balmer Jump $c_0$ is clearly seen in Figure \ref{fc1}, but for spectral types later than A2 V and earlier than G0 V. However, $c_0$ has been known to be sensitive to the effective temperature of OB type stars, see also Fig. \ref{bym1c1plot}.

\begin{figure}
\centering
\includegraphics[width=\columnwidth, scale=0.4]{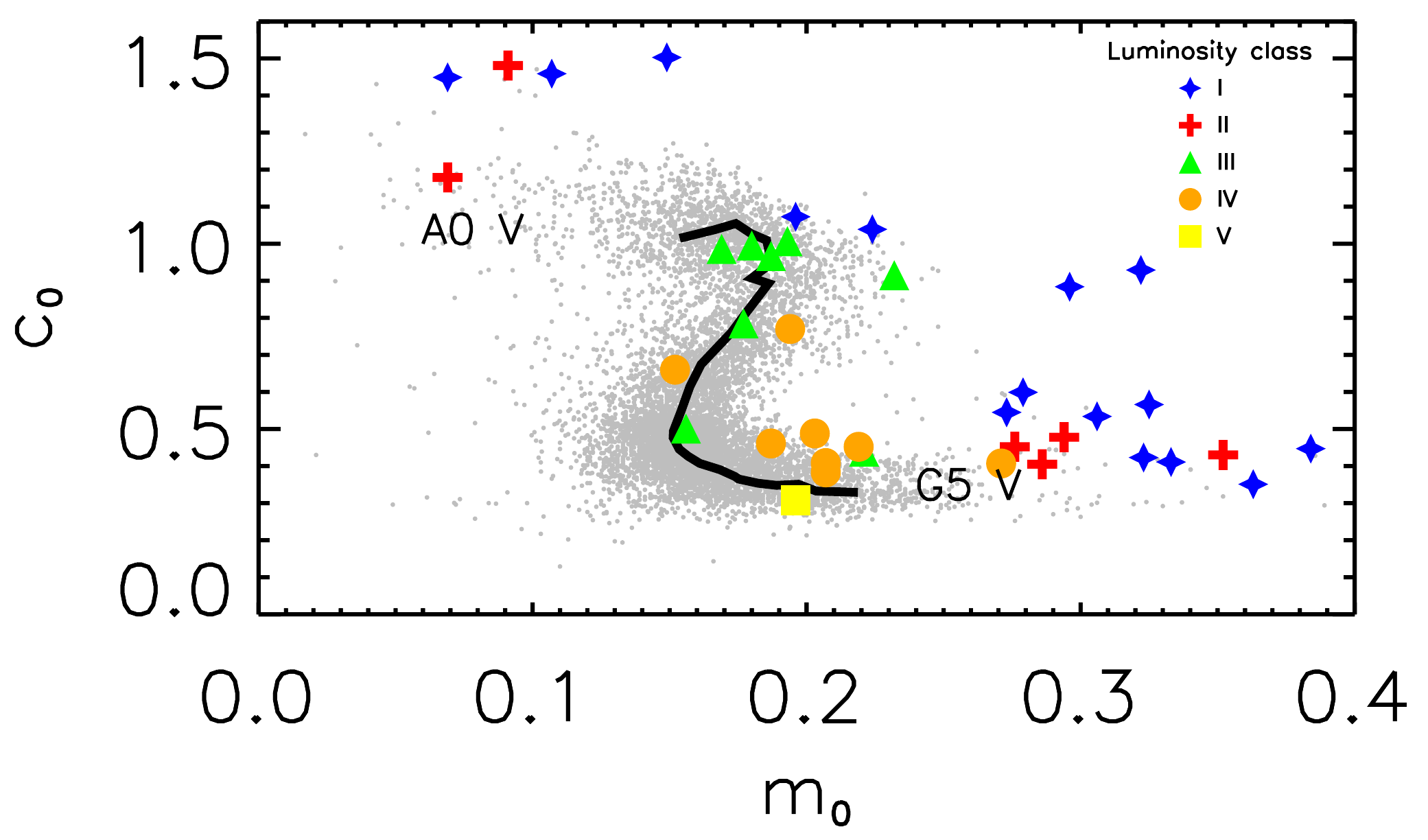}
\caption{The locus of A0-G5 dwarfs in the $m_0$, $c_0$ plane. Standard stars from \citet{LC91} (Table \ref{lctab}) are also shown in the plot using blue stars for luminosity class I, red crosses for II, green triangles for III, orange circles for IV and yellow squares for V. The black, narrow band is determined by using the mean $m_0$ and $c_0$ indices for each spectral type of the main sequence. The location of stars from MK types A0 V to G5 V is shown schematically in the diagram.}
\label{m1c1plot}
\end{figure}

\begin{figure}
\centering
\includegraphics[width=\columnwidth, scale=0.28]{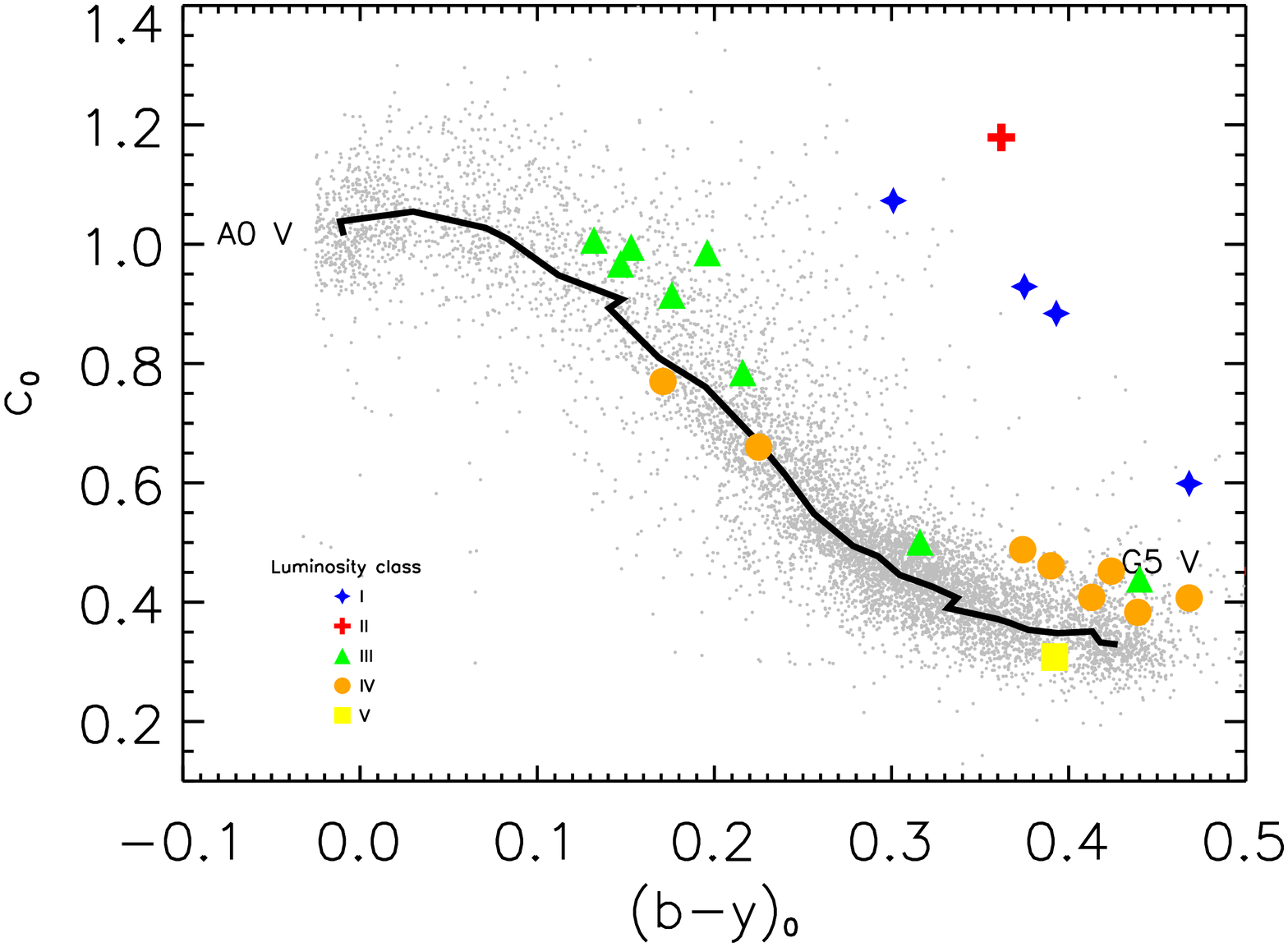}
\includegraphics[width=\columnwidth, scale=0.28]{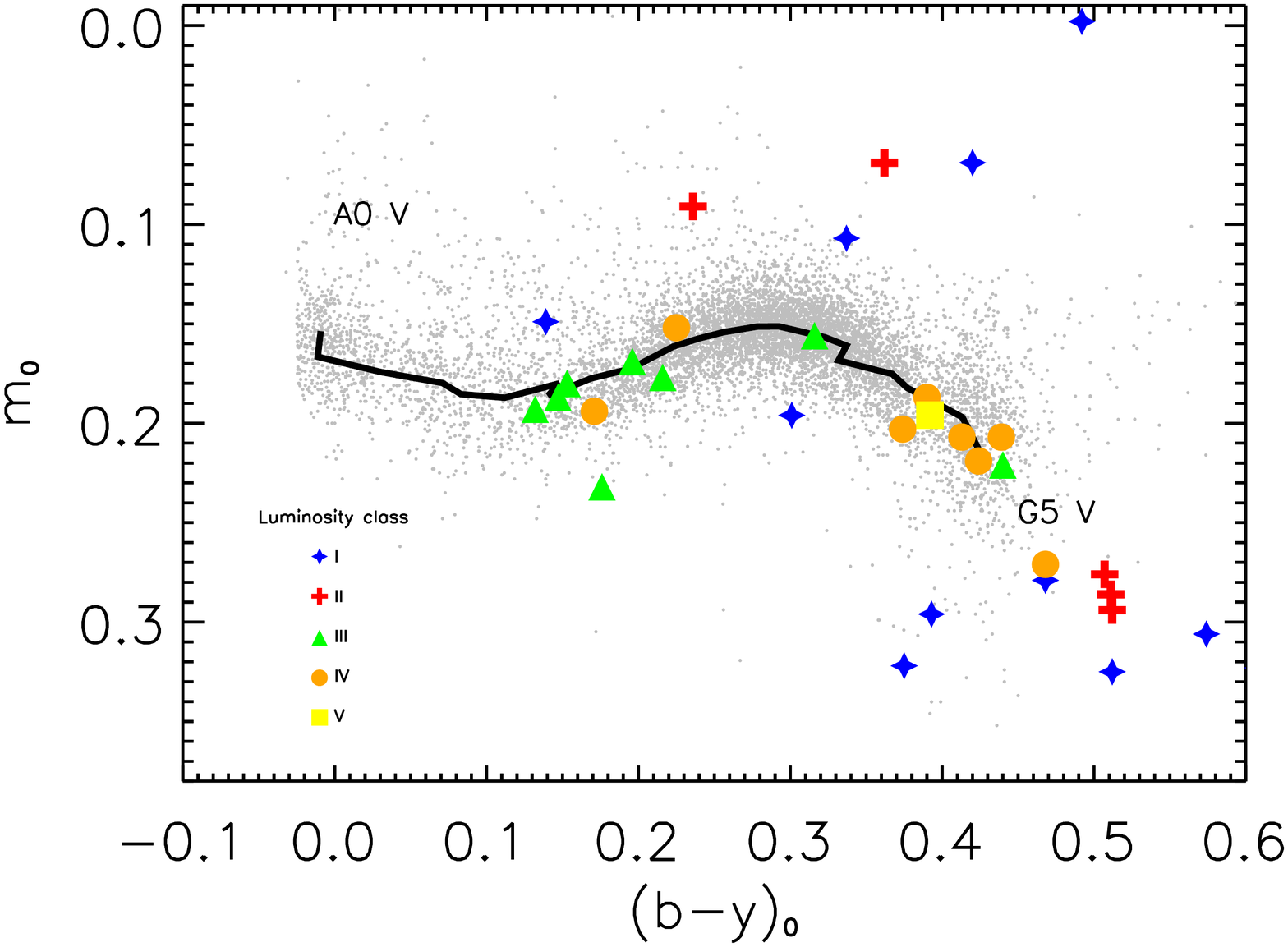}
\caption{Upper panel: the locus of A0-G5 dwarfs in the $c_0$, $(b-y)_0$ plane. $c_0$ measures the Balmer jump (surface gravity) and $(b-y)_0$, the temperature. We expect to find a maximum in $c_0$ at $(b-y)_0$ $\sim 0$, corresponding to early A-types, and, if observed values of $(b-y)$ are used, some scatter toward larger values of this index due to interstellar reddening. Lower panel: the same in the $m_0$, $(b-y)_0$ plane; there is a little variation in $m_0$ due to near-solar metallicities. Symbols as in Figure \ref{m1c1plot}.}
\label{bym1c1plot}
\end{figure}

In Figure \ref{m1c1plot} we can see the locus of A0-G5 dwarfs in the $m_0$-$c_0$ plane, first proposed by \citet{strom_1966}. Stars from Table \ref{lctab} are shown in the plot using blue stars for luminosity class I, red crosses for II, green triangles for III, orange filled circles for IV, and yellow filled squares for V. These stars are MK standards taken from Table 1, the photometry was taken from \citet{paunzen} or SIMBAD, the distances from HIPPARCOS, and as explained in \S\ref{red}, no reddening correction was necessary (d $<$ 70 pc), and M$_V$ was obtained from the $V$ photometry provided by SIMBAD. These are consistent MK standards covering all luminosity classes. They were used to show their correspondence with surface gravity. The black thick line, depicts the locus of the mean $m_0$ and $c_0$ indices for each spectral type (Figures \ref{fm1} and \ref{fc1}). Luminosity class V stars are concentrated in the grey band, and the position along this band determines the spectral class unambiguously. Supergiant stars are well separated from dwarf stars. Separation between class II and class I does not exist everywhere, but class II stars are also segregated from the main-sequence. Lastly, class III is not well distinguished from class IV and class V \citep[cf.,][]{gray2}. In Figure \ref{m1c1plot}, we can appreciate that the original definition of the Str\"omgren photometric system did not include stars later than G5, as the separation between main sequence and evolve stars  blurs for later types.

Figure \ref{bym1c1plot} shows the distribution of the stars in the $(b-y)_0$-$c_0$ plane (upper panel). The unreddened $(b-y)_0$ index, indicates the effective temperature; the $c_0$ index, the Balmer jump, i.e. the surface gravity and temperature. Note that the $c_0$ index comes to a maximum near $(b-y)_0=0$, corresponding to the early A stars, where the Balmer lines and Balmer discontinuity reach their maximum excitation. The lower panel of Figure \ref{bym1c1plot} shows the variation of $m_0$ with $(b-y)_0$; coloured symbols stand for the same as in Figure \ref{m1c1plot}. There is a slight variation of $m_0$ about the solar metallicity. 

    \begin{figure}
\centering
\includegraphics[width=\columnwidth, scale=0.28]{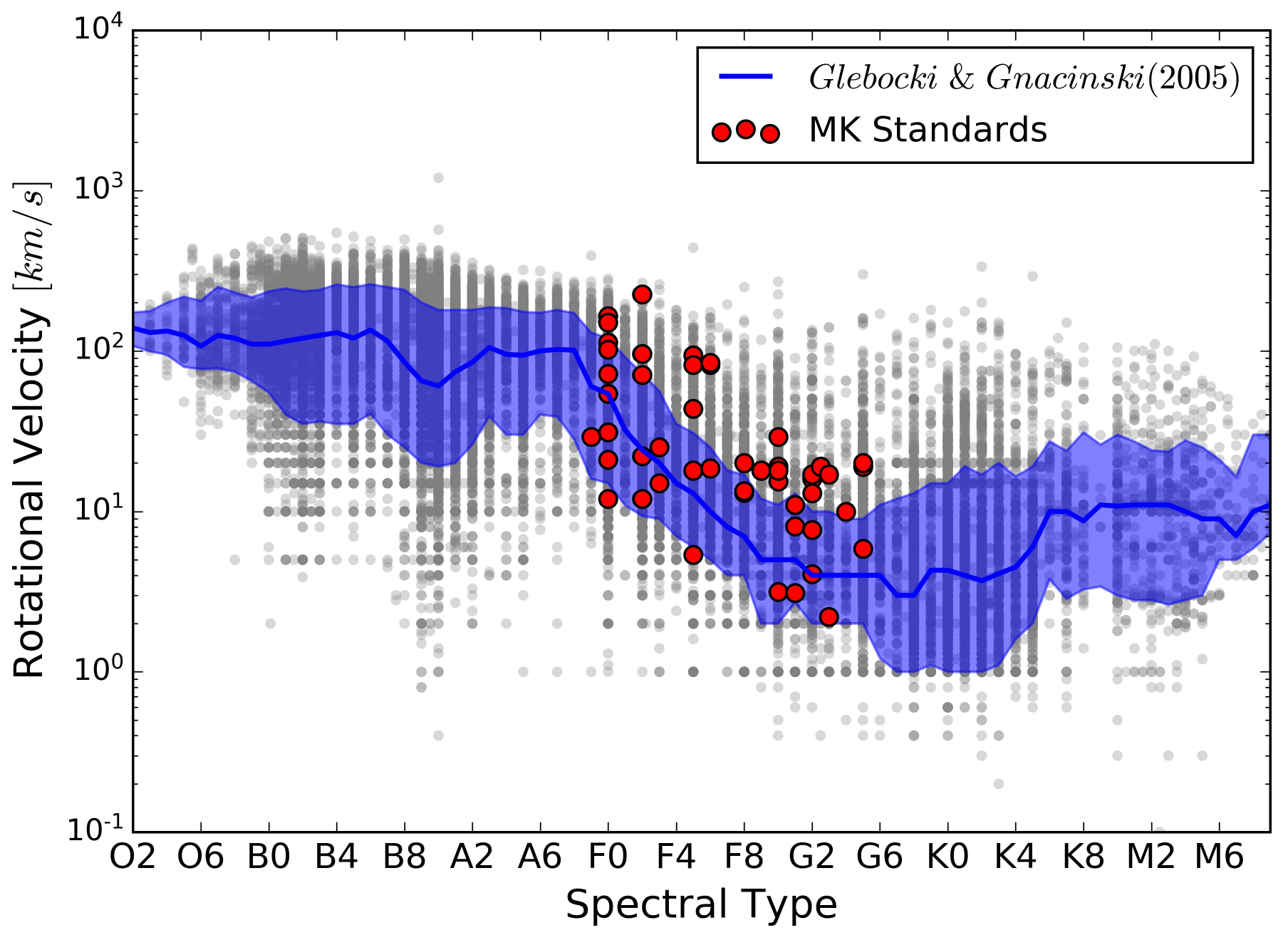}
\caption{Rotational velocities taken from \citet{glebo} (gray circles) for each spectral type, the blue curve is the median equatorial velocity . Red circles are the velocity for MK standard stars listed in Table \ref{lctab}.}
\label{vels}
\end{figure}

\section{Modelling of Physical Parameters}
\label{physicalparameters}

We used CHORIZOS: CHi-square cOde for parameterRized modeling and characterIZation of phOtometry and Spectrophotmetry \citep{maiz_2004} to obtain the effective temperature ($T_{eff}$)  and surface gravity ($\log g$). This code compares photometric data of a star with spectral energy distributions (SED) of model stellar atmospheres. The code calculates the likelihood for the full specified parameter ranges (interstellar extinction, $T_{eff}$, $\log~ g$, metallicity), thus allowing for the identification of multiple solutions and the evaluation of the full correlation matrix for the derived parameters of a single solution. 

In order to obtain the physical parameters, $T_{eff}$ and $\log~ g$ correspondent for each spectral type from A0 to G0 V, we should find the SED that better fit  our photometric data. For this aim, we used Str\"omgren color indices for each spectral type (e.g. Figure \ref{f5} for the mean $(b-y)_0$) with their respective uncertainties (the probable error derived from the fit to the color distribution by spectral type) as input parameters which are the mean $(u-v)_0$, $(v-b)_0$ and $(b-y)_0$ and; a family of SED of \citet{lej97,lej98} with the following ranges: $T_{eff}=[3500,50000]K$, $\log~ g=[0.0,5.0]$.

\begin{flushleft}
\begin{figure}
\includegraphics[width=\columnwidth, scale=0.3]{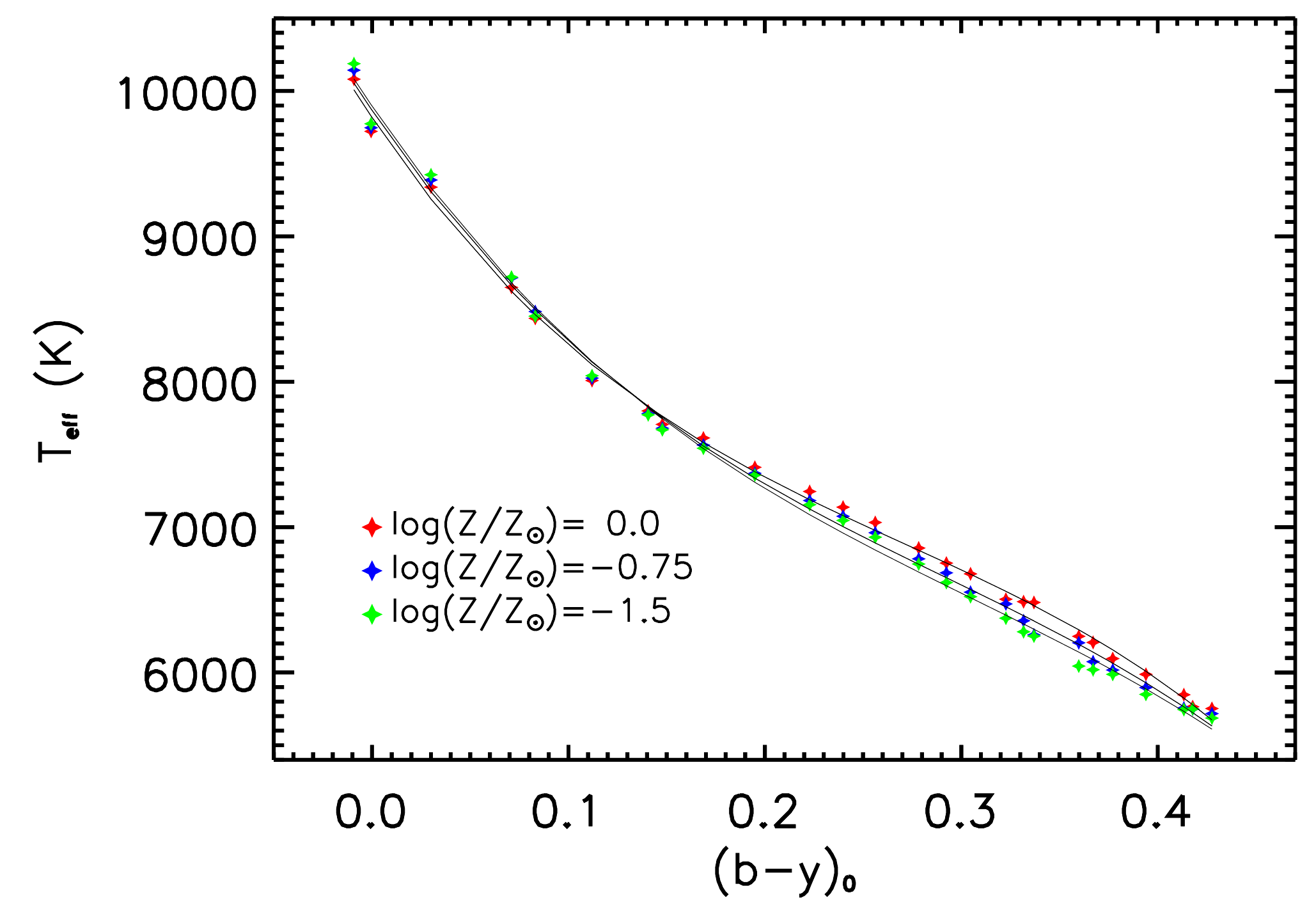}
\caption{Comparison of results using different metallicities. The points are the resulting effective temperature for each spectral type (from A0 V $\sim$10000 K to G5 V $\sim$ 5500 K). We have fixed as an input to CHORIZOS three different metallicities, $\log (Z/Z_{\odot})=$0.0, -0.75 and -1.5. The differences become noticeable for late type stars. The lines are the cubic polynomial fit for each of the three results.}
\label{metaltef}
\end{figure} 
\end{flushleft}

The range of metallicity considered by the code is  $\log (Z/Z_{\odot})=[-1.5,0.0]$, hence we tested for three different metallicities, from solar to subsolar. Figure \ref{metaltef} shows the effects of metallicity are mildly noticeable for stars later then F0 V. However, given the magnitude limit of  $V=11$ mag, we find that all the the stars included in this study are closer than 950 pc; hence, they all  belong to the solar neighbourhood. Therefore, fixing the metallicity to solar is justified. Besides, our choice is supported by the lack of strong variations of  $m_0$ for the spectral types consider in this study, see Figure \ref{fm1}. Hence, for each spectral type the corresponding  effective temperature and $\log g$ were obtained while keeping $\log (Z/Z_{\odot})=0.0$, fixed.  

The fourth parameter we varied is the colour excess with the following range: $E(B-V)= [-0.5, 5.0]$, and the mean total to selective extinction ratio for the interstellar medium $R_{V}=3.1$. As we already have a reddening-free sample of stars, therefore we fixed $E(B-V)= 0$ in CHORIZOS. We also conducted test considering the extinction as free parameter, this resulted in inconsistent outcomes.  

We have tested both \citet{kurucz} and \citet{lej97,lej98} libraries for this study. In the results we obtain a discrepancy of a few hundreds of Kelvins in the effective temperature, we then have compared with fundamental temperatures of stars (see \S \ref{fundamental}). The temperatures derived using \citeauthor{lej97} libraries are closer to the fundamental temperatures; this transformation is consistent with the one suggested by \cite{trevor}. Therefore we adopted the \citeauthor{lej97} libraries for the rest of this paper.

The effects of rotation have been ignored  in the modeling described above. Nevertheless, the average rotational velocity decreases with spectral type (temperature), our sample covers stars from A0 to G5 type stars. Early type stars are early rotators, with rotational velocities of $v\ sin\ i = 150 \;\mathrm{km\ s^{-1}}$ being typical. For a rotating star, both surface gravity and effective temperature decrease from the poles to the equator, changing the mean gravity and temperature of a rapid rotator relative to a slower rotator. The rotation of a star changes the width of a line in stellar spectra, hence resulting on a different classification. 

In Figure \ref{vels} we show a plot of the rotational velocities taken from \citet{glebo} (gray circles) for each spectral type where the blue curve is the median equatorial velocity, also in red circles are the velocity for MK standard stars listed in Table \ref{lctab}. The locus of the rotational velocity for the MK standard stars is in the same distribution for normal stars given the spectral type. Hence, in this work we have used stars whose spectra have been classified against  normally rotation standard stars. This allowed us to exclude rapid rotators, at least statiscally. Moreover, slow rotators have been excluded to a first order by rejecting peculiar A stars (see \S\ref{ss}).

 \citet{FB98}, used Monte-Carlo simulations to investigate the effect of rapid rotation on the measured $uvby-\beta$ indices, derived parameters, and hence, isochronal ages of early-type star. Those authors concluded that excluding the effects of stellar rotation overestimates  isochronal ages derived through $uvby -\beta$ photometric methods by 30/50 \% on average. 

 \citet{gray2} and \citet{gray} suggested that stellar  rotation and microturbulence velocity produces discrepancies between spectroscopic lumi\-no\-si\-ty class and  photometric indices. Some egregious discrepancies have been known among stars classified as dwarfs whose  indices correspond  to giant stars and vice versa. For example, $\rho$ And was classified  F5 IV/V, but its measured index  $c_1$ corresponds to a giant star \citep{gray2}.  
 
Therefore, we have tried to compensate the effects produced by not including  stellar rotation by considering  the observed  average properties of stars whose spectral properties resemble those of the MK standard stars in Table \ref{lctab}.

\subsection{Correction by Confronting with Fundamental Parameters}\label{fundamental}

\begin{figure*}
\centering
  \includegraphics[width=\columnwidth]{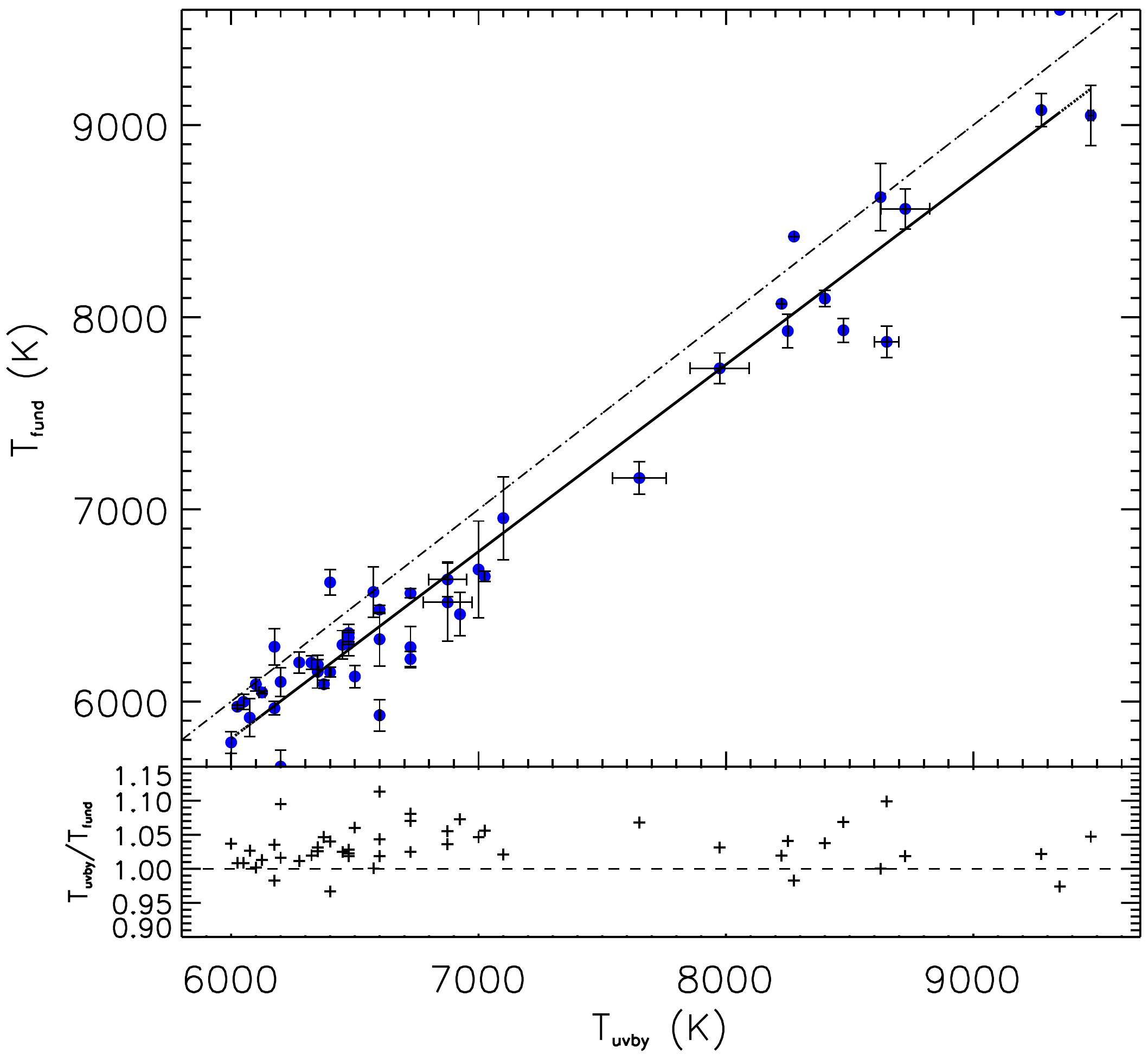}
  \caption{Comparison of the theoretical effective temperature ($T_{uvby}$) resulting after applying CHORIZOS \citep{maiz_2004} to a sample of stars with  Str\"omgren-Crawford photometry having   effective temperatures ($T_{fund}$) derived from fundamental methods, given by \citet{trevor}. The dashed line is the one-to-one relation, while  the solid line is a fit to distribution of effective temperatures, we fund  $T_{fund}=(0.9735)T_{uvby}-30.88$ (Equation \ref{equ1}). The systematic offset between the effective temperatures derived from stellar atmosphere are hotter than the fundamentally derived effective temperatures, this is seen clearer in the lower inset. This discrepancy has been found in other independent studies \citep[e.g.,][]{trevor, 2009A&A...498..527O}, it might be related to incompleteness of the line lists, the geometry,   non-LTE effects, and other physical process not accounted in the stellar atmospheres models.}
  \label{trev}
\end{figure*}

\begin{figure*}
\centering
  \includegraphics[width=\columnwidth]{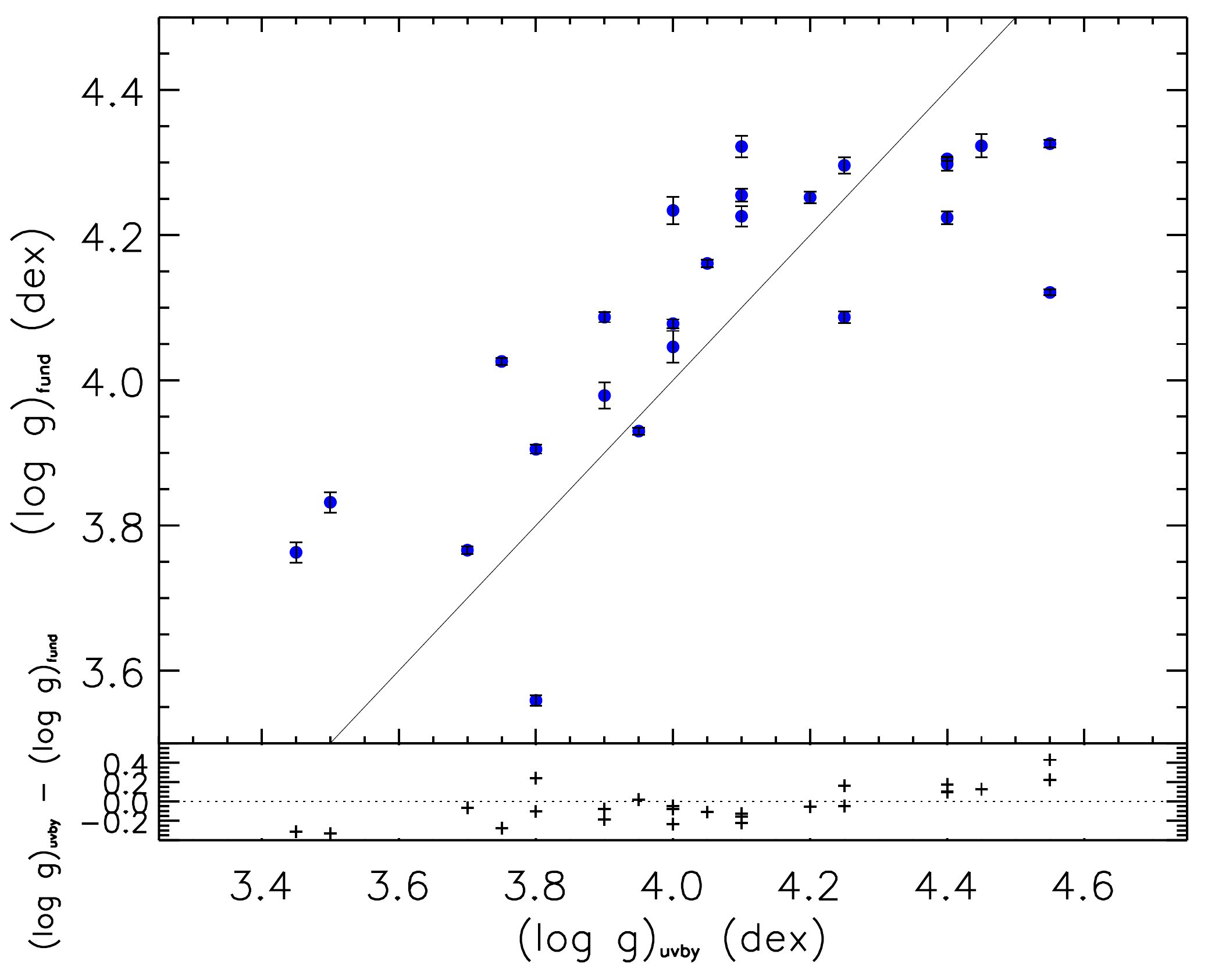}
  \caption{Comparison of the theoretical surface gravity ($\log g)_{uvby}$ resulting after applying CHORIZOS \citep{maiz_2004} to a sample of eclipsing binaries stars with Str\"omgren-Crawford photometry and  measured  $(\log~ g)_{fund}$ derived by fundamental methods. The solid line is the one-to-one relation.}
  \label{loggcom}
\end{figure*}

Interferometric measurements of the stellar angular diameters and parallax measurements, allows the direct determination of effective temperature and the total integrated flux from a star. \citet{trevor} compiled parameters derived through interferometric radii of 69 stars (listed in their table 1) whose values were taken from \citet{boyajian} and \citet{napi}. Hence, using the $uvby$ photometry of main sequence stars, we have derived $T_{eff}$ and $\log~ g$ using CHORIZOS, using  the same input setup as in \S\ref{physicalparameters}.

In Figure \ref{trev} we show a comparison of effective temperatures as provided by CHORIZOS with the compilation of fundamental effective temperatures of stars through interferometry listed in table 1 in \citet{trevor}. We found that the models predict ($T_{uvby}$) hotter effective temperatures than the ones provided by fundamental methods ($T_{fund}$). The dashed line is a one-to-one relation, most of the values of $T_{fund}$  are below the dashed line. As a result from this comparison, we performed a linear fit (solid line) to the data. This fit is used as a correction  to match the effective temperatures derived using stellar atmospheres models to those derived using  fundamental methods:
\begin{equation}\label{equ1}
T_{fund}=(0.9735)T_{uvby}-30.88.
\end{equation}
This corrections is only valid for A0-G5 main sequence stars. This is in agreement with the transformations introduced by \citet{trevor}. Hence, the effective temperatures provided by CHORIZOS were corrected using Equation \ref{equ1}. Hereafter, we will only consider the corrected $T_{eff}$.   

The slight discrepancy seen in Figure \ref{trev}  was reported  in previous studies \citep[e.g.,][]{salva}. It might be the result of incompleteness of line lists, the geometry, or non-LTE effects not accounted in  the stellar atmosphere models \citep[e.g.,][]{2009A&A...498..527O}.

We have also compared the surface gravity $(\log~ g)_{uvby}$, obtained with CHORIZOS with the corresponding stars values derived using fundamental methods $(\log~ g)_{fund}$ (using eclipsing binarires) given in \citet{trevor}. In agreement with \citeauthor{trevor}, we only find a dispersion around the one-to-one relation, see Figure \ref{loggcom}. Hence, we did not introduce any corrections to the derived surface gravity. Hereafter, we will refer to $(\log~ g)_{uvby}$ as $\log~ g$. 

\subsection*{Results}
\label{results}

For each spectral type, from A0 V to G5 V (26 subclasses), we have derived $T_{eff}$ and $\log g$ for two separated samples: stars closer than 70 pc (1496 stars) and the total sample which also includes derredened stars farther than 70 pc (7054 total), Tables \ref{tab70} and \ref{tabtodas}, respectively.  The errors in $T_{eff}$ and $\log g$ were provided by CHORIZOS. There are slight differences for the parameters for each spectral class in these tables. As we discussed before, star earlier than A6 show a systematic shift when they are corrected assuming a regular extinction law (see Figure \ref{7070}); this is expected, as young stars are embedded in the spiral arms. Later types (F and G) are distributed more isotropically, hence the effects of extinction are minimized. Nevertheless, the differences  are within the errors. 

\begin{figure}
\centering
\includegraphics[scale=0.35,angle=0]{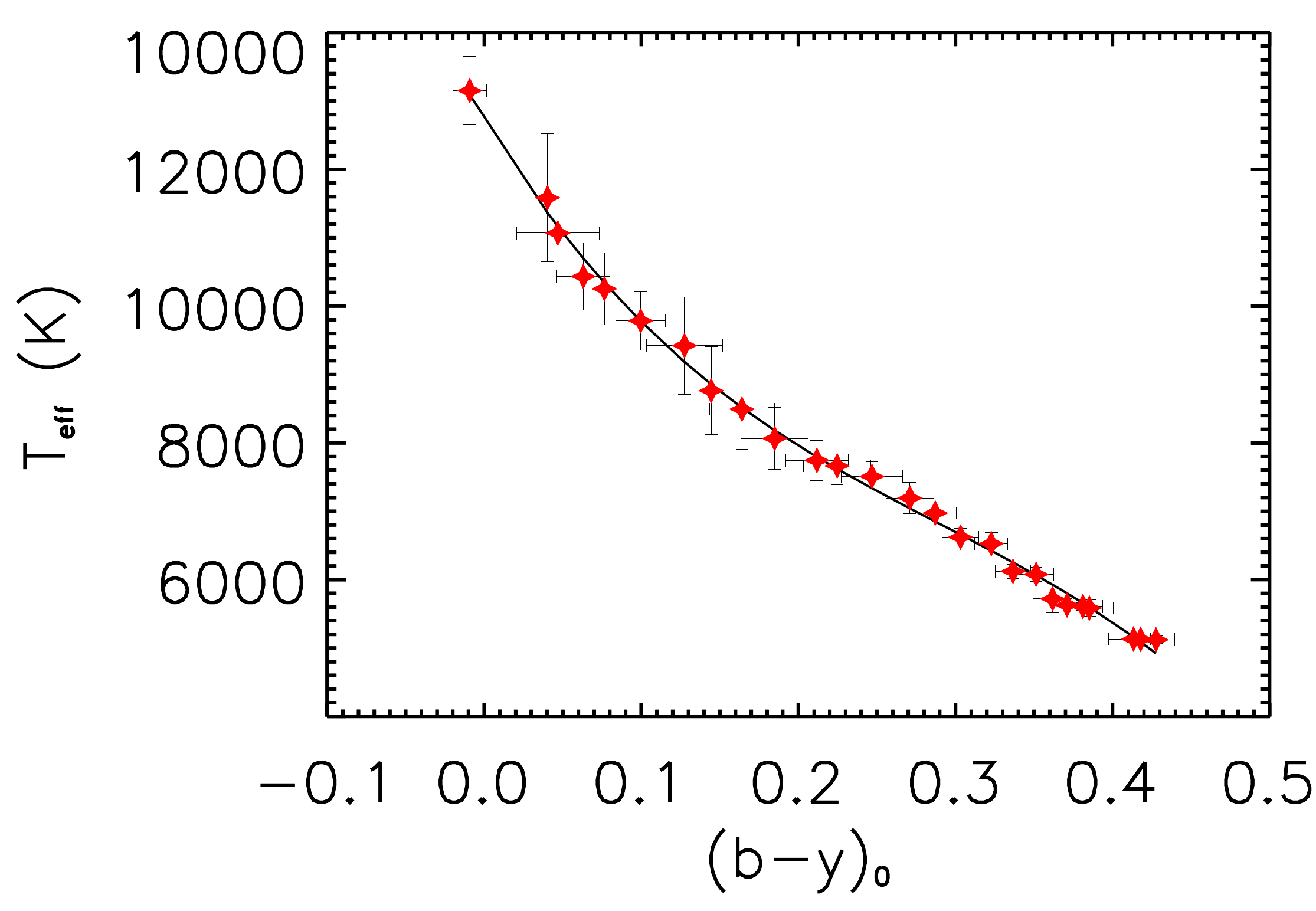}
\caption{Relation between the derived $T_{eff}$ and the $(b-y)_0$ color for A0 V to G5 V stars of Table \ref{tabtodas}. Each point represents a given spectral type, errors in the calibrated temperature are given by \textsc{CHORIZOS} \citep[cf.,][]{clem_2004}, and the solid line is a cubic polynomial fit: $T_{eff}=9369(\pm 51) - 18732(\pm 991)\left(b-y\right)_0 + 44026(\pm 5282)\left(b-y\right)_0^2-50504(\pm 7968)\left(b-y\right)_0^3$.}
\label{teff}
\end{figure}

 We can see a clear progression for $(b-y)_0$ and $T_{eff}$ which is shown in Figure \ref{teff} where a plot of $(b-y)_0$ versus the derived $T_{eff}$ is shown for all the stars of our final sample. A third order linear polynomial accounts for temperature variations in the range between A0 V and G5 V stars, the resulting fit is:
\begin{eqnarray}\label{eq2}
T_{eff}= 9369(\pm 51) - 18732(\pm 991)\left(b-y\right)_0 +  \nonumber \\ 
+ \, 44026(\pm 5282)\left(b-y\right)_0^2-50504(\pm 7968)\left(b-y\right)_0^3.
\end{eqnarray}
The caveat in this calibration is that our statistical approach overlooks the effects of rotation, magnetic fields, metallicity variations, among others; besides, the models by \citet{lej97,lej98} are unidimensional and considered local thermal equilibrium (LTE). Nevertheless, we have found a tight correlation between spectral types, $(b-y)_0$, and $T_{eff}$. Equation \ref{eq2} is in agreement with previous calibrations such as \citet[]{alonso, gray} and tables 6 and 7 in \citet{clem_2004}.


\begin{table*}

 \begin{center}
 \caption{Mean photometric Str\"omgren-Crawford colors, indices and physical parameters as a function of MK spectral types for stars within 70 pc. See \S \ref{physicalparameters} of the derivation of $T_{eff}$ (column 10) and  $\log g$ (column 11).}\label{tab70}
 \begin{tabular}{lccccccccccc}
\hline
\hline
  \multicolumn{1}{c}{ Spectral} &
  \multicolumn{1}{c}{ $(b-y)_0$}&
  \multicolumn{1}{c}{ $\sigma (b-y)_0$}&
  \multicolumn{1}{c}{ $m_0$} &
  \multicolumn{1}{c}{ $\sigma m_0$ }&
  \multicolumn{1}{c}{ $c_0$}&
  \multicolumn{1}{c}{ $\sigma c_0$}&
    \multicolumn{1}{c}{ H$\beta$}&
  \multicolumn{1}{c}{ $\sigma$ H$\beta$}&
  \multicolumn{1}{c}{ $T_{eff}$}&
  \multicolumn{1}{c}{ $\log g$}&
    \multicolumn{1}{c}{ $M_{V}$}\\
 Type & & & & & & & & &  (K)&  $cm/s^2$&\\
\hline

 A0 V & 0.0004  &  0.0140  &  0.1609  &  0.0256  &  1.0317  &  0.0646  & 2.8917   &  0.0262&  	9326	$\pm$	285	&	4.33	$\pm$	0.24	 & 1.11 $\pm$ 0.58\\  
 A1 V & 0.0177  &  0.0205  &  0.1786  &  0.0140  &  1.0043  &  0.0431  & 2.8997&     0.0152&  	8984	$\pm$	258	&	4.43	$\pm$	0.15	 & 1.42 $\pm$ 0.49\\  
 A2 V & 0.0306  &  0.0261  &  0.1802  &  0.0222  &  1.0241  &  0.0333  & 2.8918  &   0.0183& 	8701	$\pm$	275	&	4.35	$\pm$	0.21	 & 1.48 $\pm$ 0.56\\  
 A3 V & 0.0559  &  0.0186  &  0.1835  &  0.0150  &  1.0114  &  0.0837  & 2.8799  &  0.0189& 	8321	$\pm$	219	&	4.41	$\pm$	0.22	 & 1.46 $\pm$ 0.64\\  
 A4 V & 0.0705  &  0.0131  &  0.1928  &  0.0119  &  0.9432  &  0.0739  & 2.8559    & 0.0277&  	8232	$\pm$	89	&	4.49	$\pm$	0.05	 & 1.89 $\pm$ 0.45\\  
 A5 V & 0.0957  &  0.0301  &  0.1919  &  0.0162  &  0.9514  &  0.0331  & 2.8452    & 0.0268&  	7979	$\pm$	276	&	4.58	$\pm$	0.31	 & 1.79 $\pm$ 0.51\\  
 A7 V & 0.1136  &  0.0116  &  0.1997  &  0.0147  &  0.9054  &  0.0351  & 2.8286     &0.0206& 	7513	$\pm$	36	&	4.01	$\pm$	0.10	 & 1.86 $\pm$ 0.55\\  
 A8 V & 0.1502  &  0.0115  &  0.1825  &  0.0081  &  0.8326  &  0.0856  & 2.7835 &    0.0320& 	7505	$\pm$	52	&	4.17	$\pm$	0.26	 & 1.96 $\pm$ 0.63\\  
 A9 V & 0.1781  &  0.0488  &  0.1717  &  0.0148  &  0.7605  &  0.1067  & 2.7525   &  0.0426& 	7201	$\pm$	375	&	4.72	$\pm$	0.31	 & 2.49 $\pm$ 0.54\\  
 F0 V & 0.1980  &  0.0387  &  0.1690  &  0.0236  &  0.7133  &  0.1048  & 2.7428    & 0.0397& 	6987	$\pm$	230	&	4.52	$\pm$	0.39	 & 2.55 $\pm$ 0.50\\  
 F1 V & 0.2221  &  0.0552  &  0.1619  &  0.0121  &  0.6144  &  0.0897  & 2.7266   &  0.0220&  	6922	$\pm$	277	&	4.60	$\pm$	0.35	 & 2.90 $\pm$ 0.30\\  
 F2 V & 0.2551  &  0.0340  &  0.1531  &  0.0137  &  0.5412  &  0.0793  & 2.6944  &   0.0298& 	6712	$\pm$	143	&	4.54	$\pm$	0.35	 & 3.02 $\pm$ 0.41\\  
 F3 V & 0.2715  &  0.0147  &  0.1492  &  0.0112  &  0.4800  &  0.0385  &       2.6708   &  0.0187& 	6566	$\pm$	82	&	4.43	$\pm$	0.25	 & 3.27 $\pm$ 0.37\\  
 F4 V & 0.2899  &  0.0218  &  0.1563  &  0.0252  &  0.4644  &  0.0466  &       2.6625   &  0.0274& 	6419	$\pm$	128	&	4.52	$\pm$	0.34	 & 3.13 $\pm$ 0.55\\  
 F5 V & 0.2988  &  0.0153  &  0.1538  &  0.0103  &  0.4324  &  0.0431  &       2.6518   &  0.0194& 	6341	$\pm$	100	&	4.42	$\pm$	0.32	 & 3.44 $\pm$ 0.47\\  
 F6 V & 0.3196  &  0.0166  &  0.1546  &  0.0123  &  0.4149  &  0.0523  &       2.6397    & 0.0179& 	6285	$\pm$	45	&	4.47	$\pm$	0.31	 & 3.50 $\pm$ 0.52\\  
 F7 V & 0.3345  &  0.0165  &  0.1587  &  0.0134  &  0.3881  &  0.0432  &        2.6280   &  0.0157& 	6078	$\pm$	80	&	4.42	$\pm$	0.34	 & 3.64 $\pm$ 0.55\\  
 F8 V & 0.3499  &  0.0152  &  0.1678  &  0.0132  &  0.3755  &  0.0472  &       2.6195   &  0.0177& 	6047	$\pm$	30	&	4.42	$\pm$	0.34	 & 3.87 $\pm$ 0.52\\  
 F9 V & 0.3600  &  0.0164  &  0.1687  &  0.0199  &  0.3477  &  0.0480  &        2.6053  &   0.0165& 	5887	$\pm$	118	&	4.53	$\pm$	0.35	 & 4.08 $\pm$ 0.62\\  
 G0 V & 0.3733  &  0.0172  &  0.1757  &  0.0136  &  0.3429  &  0.0413  &       2.6045   &  0.0212& 	5808	$\pm$	19	&	4.42	$\pm$	0.37	 & 4.20 $\pm$ 0.61\\  
 G1 V & 0.3832  &  0.0161  &  0.1862  &  0.0220  &  0.3417  &  0.0394  &        2.6066  &   0.0195& 	5804	$\pm$	28	&	4.47	$\pm$	0.39	 & 4.14 $\pm$ 0.61\\  
 G2 V & 0.3927  &  0.0190  &  0.1901  &  0.0143  &  0.3302  &  0.0388  &        2.5977  &   0.0126& 	5744	$\pm$	110	&	4.48	$\pm$	0.38	 & 4.37 $\pm$ 0.49\\  
 G3 V & 0.4042  &  0.0171  &  0.1979  &  0.0179  &  0.3336  &  0.0417  &       2.5936   &  0.0189& 	5565	$\pm$	23	&	4.45	$\pm$	0.39	 & 4.55 $\pm$ 0.55\\  
 G4 V & 0.4130  &  0.0153  &  0.2099  &  0.0186  &  0.3284  &  0.0385  &       2.5881   &  0.0151& 	5563	$\pm$	21	&	4.42	$\pm$	0.41	 & 4.51 $\pm$ 0.48\\  
 G5 V & 0.4253  &  0.0175  &  0.2220  &  0.0229  &  0.3168  &  0.0347  &        2.5897  &   0.0151&	5562	$\pm$	19	&	4.44	$\pm$	0.44	 & 4.89 $\pm$ 0.49\\  
\hline	
\end{tabular}
\end{center}
\end{table*}

\begin{table*}
 \begin{center}
\begin{center}
  \caption{Mean photometric Str\"omgren-Crawford colors, indices and physical parameters for the complete sample as a function of MK spectral types.} \label{tabtodas}
\end{center}
 \begin{tabular}{lccccccccccc}
\hline
\hline

  \multicolumn{1}{c}{ Spectral} &
  \multicolumn{1}{c}{ $(b-y)_0$}&
  \multicolumn{1}{c}{ $\sigma (b-y)_0$}&
  \multicolumn{1}{c}{ $m_0$} &
  \multicolumn{1}{c}{ $\sigma m_0$ }&
  \multicolumn{1}{c}{ $c_0$}&
  \multicolumn{1}{c}{ $\sigma c_0$}&
    \multicolumn{1}{c}{ H$\beta$}&
  \multicolumn{1}{c}{ $\sigma$H$\beta$}&
  \multicolumn{1}{c}{ $T_{eff}$}&
  \multicolumn{1}{c}{ $\log g$}&
    \multicolumn{1}{c}{ }\\
 Type & & & & & & & & & (K)&  $cm/s^2$&\\
\hline

 A0 V & -0.0092 &0.0108 & 0.1537 &0.0209 & 1.0148 &0.0627  	&       2.8788  &   0.0341 &	9575	$\pm$ 	250	&	4.26	$\pm$ 	0.27	\\
 A1 V & 0.0403 &0.0335& 0.1666 &0.0182 & 1.0382 &0.0670  	&       2.8841   & 0.0256 &	8792	$\pm$ 	468	&	4.46	$\pm$ 	0.32	\\
 A2 V & 0.0469 &0.0326 & 0.1742 &0.0180 & 1.0547 &0.0820  	&       2.8834   &  0.0265 &	8534	$\pm$ 	425	&	4.50	$\pm$ 	0.35	\\
 A3 V & 0.0631 &0.0168 & 0.1850 &0.0201 & 1.0290 &0.0917  	&       2.8693  &   0.0249 &	8217	$\pm$ 	247	&	4.40	$\pm$ 	0.27	\\
 A4 V & 0.0765 &0.0278 & 0.1898 &0.0126 & 0.9957 &0.1147  	&       2.8532  &   0.0302 &	8126	$\pm$ 	261	&	4.49	$\pm$ 	0.30	\\
 A5 V & 0.0996 &0.0235 & 0.1890 &0.0189 & 0.9422 &0.0860  	&       2.8310  &   0.0363 &	7892	$\pm$ 	213	&	4.46	$\pm$ 	0.26	\\
 A6 V & 0.1275 &0.0538 & 0.1821 &0.0141 & 0.8950 &0.1319  	&       2.8086  &   0.0392 &	7711	$\pm$ 	356	&	4.71	$\pm$ 	0.30	\\
 A7 V & 0.1445 &0.0243 & 0.1928 &0.0146 & 0.8851 &0.0954  	&       2.8078  &   0.0308 &	7382	$\pm$ 	320	&	4.54	$\pm$ 	0.37	\\
 A8 V & 0.1641 &0.0308 & 0.1842 &0.0157 & 0.8085 &0.1294  	&       2.7840   &  0.0365 &	7246	$\pm$ 	292	&	4.57	$\pm$ 	0.36	\\
 A9 V & 0.1849 &0.0318 & 0.1777 &0.0183 & 0.7568 &0.1165  	&       2.7578   &  0.0350 &	7032	$\pm$ 	226	&	4.48	$\pm$ 	0.42	\\
 F0 V & 0.2118 &0.0295 & 0.1665 &0.0169 & 0.6716 &0.1040  	&       2.7282  &   0.0325 &	6872	$\pm$ 	146	&	4.40	$\pm$ 	0.42	\\
 F1 V & 0.2247 &0.0318 & 0.1634 &0.0173 & 0.6123 &0.0985  	&       2.7168  &   0.0336 &	6832	$\pm$ 	138	&	4.49	$\pm$ 	0.39	\\
 F2 V & 0.2468 &0.0287 & 0.1574 &0.0141 & 0.5469 &0.0700  	&       2.6951  &   0.0266 &	6754	$\pm$ 	109	&	4.49	$\pm$ 	0.34	\\
 F3 V & 0.2710 &0.0227 & 0.1551 &0.0122 & 0.4926 &0.0452  	&       2.6739  &   0.0226 &	6596	$\pm$ 	115	&	4.46	$\pm$ 	0.29	\\
 F4 V & 0.2871 &0.0201 & 0.1533 &0.0134 & 0.4765 &0.0421  	&       2.6607  &   0.0207 &	6487	$\pm$ 	105	&	4.43	$\pm$ 	0.27	\\
 F5 V & 0.3032 &0.0174 & 0.1540 &0.0116 & 0.4436 &0.0444  	&       2.6507  &   0.0195 &	6310	$\pm$ 	64	&	4.35	$\pm$ 	0.33	\\
 F6 V & 0.3228 &0.0156 & 0.1553 &0.0133 & 0.4224 &0.0463  	&       2.6396  &   0.0171 &	6263	$\pm$ 	82	&	4.43	$\pm$ 	0.31	\\
 F7 V & 0.3366 &0.0166 & 0.1601 &0.0126 & 0.4030 &0.0500  	&       2.6291  &   0.0164 &	6061	$\pm$ 	52	&	4.33	$\pm$ 	0.37	\\
 F8 V & 0.3513 &0.0164 & 0.1678 &0.0144 & 0.3844 &0.0469  	&       2.6219  &   0.0170 &	6039	$\pm$ 	53	&	4.41	$\pm$ 	0.35	\\
 F9 V & 0.3618 &0.0185 & 0.1692 &0.0239 & 0.3637 &0.0601  	&       2.6102  &   0.0200 &	5859	$\pm$ 	103	&	4.55	$\pm$ 	0.37	\\
 G0 V & 0.3710 &0.0196 & 0.1756 &0.0134 & 0.3530 &0.0510  	&       2.6098 &    0.0211 &	5814	$\pm$ 	43	&	4.44	$\pm$ 	0.39	\\
 G1 V & 0.3811 &0.0185 & 0.1842 &0.0233 & 0.3437 &0.0446  	&       2.6071  &   0.0190 &	5804	$\pm$ 	29	&	4.52	$\pm$ 	0.38	\\
 G2 V & 0.3852 &0.0225 & 0.1867 &0.0184 & 0.3427 &0.0495  	&       2.5998   &  0.0168 &	5792	$\pm$ 	62	&	4.55	$\pm$ 	0.38	\\
 G3 V & 0.4133 &0.0233 & 0.1970 &0.0203 & 0.3510 &0.0491  	&       2.5941   &  0.0194 &	5565	$\pm$ 	23	&	4.49	$\pm$ 	0.42	\\
 G4 V & 0.4178 &0.0094 & 0.2035 &0.0190 & 0.3325 &0.0398  	&       2.5891  &   0.0156 &	5563	$\pm$ 	24	&	4.06	$\pm$ 	0.42	\\
 G5 V & 0.4276 &0.0178 & 0.2187 &0.0239 & 0.3291 &0.0430  	&       2.5896   &  0.0162 &	5560	$\pm$ 	28	&	4.40	$\pm$ 	0.46	\\

 \hline	
\end{tabular}
\end{center}
\end{table*}

\begin{figure}
\centering
  \includegraphics[width=\columnwidth, scale=0.32,angle=0]{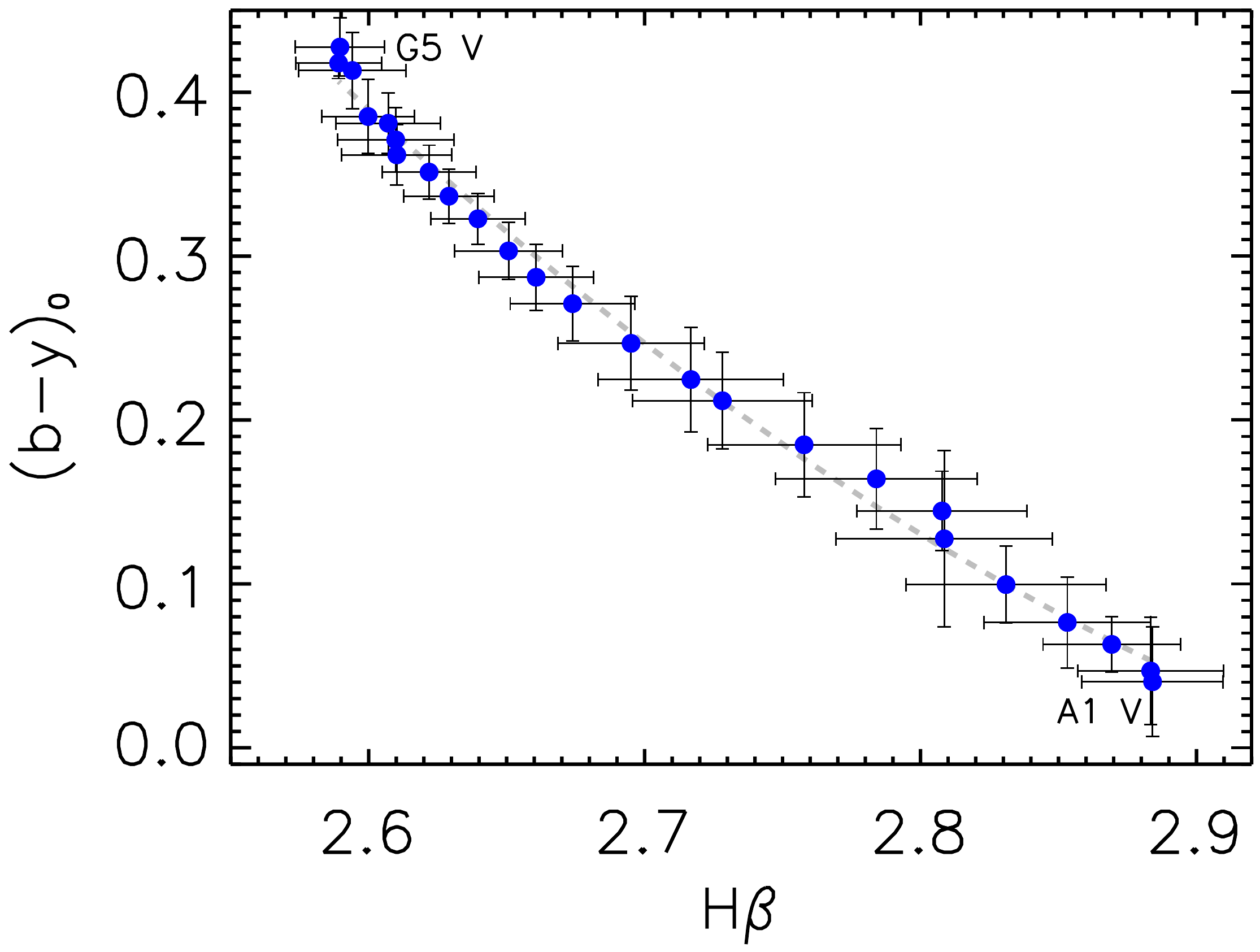}
  \caption{Comparison of the mean $(b-y)_0$ color for 4322 stars with H$\beta$ value used for the intrinsic-color calibration. Each point represents a spectral type from A1 V to G5 V, the dashed line is second order polynomial fit:
  $(b-y)_0=13.312(\pm 2.205)-8.382(\pm 1.617){\mathrm H}_\beta + 1.312(\pm 0.296){\mathrm H}_\beta^2$}
  \label{byhb}
\end{figure}

We also show the tight correspondence between $(b-y)_0$ and $H\beta$. For this aim we use the sample of stars within 70 pc (Table \ref{tab70}). Figure \ref{byhb} that shows $H\beta$ vs. $(b-y)_0$. A second order polynomial can describe such correspondence: 

\begin{equation}
(b-y)_0=13.312(\pm 2.205)-8.382(\pm 1.617){\mathrm H}_\beta + 1.312(\pm 0.296){\mathrm H}_\beta^2
\end{equation}
From this we can also confirm that $(b-y)_0$ and $H\beta$ vary with  $T_{eff}$.

Table \ref{tab70} is in close agreement with the results of \citet{Oblak}; however, we have improved on the sampling of some spectral types and on the  handle of  interstellar reddening corrections. Tables \ref{tab70} and \ref{tabtodas} complement the calibrations presented in Apendix B of \citet{grayc}. Hence, we suggest that  the discrepancies in spectral types discussed in \S\ref{ss} are inconsequential to the main purpose of this work.

\subsection{Comparison with Broad Band Photometric Systems}\label{stat}

For a subsample of stars, we have consider broad band filters covering from the optical to the near infrared, we used UBVRIJHK photometry available in the literature. We run CHORIZOS for individual stars using UBVRIJHK and their respective Str\"omgren photometry. 

\begin{figure}
\centering
  \includegraphics[width=\columnwidth]{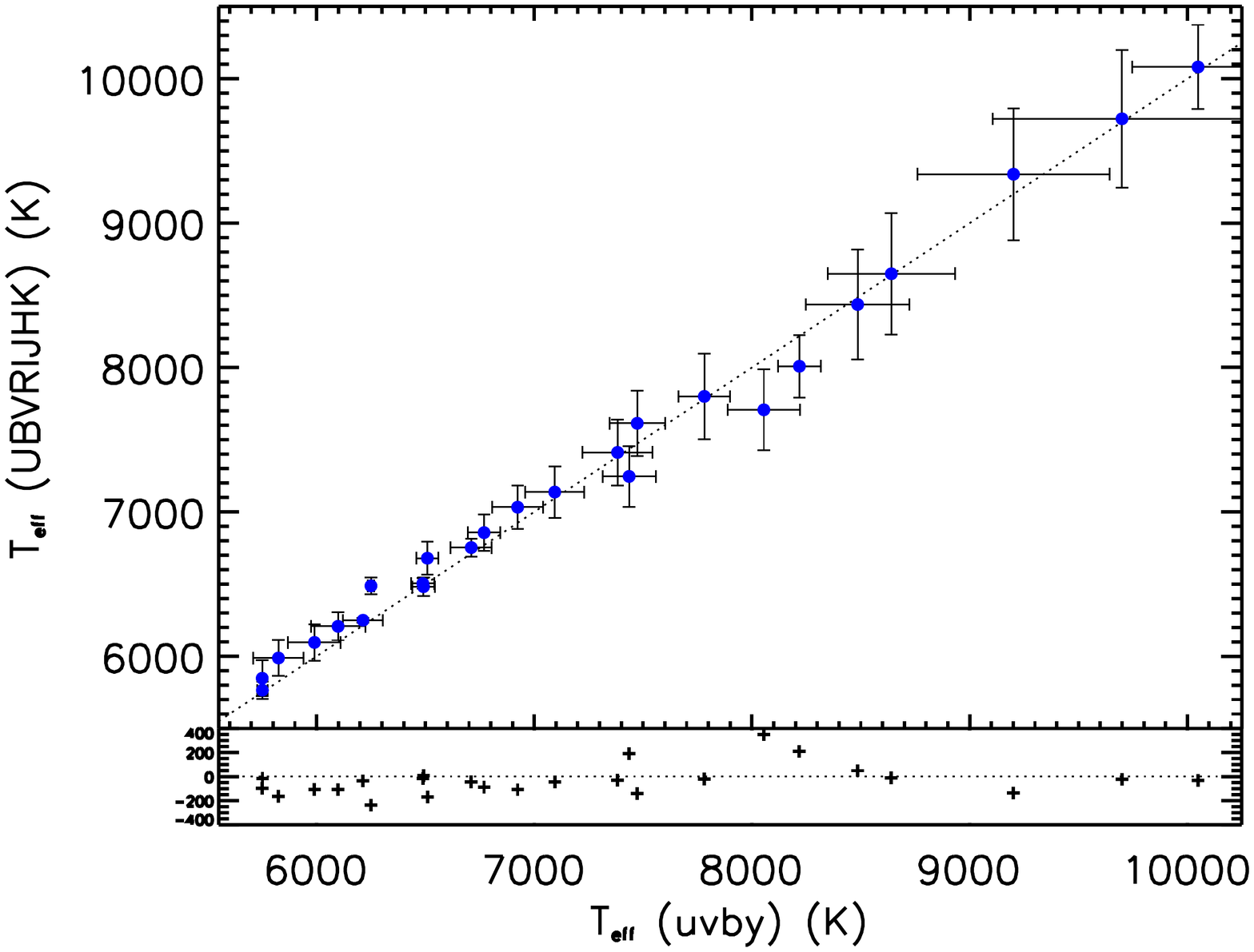} 
  \caption{Relation between the obtained $T_{eff}$ using Str\"omgren-Crawford photometry and that obtained by using  UBVRIJHK photometry available for each star.}
  \label{difcolors}
\end{figure}

We have compared the resulting effective temperatures in Figure \ref{difcolors} wih the $T_{eff}$. We can conclude from this, that broad band filters covering a much wider range than the Str\"omgren photometric system, provide only  modest improvement in the determination of physical parameters. This result shows the effectiveness of the {\it uvby}-${\rm H}_\beta$ Str\"omgren-Crawford Photometric System over broad band photometry. 

\begin{table}
\caption{Empirical Definition for  Solar-Twin Candidates}
\centering
\begin{tabular}{l l l}
\hline\hline
Spectral Type G2V&&\\
$(b-y)_0 =  0.3852^{+0.0225}_{-0.0225}$ &$m_0= 0.1867^{+0.0184}_{-0.0184}$ &  $c_0=0.3427^ {+0.0495}_{-0.0495}$ \\
&&\\
${\mathrm H}_\beta= 2.5998^{+0.0168}_{-0.0168}$ & SM1= 1.684 (see \S\ref{pcasec})&\\
&&\\
 $T_{eff}=5792^{+62}_{-62}$ & $ \log~ g = 4.55^{+0.38}_{-0.38}$&\\
  This work&&\\

\hline
$(b-y)=0.4105^{+0.0015}_{-0.0015}$& $m_1= 0.2122^{+0.0018}_{-0.0018}$&$c_1=0.3319^{+0.0054}_{-0.0054}$\\
&&\\
${\mathrm H}_\beta=2.5915^{+0.0024}_{-0.0024}$&&\\
\citet{melendez}&&\\
\hline
$(b-y)=0.403^{+0.013}_{-0.013}$ &$m_1= 0.200^{+0.026}_{-0.026}$&
$c_1= 0.370^{+0.068.}_{-0.068}$\\
\citet{holm06}&\\
\hline
$(b-y)=0.4089^{+0.0100}_{-0.0100}$&&\\
\citet{Casagrande}&&\\
\hline

\end{tabular}
 \label{solartwin}
\end{table}

\subsection{An Empirical Definition of Solar-Twin Candidates}

A solar twin is a star with properties identical to the Sun. We can take the  properties of the G2 V class in Table \ref{tabtodas} to define the properties for solar-twin candidates \citep{Ga79}. The properties of solar-twin candidates are given in Table \ref{solartwin}. Our definition agrees with the empirical definitions of \citet{holm06, Casagrande}, but differs slightly with \citet{melendez} who defined a very narrow range of variation in the photometric properties. We should also remark that the physical properties of our solar-twin candidates agree with those of the Sun: $T_{eff} = 5777$ K, $\log~ g = 4.44$, and ${\rm [Fe/H]}= 0.00$ \citep[e.g.,][]{G08}. We can also add to Table \ref{solartwin} the value of the metaindex SM1 for G2 V stars, see \S\ref{pcasec}, below. 

We found 80 solar-twin candidates \citep[cf.,][]{Ra14}. The 42 G2 V stars in Table \ref{onlinetab} could also be considered as solar-twin candidates (\S\ref{hvs}). The study of the properties of solar twins is very important in the study of the evolution of the Sun and the formation of exoplanets.

 The results presented in this section show that exploring the interplay among the MK system, an astrophysically motivated photometric system, and theory is very useful to broaden our  insight on stellar structure, this was remarked earlier by  \citet{C84}.

\section{Automatic Classification Based on Str\"omgren Metaindices and Applications}\label{pcasec}

We have recovered the physical basis of the MK spectroscopic classification. We can, then, search for alternative schemes for the classification and  advance over the original scheme proposed by Str\"omgren. In this section, a principal component analysis (PCA) was applied to derive to established a new classification classification scheme. Our aim, is to derive a scheme that can be applied easily over a large spectroscopic range. Most of the reported photometry have been confined to  $(b-y)_0$, $c_0$ and $m_0$; hence, we limit our analysis to those parameters. 
 
\begin{figure}
\centering
  \includegraphics[width=\columnwidth]{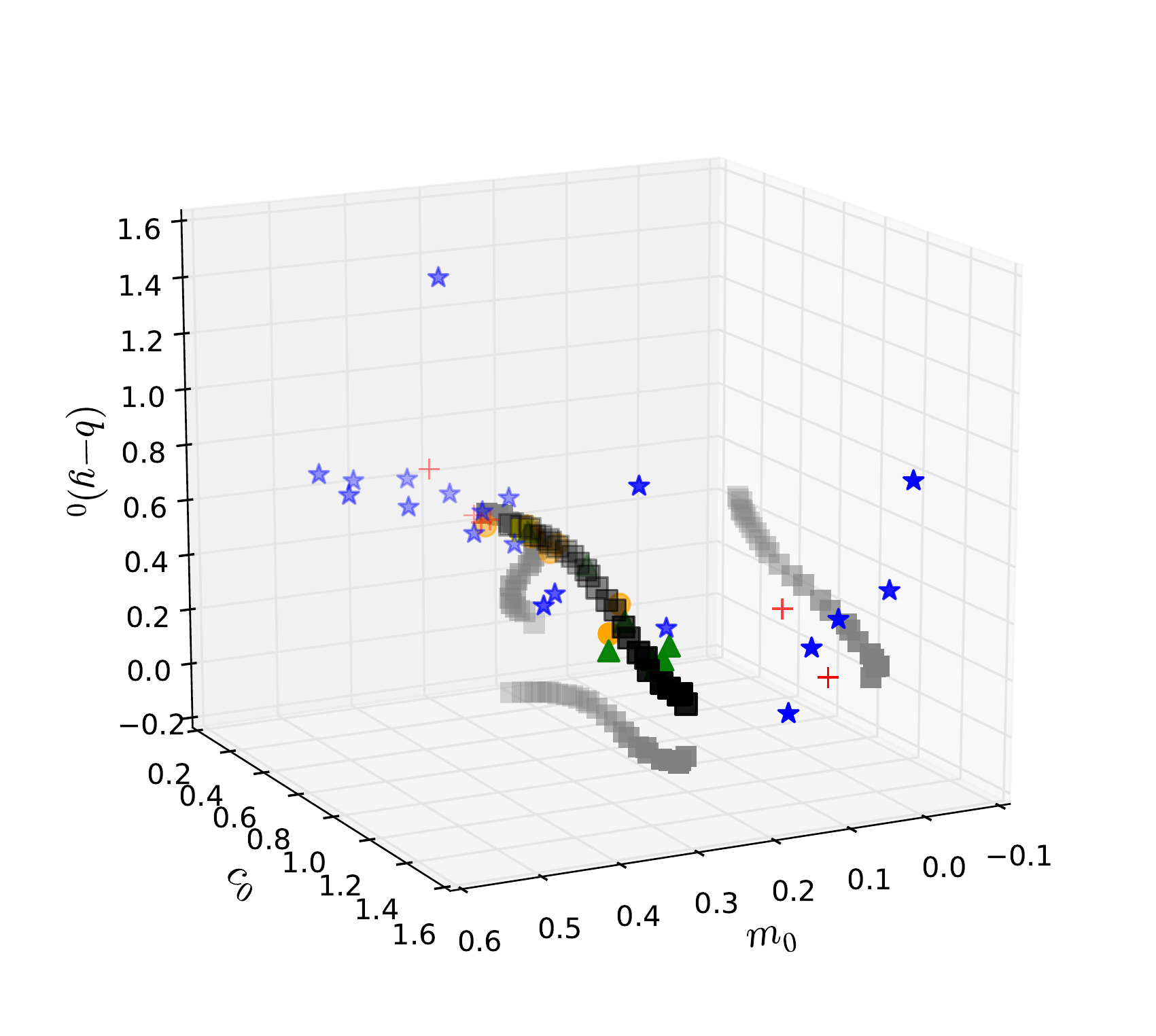}
  \caption{The three-dimensional view of the average photometric color and indices for the complete sample of dwarf stars by MK spectral types in classical Str\"omgren color-indices space (squares). Standard stars from \citet{LC91} with the same symbols as in Figure \ref{m1c1plot}. The projections on the $m_0$-$c_0$, $m_0$-$(b-y)_0$ and $c_0$-$(b-y)_0$ planes are also shown in gray. Standard stars from \citet{LC91} with the same symbols as in Figure \ref{m1c1plot}.}
    \label{plot3d}
\end{figure}

In Figure \ref{plot3d} we show the three-dimensional view of the average pho\-to\-me\-tric color and indices for the complete sample of dwarf stars by MK spectral types in classical Str\"omgren color-indices space (squares). Standard stars from \citet{LC91} listed in Table \ref{lctab}; are shown in the plot using blue stars for luminosity class I, red crosses for II, green triangles for III, orange circles for IV and yellow squares for V. Dwarfs stars occupy a well-constrained subspace. The projections of this sequence are also shown in gray squares in the three planes. The details of the values of the dwarf sequence (black squares) are given in Table \ref{tabtodas}.

The Principal Component Analysis (PCA) is a statistical multivariate analysis method, proposed by \citet{pearson} to transform a set of $n$ possible correlated variables into a set of reduced uncorrelated variables. This method uses an orthogonal transformation to find  a linear combination of its first $n$ variables with the condition to have the largest possible variance, this first linear combination is known as the first principal component (PC1). Then, the method finds a second linear combination that also has the largest variance with the additional condition to be orthogonal to the PC1, this second linear combination is known as the second principal component (PC2); and so on. Therefore we can represent the data into a new orthogonal space defined by the total number of principal components as their new axis; the maximum number of principal components is equal to the initial number of variables of the sample. One of the advantages of the PCA is that we can reduce the variable space into a new space of $n-m$ dimensions where $m$ is the number of principal components whose variances are less than the minimum variance of the data. This method is widely used in an astronomical context \citep[e.g.,][]{Sheth:2012aa,stein}. \citet{stein} applied a  PCA analysis to Str\"omgren photometric measurments of galaxies of different morphologies and nuclear activity. We follow \citeauthor{stein} to form a better parametric description of  our data. 

\subsection{From Str\"omgren colors to PCA Space: Definition of the ``Dwarf-Star Box" and  the ``Supergiant Sequence" }

We have a large sample of field stars for this study, all with intermediate band photometry. From a three-dimensional view of the $(b-y)$ color and the $m_0$ and $c_0$ indices we can see the position of the dwarf stars in a PCA space. We use the base vectors of this new coordinate system to define the following {\em metaindices}: 

\begin{eqnarray}\label{pcaeq}
{\rm SM1}&=&5.276(b-y)_0+13.641m_0-2.525c_0-2.047  \\
{\rm SM2}&=&-0.588(b-y)_0+56.997m_0+0.927c_0-10.348\\
{\rm SM3}&=&-5.390 (b-y)_0 +7.139m_0-2.573c_0+1.677 
\end{eqnarray}

\begin{figure}
\centering
 \includegraphics[width=\columnwidth]{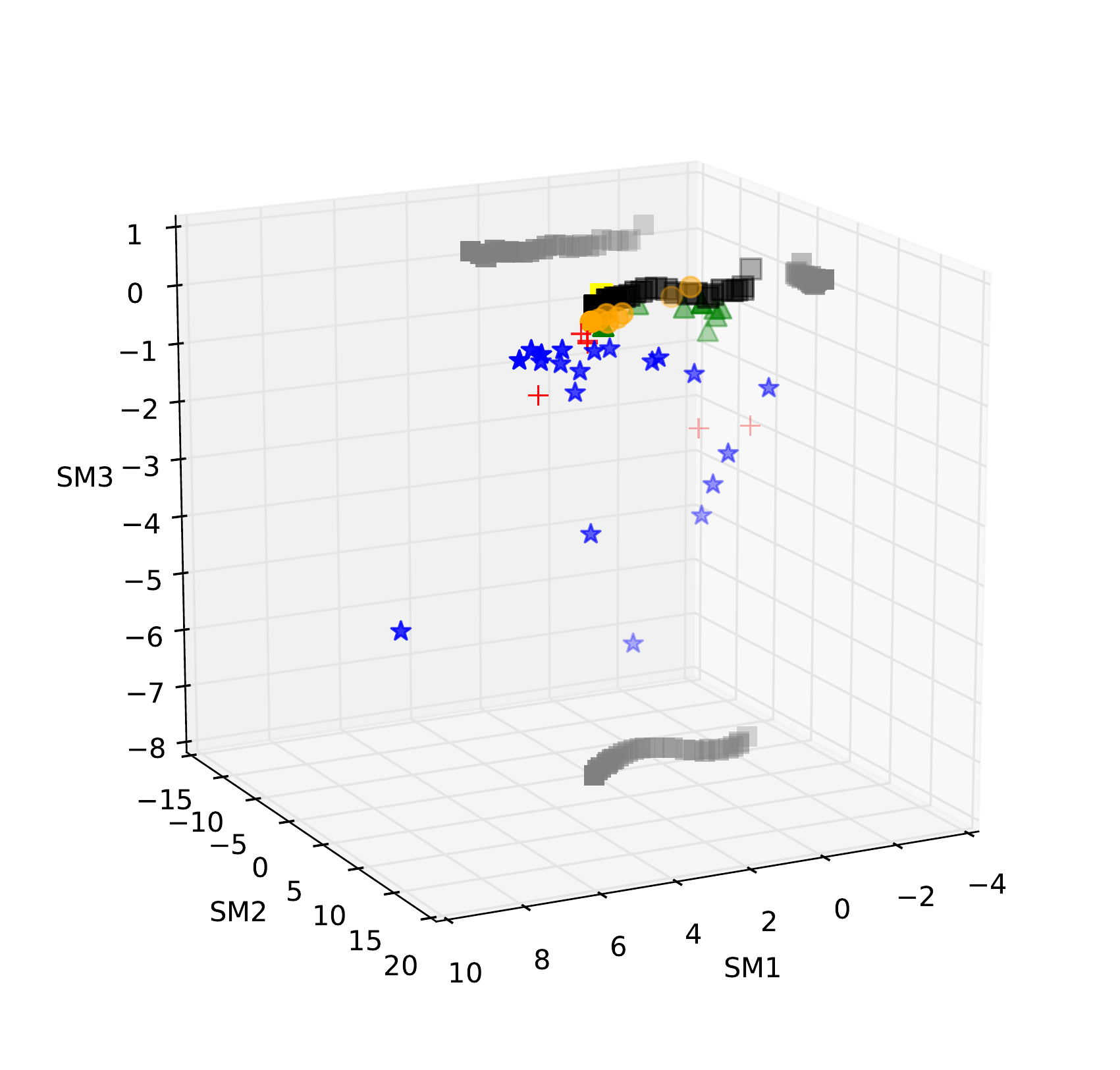}
 \caption{PCA-space of the average photometric color and indices for the complete sample of dwarf stars by MK spectral types in classical Str\"omgren color-indexes (squares), the projections of the planes are also shown in gray. Standard stars from \citet{LC91} with the same symbols as in Figure \ref{m1c1plot}.}
 \label{3dpca}
\end{figure}

\begin{figure}
\centering
  \includegraphics[ scale=0.45]{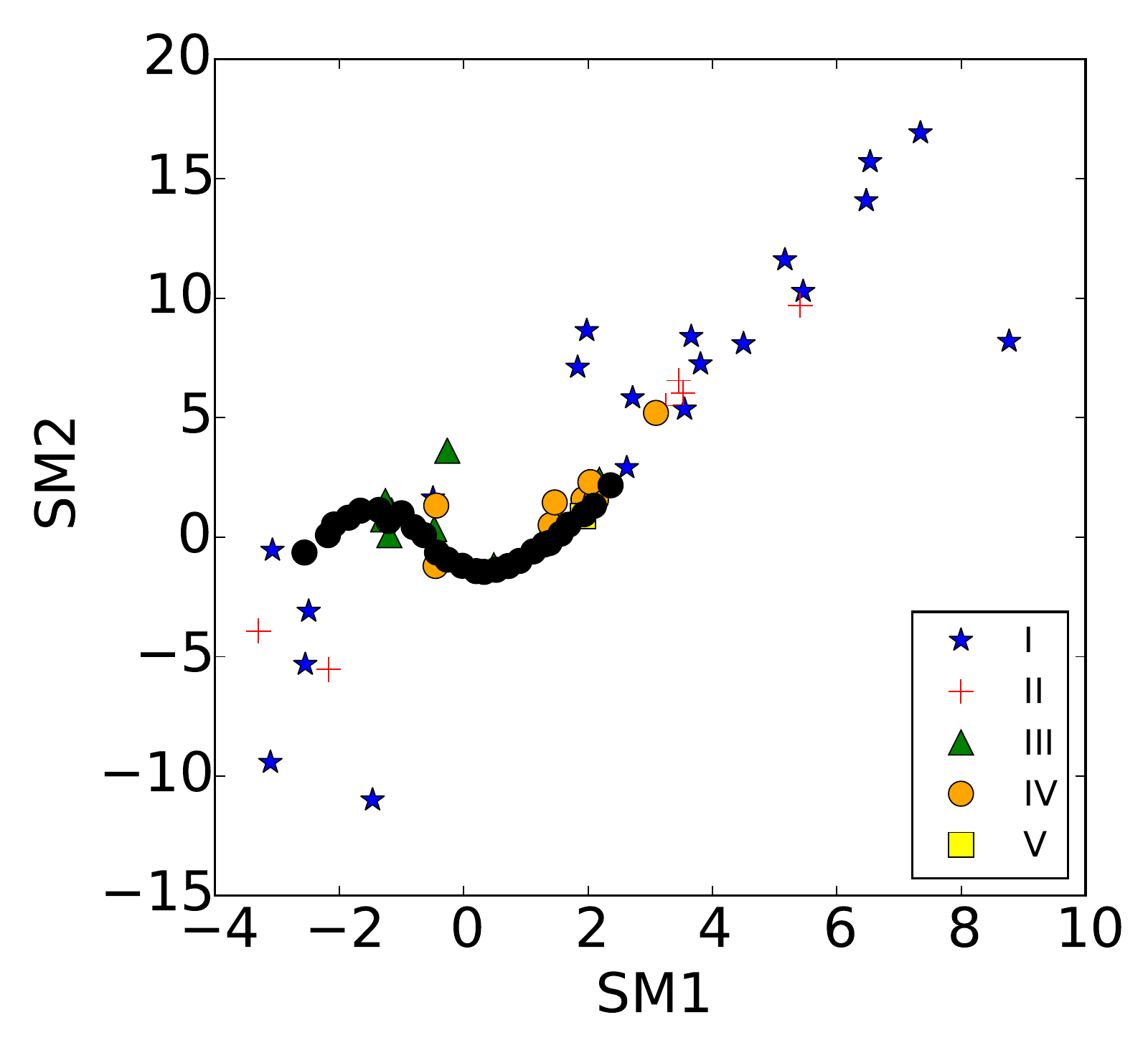}
  \includegraphics[ scale=0.45]{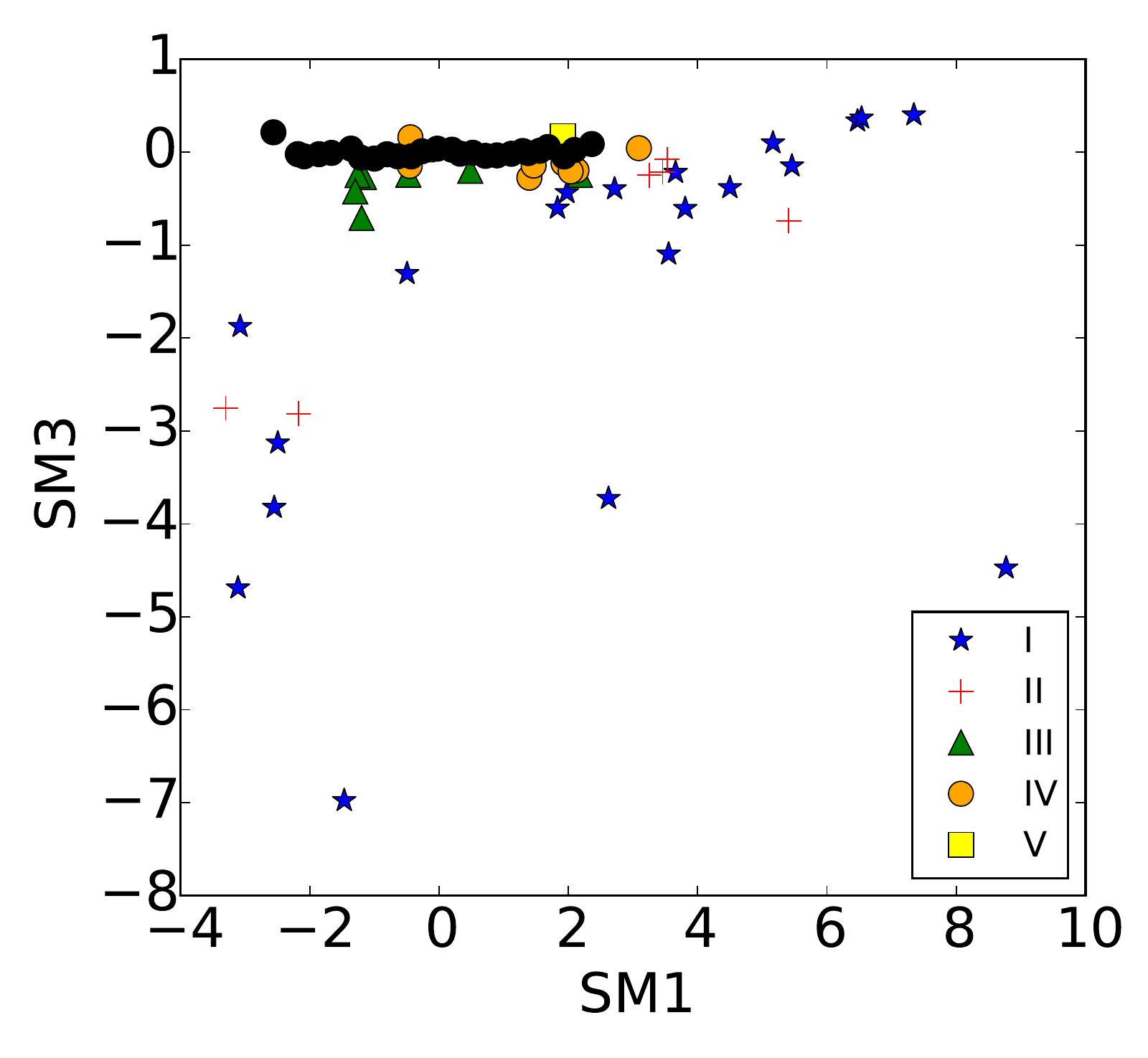}
  \includegraphics[ scale=0.45]{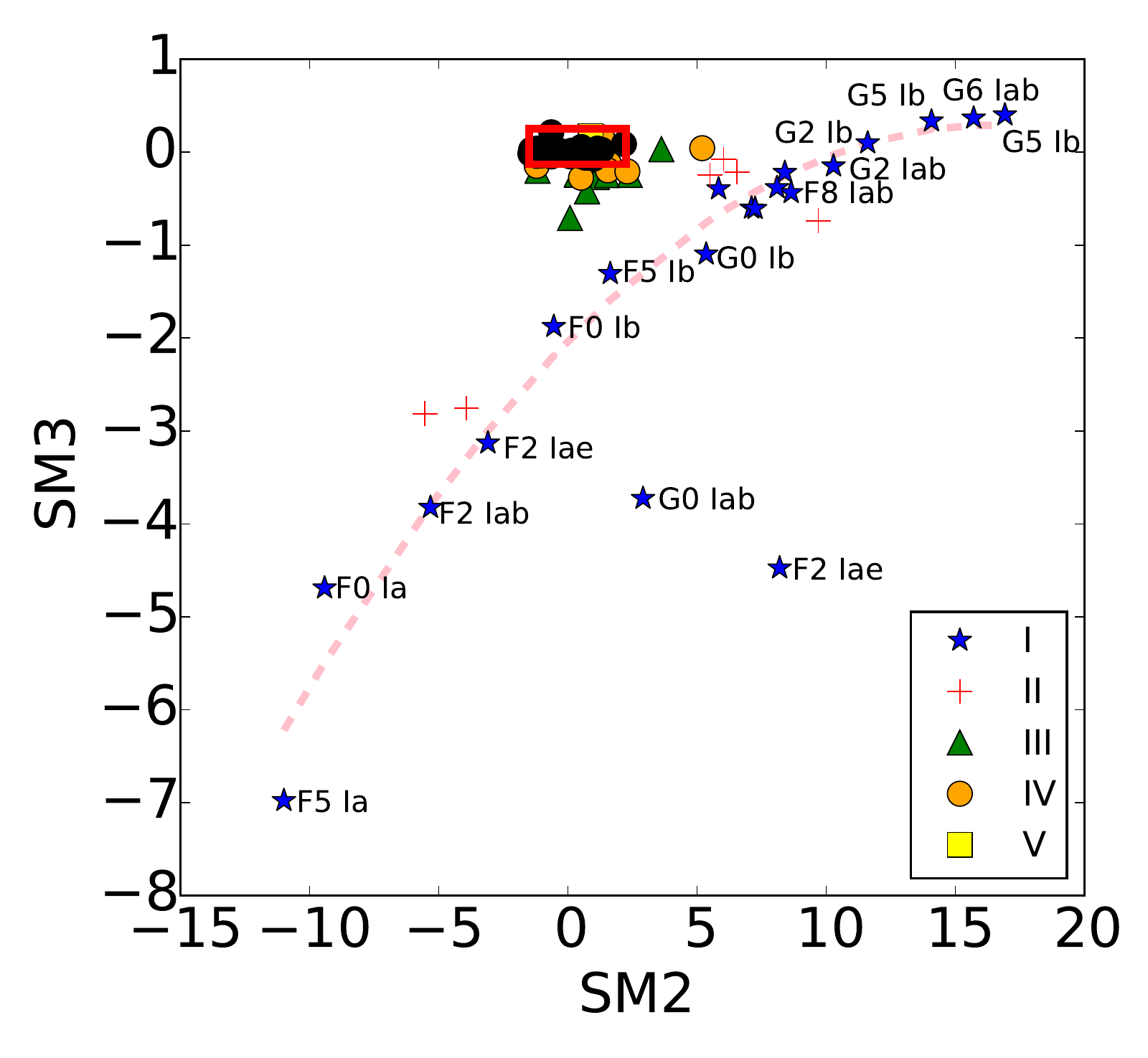}
  \caption{The projections of the dwarf sequence are shown in these three panels. Upper panel is formed by SM1-SM2 plane; central panel, by SM1-SM3 plane; and the lower panel, by SM2-SM3 plane. In the plane SM2-SM3 we define a ``Dwarf-Star box" where stars located within $-1.50< SM2<2.25$ and $-0.13<SM3<0.25$ are automatically classified as luminosity class V stars. The pink dashed line in the plane SM2-SM3 defines the locus of the supergiants stars and the fit is given by Equation \ref{pc3fit} and symbols as in Figure \ref{m1c1plot}.}
  \label{projections}
\end{figure}

\begin{figure}
\centering
  \includegraphics[width=\columnwidth]{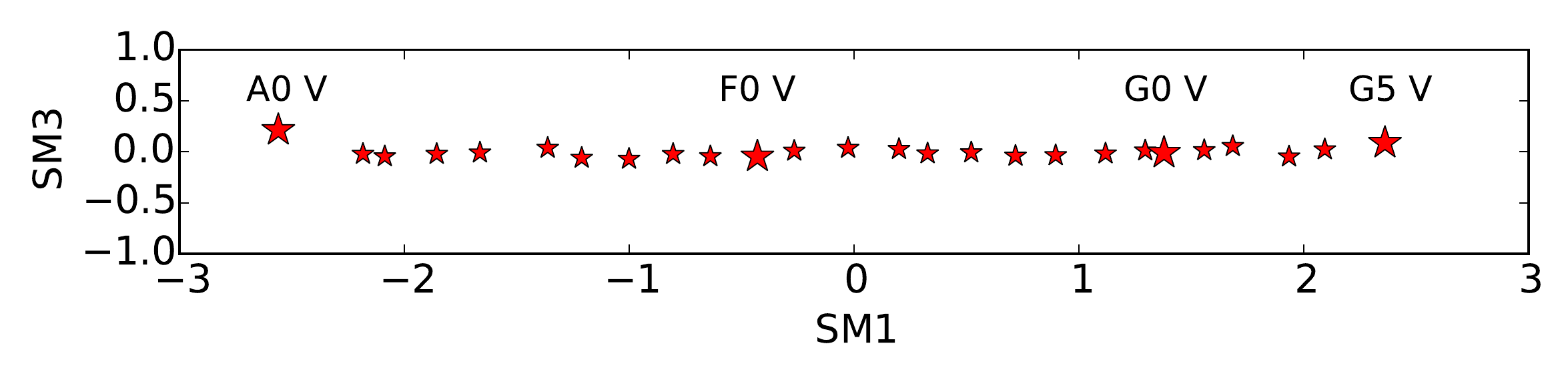}
  \caption{In this SM1-SM3 plane we see a clear separation among spectral types from A0 V to G5 V stars. Negative SM1 values correspond to stars earlier than F2 V, while the positive values correspond to later types, this is a one-to-one correspondence. A linear parametrization for the main sequence for this spectral range can be directly derived from SM3, it is given by $5.390 (b-y)_0 -7.139m_0+2.573c_0=1.677$. The larger star symbols in the graph are used to guide the eye among the spectral types.}
  \label{mspca}
\end{figure}

Figure \ref{3dpca} gives a view of the average photometric color and indices for the complete sample of dwarf stars by MK spectral types in Str\"omgren color-indexes (squares) projected onto a PCA space; the projections of this sequence are also shown in gray squares in the three planes, the projections of the planes are also shown in gray. Standard stars from \citet{LC91} with the same symbols as in Figure \ref{m1c1plot}. In Figure \ref{projections} we see these projections, it clearly defines the sequence for dwarf stars in planes SM1-SM2 and SM1-SM3. From plane SM1-SM3 we determine that vector SM1 follows a clear separation between spectral types of the dwarf sequence from A0 V to G5 V stars. Lower values of SM1 correspond to early type stars while the higher ones correspond to late type stars. A zoom of this sequence is shown in Figure \ref{mspca} where bigger symbols are only to guide the eye to the labels. It is evident that the spectral type is determined by the values of SM1 which are presented in Table \ref{pc1tab}. With the value of SM1 of any given star, we can assign a spectral type applying the nearest neighbour method. Hence, we can separate dwarfs from supergiants and bright giants and provide synthetic spectral types for A0V to G5 V stars. 

\begin{table}\centering
  \caption{Spectral type as a function with SM1} 
  \label{pc1tab}
  \begin{tabular}{llclcl}
 \hline
 \hline
  \multicolumn{1}{l}{Spectral} &
  \multicolumn{1}{l}{SM1} &
  \multicolumn{1}{c}{Spectral}&
  \multicolumn{1}{l}{SM1} &
  \multicolumn{1}{c}{Spectral} &
  \multicolumn{1}{l}{SM1} \\
  type  & &type & &type & \\
\hline
A0 V & -2.562 & F0 V & -0.430 & G0 V & 1.379 \\
A1 V & -2.183 & F1 V & -0.266 & G1 V & 1.558  \\
A2 V & -2.087 & F2 V & -0.026 & G2 V & 1.684  \\
A3 V & -1.856 & F3 V & 0.200 & G3 V & 1.935  \\
A4 V & -1.663 & F4 V & 0.327 & G4 V & 2.094  \\
A5 V & -1.362 & F5 V & 0.525 & G5 V &  2.361 \\
A6 V & -1.211 & F6 V & 0.718 &  &   \\
A7 V & -1.000 & F7 V & 0.897 &  &   \\
A8 V & -0.804 & F8 V & 1.119 &  &   \\
A9 V & -0.639 & F9 V & 1.295 &  &   \\
\hline

\end{tabular}
\end{table}

Panel the right panel defines a ``Dwarf-Star box" where stars located within:

\begin{equation} \label{box}
\left\{
\begin{array}{c}
-1.50< SM2<2.25 \\
-0.13<SM3<0.25 
\end{array}
\right.
\end{equation}

This box is partially contaminated by subdwarfs; however, none of the higher luminosity classes contaminate this box.
 Hence, stars falling in this box can be  classified automatically as luminosity class V stars. We can also define the main sequence for A0-G5 stars, directly from SM3 (Equation 6):
\begin{equation} \label{zams}
5.390 (b-y)_0 -7.139m_0+2.573c_0=1.677.
\end{equation}

The pink dashed line in the SM3 vs. SM2 diagram (see Figure \ref{projections}, lower panel) is a  parametrization to the  supergiant stars. A second order polynomial defined by:
\begin{equation}\label{pc3fit}
SM3= - 2.003 (\pm 0.012) + 0.284 (\pm 0.012) SM2 - 0.008 (\pm 0.001) (SM2)^2, 
\end{equation}
 provides an adequate fit, we call this  the Supergiant Sequence. The progression doesn't follow the distribution of spectral types as cleanly as the dwarfs.  Supergiants with negative values of SM2 and SM3 are earlier than G2 I; however do not follow a clear progression with spectral type. Nevertheless, for types later than G2 the supergiants hint a progression with spectral type. Also, bright giants (II)  seem to scattered around Equation \ref{pc3fit}.  More work is needed to establish the  metaindex distribution for supergiants and bright giants.  We should also take into account the effects of microturbulence \citep[][pp. 226-227]{gray,grayc}, winds and the intrinsic instabilities in the structure of supergiants. Nevertheless, the resulting transformations separate dwarfs from supergiants and bright giants into perpendicular planes; therefore, we can use this information to  separate these three luminosity classes, unambiguously. We can also a attempt to segregate subgiants and giants, but there is no discernible  trend related to the spectral type. 
 
 Summarizing the properties of the metaindices: the SM2 vs. SM1 plane provides nearly the same information as the $c_0$  vs. $m_0$  diagram (Fig. \ref{m1c1plot}). The SM3 vs. SM1 plane provides the main sequence for A0-G5 stars, in this projection the main sequence can be described by straight line (Equation \ref{zams}). The SM3 vs. SM2 plane provides the dwarf star box (Equation \ref{box}) and the sequence for supergiants and bright giants (Equation \ref{pc3fit}).
 A synthetic classification scheme can be implemented using the metaindices SM1, SM2, and SM3 as given by Equations 4, 5 and 6 respectively. All that is needed is a set unreddened Str\"omgren photometric measurements. Stars falling into the Dwarf Star Box are selected as main sequence stars. Then by interpolation using  Equation \ref{zams}, we can generated a spectral type. Supergiants and bright giants can be separated using Equation \ref{pc3fit}. Given the small contamination of the Dwarf Star Box, stars not falling on the supergiant sequence, and outside the Box could be
 considered giants or subgiants. 

\subsection{Applications of the New Classification Scheme}

We present two applications of the metaindices formalism introduced above.  We consider a sample of candidate F supergiants. Separetely, we also consider a set  of high-velocity stars. Very few of those high-velocity stars had luminosity classes assigned before. 

\subsubsection{Search for Supergiants Stars in the Halo of our Galaxy} 

\citet{arellano} generated a $uvby$-H$\beta$ photometric study for a group  candidates F-type supergiants at high galatic latitude. The spectral types were provided by \citet{bartaya} from objective prism plates. \citeauthor{arellano} concluded that most of the stars had solar abundances and gravities larger than 3.0, and suggested that most of these stars were low mass stars of the old disk population.

We can test \citeauthor{arellano} result straightforwardly. We have taken the four color observations presented in Table 4 from \citet{arellano}. The reddening for each of the candidates is given in column 2 of their Table 7. The stars were dereddened according to the following equations: $E(b-y)=0.73E(B-V)$; $c_0=c_1-0.25E(b-y)$; $m_0=m_1+0.32E(b-y)$ \citep{C75}. The reddening corrections are small.

\begin{figure}
\centering
    \includegraphics[scale=0.45]{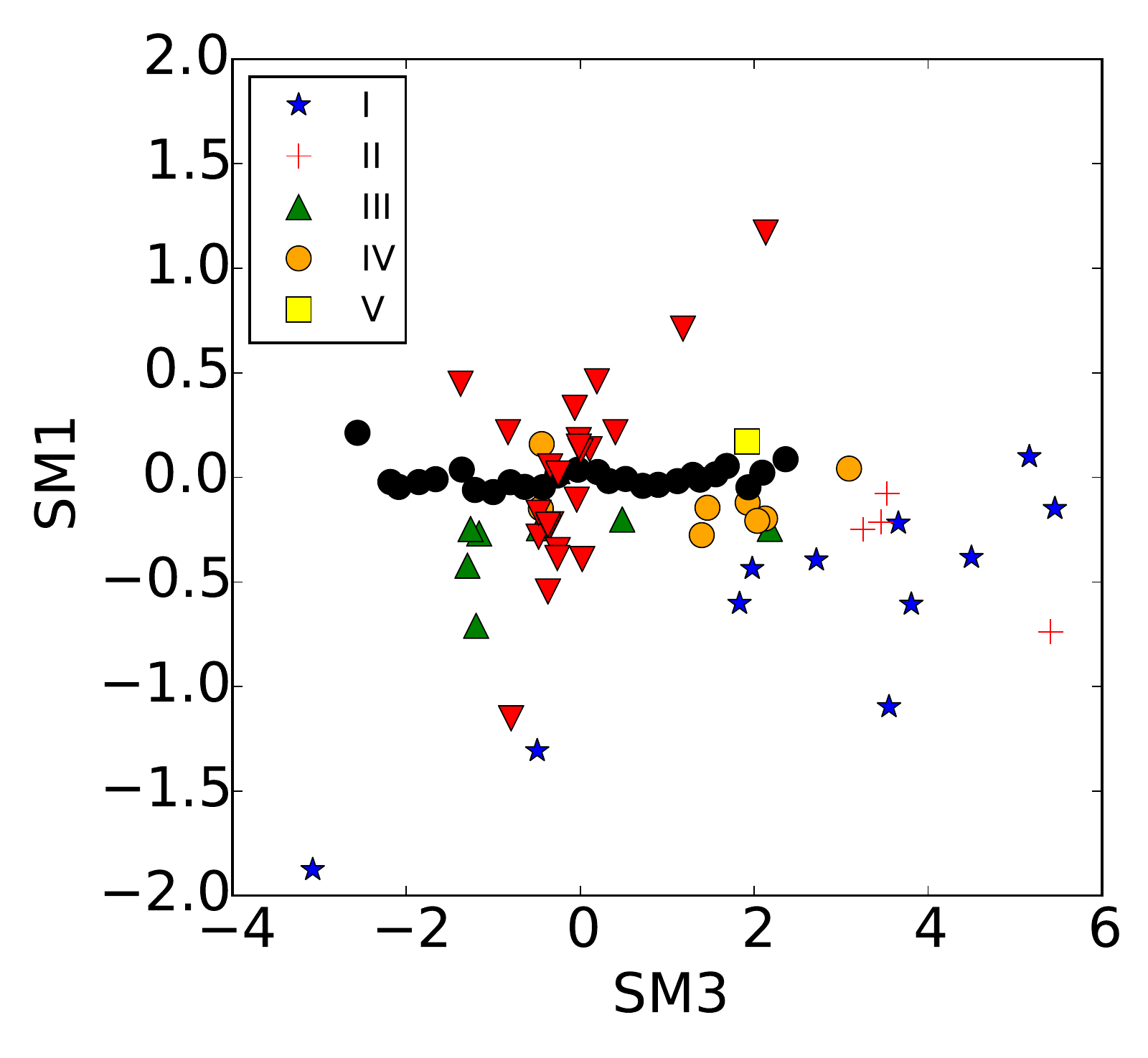}
  \includegraphics[scale=0.45]{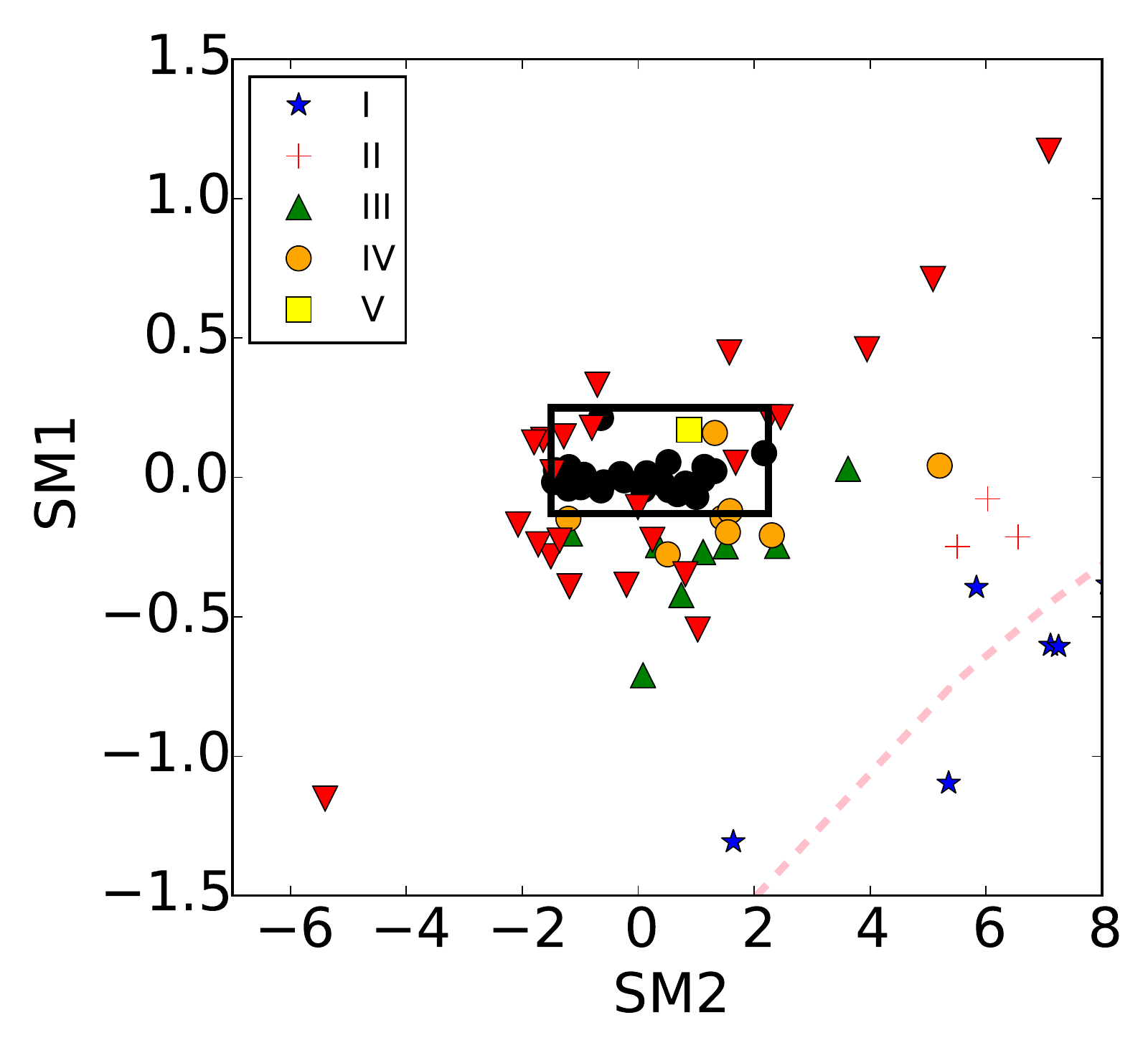}

  \caption{Candidates stars from \citet{arellano} in red triangles. These are zooms of the previous SM1-SM3 and SM2-SM3 plots from Figure \ref{projections}.}
  \label{zoomch}
\end{figure}

The metaindices were generated according to Equations 4, 5 and 6. Figure \ref{zoomch} shows the positions these stars in red downside  triangles in two of our diagnostic diagrams. From SM2-SM3 plot we can find six stars falling inside the ``Dwarf-Stars Box", while the rest of them, are distributed around the box, but away the parametrization for supergiants and bright giants. For the six dwarfs, we can provide stellar spectral classification by interpolation in Table \ref{pc1tab}, using the nearest neighbour method. The results are listed in Table \ref{newclass}. In column 3 is the classification proposed by \citet{bartaya}, in column 4 is the classification reported by SIMBAD, in column 5 we provide the spectrla  types assigned by following MK process strictly performed by  \citet{LC91} and later reported  by \citet{LC93}, in  column 6 we provide types reported in the compilation by  \citet{skiff},   and  finally in column 7 we present our new synthetical classifications. There is a general agreement with the published classifications, the uncertainty on the spectral type falls within the uncertainties (see \S\ref{distri}). However, the most discrepant is the stars BD +45 1459 (HD 60653)  classified as a giant (F5 III) by \citet{LC91} but with photometric  indices corresponding to F2 V star; this star might represent another egregious discrepancy \citep{gray2}.  The rest of the stars fall very close to the regions occupied by subdwarfs or giant stars. Moreover, stars on the upper right region outside the Dwarf-Star Box in the SM3 vs. SM2 diagram (Figure \ref{zoomch}, lower panel), could be considered late type G or early K dwarfs (see \S\ref{hvs}); therefore, we can conclude that none of the stars included in the study of \citet{arellano} are supergiants.

Our results show  that the classifications provided by \citet{bartaya} were inaccurate,  in agreement with the results of  \citet{arellano}, \citet{LC91}, and \citet{LC93}.

\begin{table*}\centering
  \caption{Synthetical classifications for six dwarf stars in \citet{arellano}}
  \label{newclass}
  \begin{tabular}{llccccc}
 \hline
 \hline
  \multicolumn{1}{l}{BD/BSD} &
    \multicolumn{1}{c}{HD} &
  \multicolumn{5}{c}{SPECTRAL TYPE}\\
     & &\citet{bartaya} & SIMBAD   &\citet{LC93} & \citet{skiff} & This work \\
\hline
+61 959 &56664& F0 I  & F0 &-&-& A8 V\\
+45 1459 &60653& F3 I  & F2  &F5 III&  F2 III/IV & F2 V \\
+45 1462 &60712& F5 I  & F2  &-&- & F2 V\\
+45 1507 &64445& F5 Ib  & G0  &-&-& F2 V\\
+45 1624 &74132& F5 I  & F5   &F2 V& - & F1 V \\
+45 1769 &85039& F5 I  & F3 III  &F3 IV& F3 Vn & F2 V \\

\hline 
\end{tabular}
\\ \raggedright SIMBAD does not report the reference of those six classifications.
\end{table*}

\subsubsection{Classification of  High Velocity Stars with High Metallicity}\label{hvs}

\begin{figure}
\centering
  \includegraphics[scale=0.45]{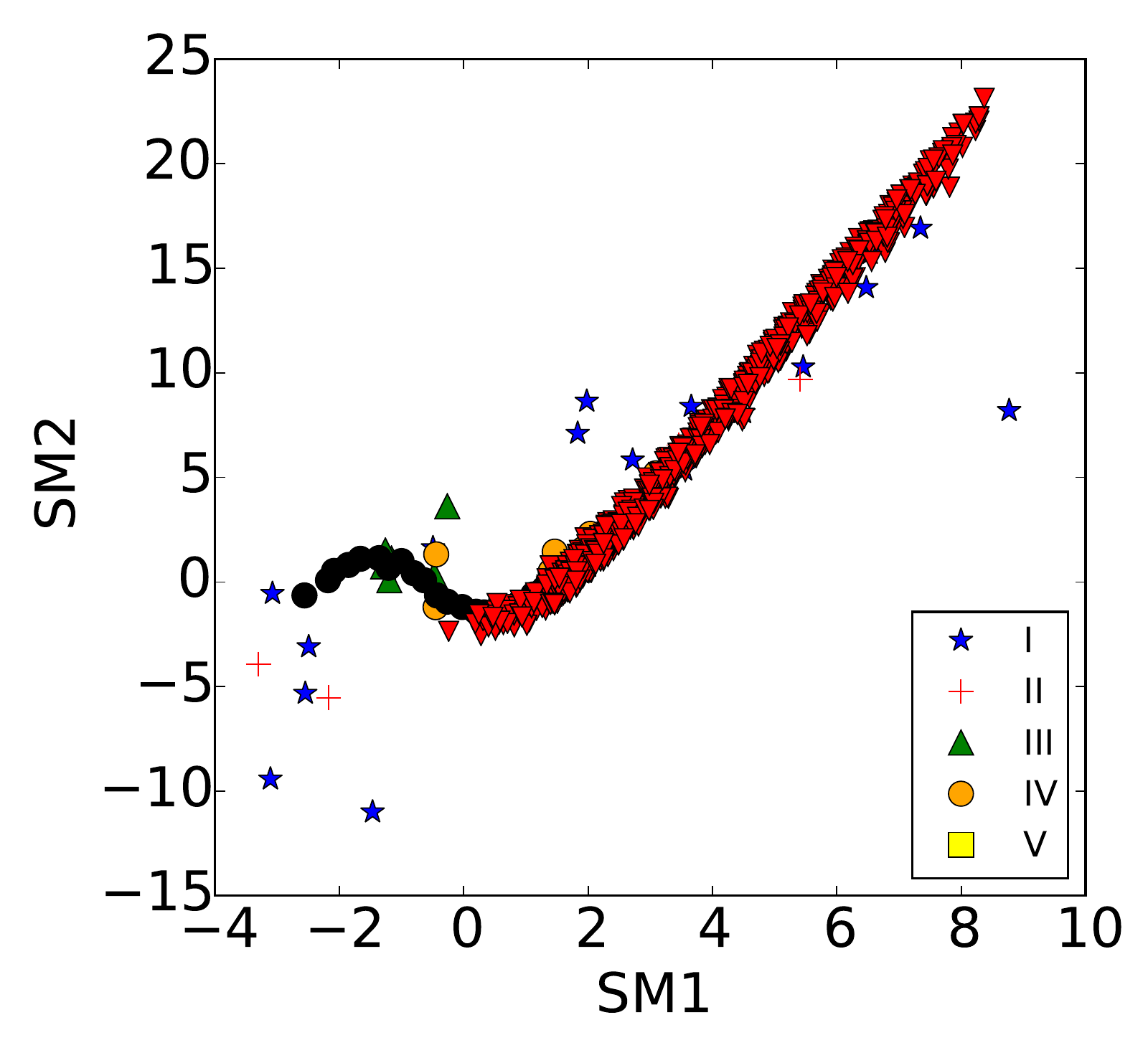}
  \includegraphics[scale=0.45]{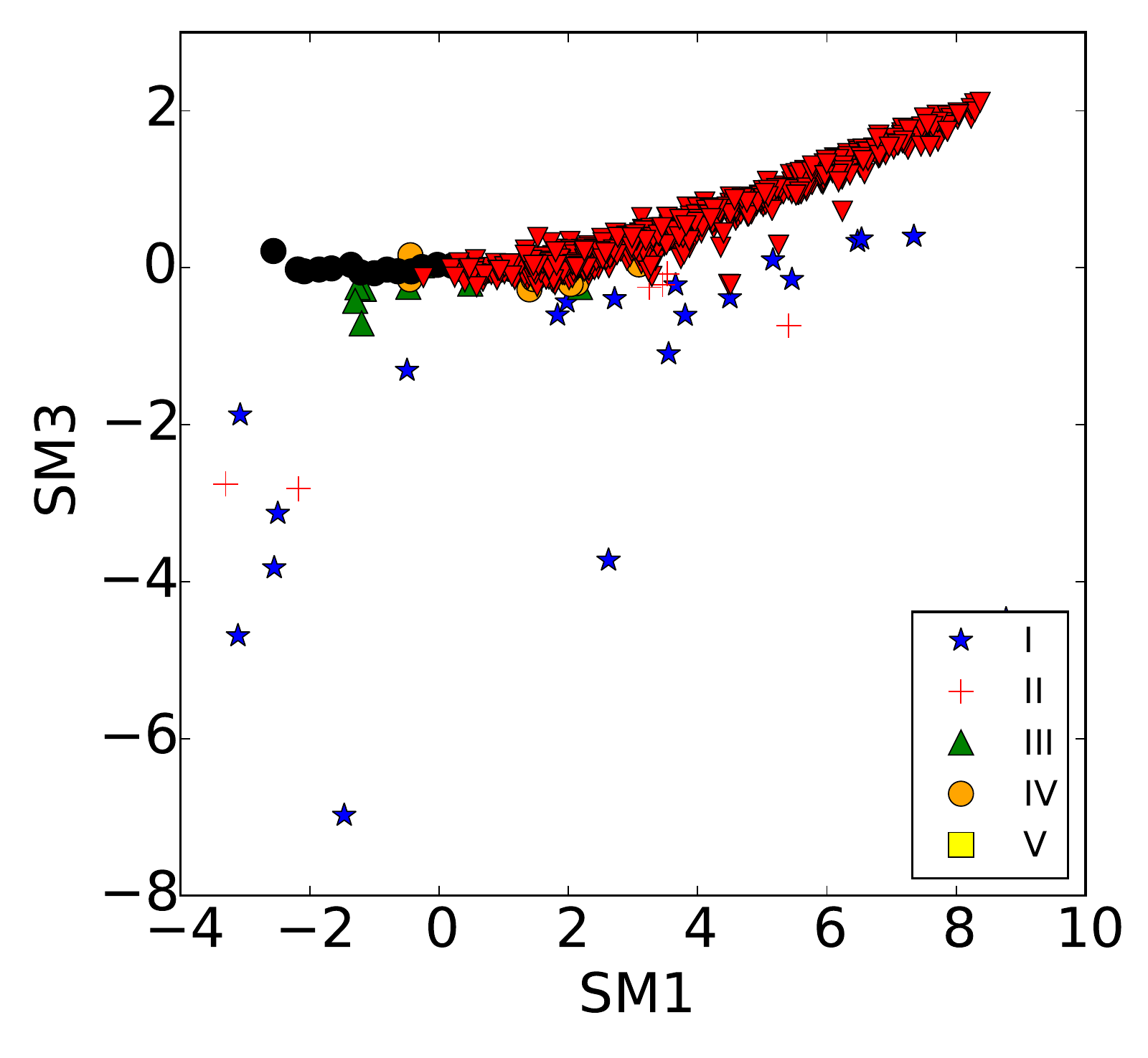}
  \includegraphics[scale=0.45]{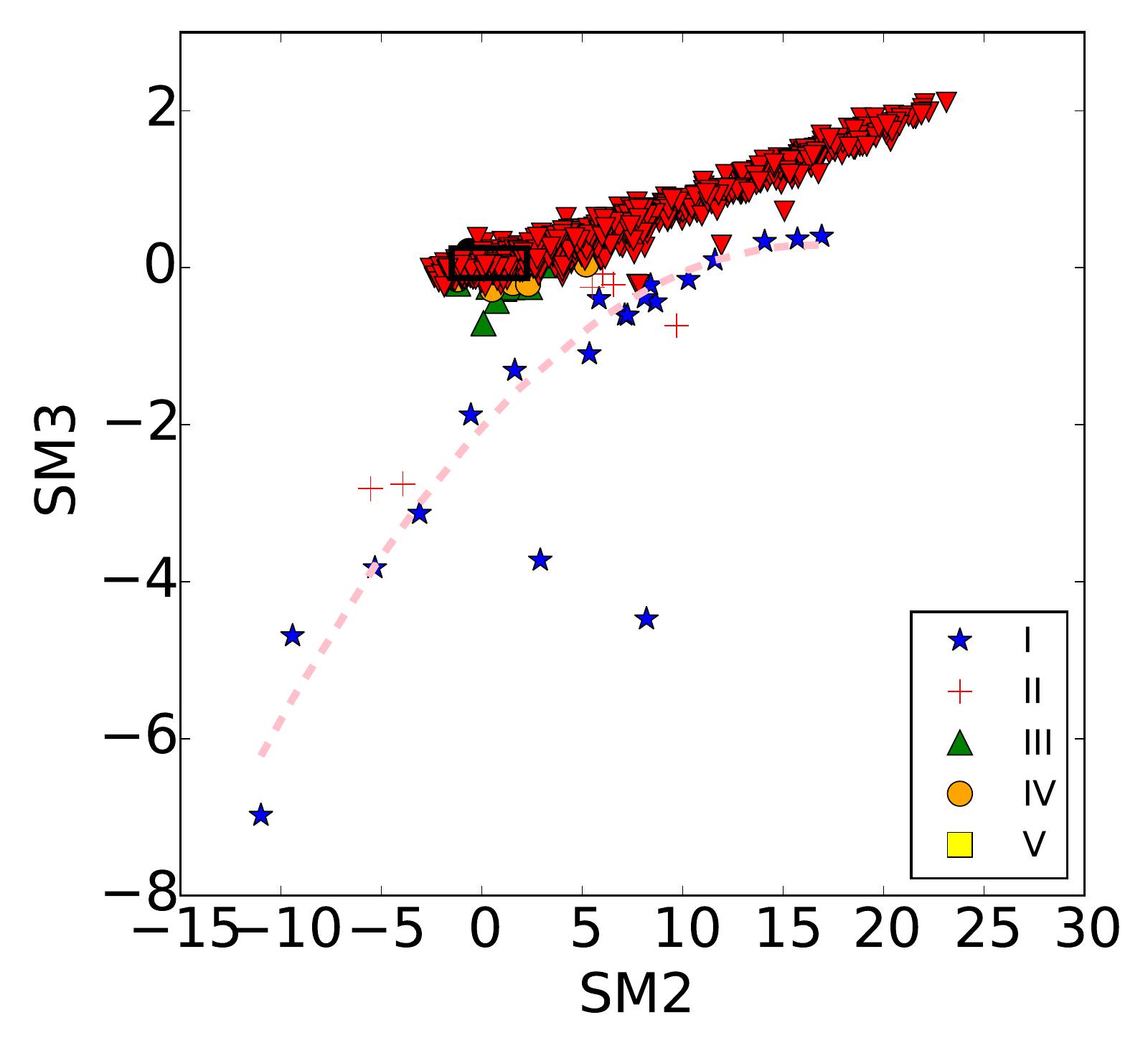}
  \caption{High velocity stars with $-0.5 \leq $[Fe/H]$ \leq +0.5$ from the publications of \citet{schuster1988} and \citet{schuster1993, schuster2006}. Many of these stars have Sp $>$ G5, and those that can be classified MK by our method are summarized in Table \ref{newclasstab}. The projections of the dwarf sequence are shown in these three panels. Same panels as in Figure \ref{projections}.}
  \label{TESTschus}
\end{figure}

\begin{figure}
\centering

  \includegraphics[scale=0.45]{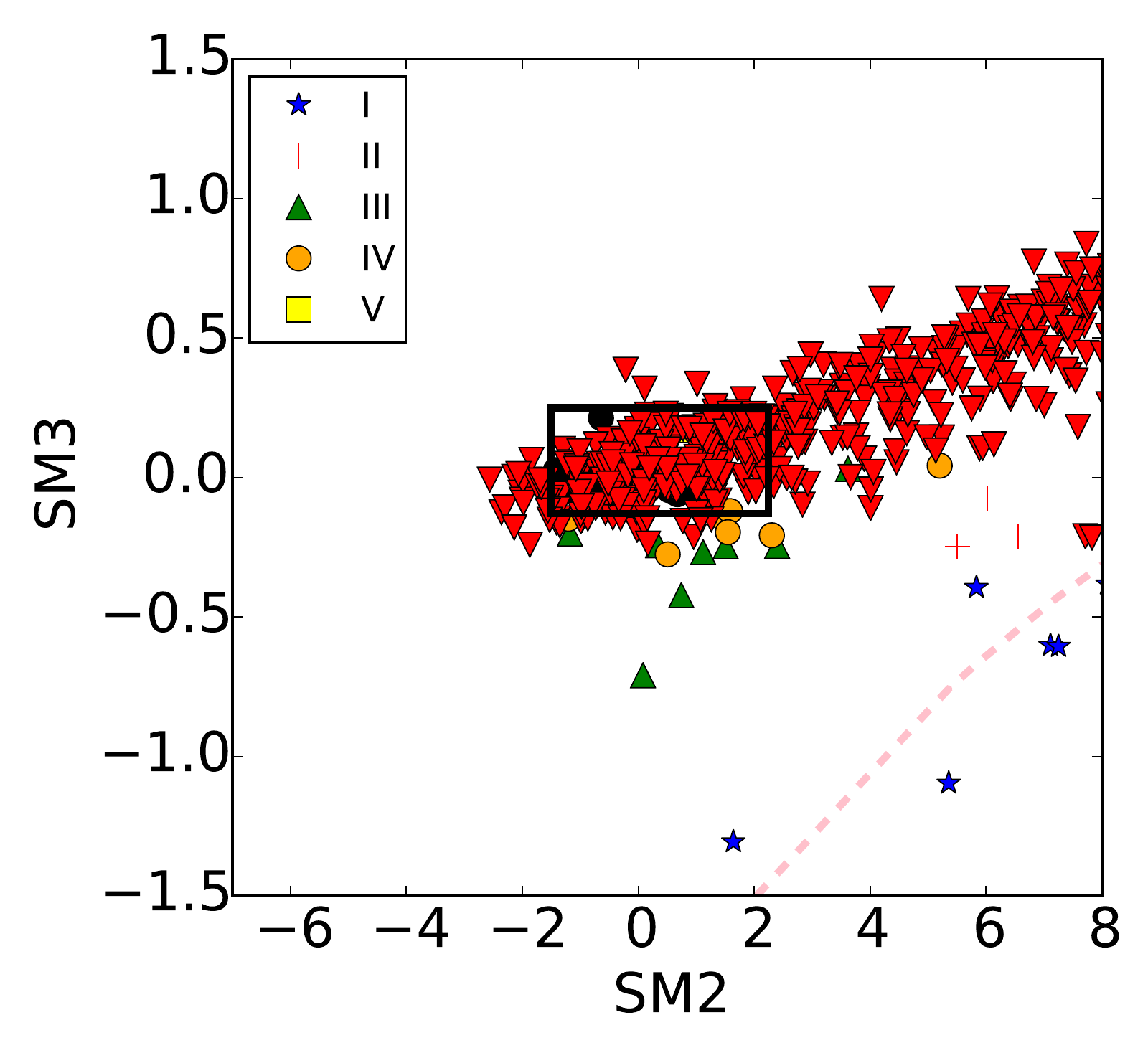}
  \caption{A zoom from the SM3-SM2 diagram  of Figure \ref{TESTschus} is shown. The 254 stars of the sample inside the ``Dwarf-Star box" were classified as main sequence stars.}
  \label{zoomschus}
\end{figure}

From a sample of high-velocity stars studied by \citet{schuster1988} and \citet{schuster1993, schuster2006}, we have selected 761 stars with metallicities close to the Sun ($-0.50 \leq [Fe/H] \leq +0.50$). These stars are distributed uniformly over the sky; that is, there are no significant north-south differences. There are F, G and early K-type stars are in the sample. We corrected for interstellar extinction when $E(b-y) \geq 0.015\;{\rm mag}$. Then, we have applied the metaindices transformation equations as in the previous example. Figure \ref{TESTschus} depicts the distribution of the selected stars, while Figure \ref{zoomschus} is a zoom around the Dwarf-Star box. There are 254 stars inside the box, we label them with the luminosity class V. We provide the synthetic spectral types for these stars in Tables \ref{onlinetab} and \ref{onlinetabb}, they are presented in the column SC. Table \ref{newclasstab} summarize our results. Column named PC in Tables \ref{onlinetab} and \ref{onlinetabb} present published  classifications. We found 57 stars with full spectral classifications (spectral type and luminosity class). We evaluated the difference between the synthetic classification and the published classification, we found that the mean, the median, and the mode are zero, while the standard deviation is SD=2.95. Only the star G060-067 (HD 114174) has been spectroscopically  classified as a subgiant, its published type is G5 IV \citep{NR55}; hence, we can say that the Dwarf-Star box has a contamination of about  2 \%. We conclude that by using our classification scheme, based on metaindices, we can provide synthetic spectral classifications with an uncertainty of about 2 MK spectral types (see \S\ref{distri}) and about 98 \% precision in the luminosity class for dwarf stars. We are limited by the accuracy of the photometric measurements and the spread of the stellar  physical properties for each MK spectral type (\S\ref{distri}). Our sample of 254 classified high-velocity stars  can be considered for further  studies of the galactic structure \citep[e.g.,][]{SSC12} or exoplanet studies, as there are 45 solar-twin candidates in Table \ref{onlinetab}.

In the SM3-SM2 digram (Figure \ref{TESTschus}, lower panel) there are three stars near to the supergiant sequence, these stars could be considered supergiant candidates. Nevertheless,  their reported classifications corresponds to late dwarf stars G014-032 (G8 V), G080-016 (K1 V) and  G016-032 (K0 IV). Hence, we suggest that these stars should be considered in the egregious discrepancy category \citep{gray2}.

The long tail traced by red downside triangles with positive values for SM1, SM2, and SM3 outside the Dwarf-Star Box in Figures \ref{TESTschus} and \ref{zoomschus}, are late type dwarfs: later than G5 V but earlier than K4 V. It seem that  the separation among dwarfs and supergiants is also  preserved, see Figures \ref{TESTschus}, \ref{zoomschus}. This looks very promising, as our formalism can be extended by generation a new set of metaindices to cover from  A0 through K4 stars. This represents an economic extension of  
St\"omgren Photometric System because it avoids the introduction of any new photometric filter \citep[e.g., the $hk$ filter,][]{1991AJ....101.1902A}. We will consider this option in a furture study. 

\begin{table*}\centering
\begin{threeparttable}

\caption{Individual synthetic MK classifications for the stars in the studies of \citet{schuster1988} and \citet{schuster1993,schuster2006}.}  
\label{onlinetab}
\tiny
  \begin{tabular}{lcccccll lcccccll}
 \hline
 \hline
  \multicolumn{1}{l}{Star} &
  \multicolumn{1}{c}{V} &
  \multicolumn{1}{c}{$(b-y)_0$} &
  \multicolumn{1}{c}{$m_0$}& 
  \multicolumn{1}{c}{$c_0$}& 
  \multicolumn{1}{c}{H$\beta$}& 
  \multicolumn{1}{c}{PC}&
    \multicolumn{1}{c}{SC}& 
      \multicolumn{1}{|l}{Star} &
  \multicolumn{1}{c}{V} &
  \multicolumn{1}{c}{$(b-y)_0$} &
  \multicolumn{1}{c}{$m_0$}& 
  \multicolumn{1}{c}{$c_0$}& 
  \multicolumn{1}{c}{H$\beta$}& 
  \multicolumn{1}{c}{PC}&
    \multicolumn{1}{c}{SC}\\
\hline
HD 82328&3.170&0.314&0.153&0.463&2.646	& F7V&	F5 V	&	
HD 94518&8.340&0.383&0.172&0.273&2.597&G2 V&	G2 V	\\
LTT15703&8.514&0.305&0.159&0.474&2.666	 &F6 &	F5 V&
HD 49409&7.922&0.385&0.176&0.298&2.580&G2&	G2 V	\\
G114-025&10.654&0.296&0.155&0.422&2.626	&F7 V & F5V&
G127-048&8.670&0.380&0.189&0.356&2.603& G5 & 	G2 V	\\
HD 298812&9.292&0.332&0.153&0.443&0.000	&F8&	F6 V	&	
HD 97507&8.590&0.392&0.176&0.307&2.571&G3 V&	G2 V	\\
HD 143333&5.457&0.333&0.154&0.448&2.620	&F8 V&	F6 V	&	
HD 143464&10.008&0.402&0.176&0.327&2.573& G5 (V)	&G2 V\\
HD 82152&8.109&0.323&0.159&0.415&0.000	& F7 V &	F6 V	&	
HD 87998&7.265&0.391&0.181&0.328&2.610& G0 V &	G2 V	\\
HD 98220&6.848&0.328&0.158&0.382&2.635	& F7 V&	F7 V	&	
LTT15705&8.060&0.410&0.175&0.335&2.581&F8&	G2 V	\\
LTT15336&6.804&0.336&0.163&0.412&2.631	&F5 V&	F7 V&
HD 20807&5.232&0.383&0.179&0.299&2.602& G1 V&	G2 V	\\
HD 20052&10.151&0.356&0.153&0.359&2.603	& F8V&	F8 V	&
HD 6517&8.959&0.402&0.178&0.333&2.574& G0 V& 	G2 V	\\
LTT15268&8.525&0.349&0.158&0.340&2.619	& F8 &F8 V & 
G027-022&8.895&0.382&0.193&0.371&2.614&	G2-3 V&G2 V	\\
LTT16139&6.573&0.364&0.162&0.379&2.618	&F8&	F8 V	&
HD 24049&9.480&0.390&0.173&0.278&2.566&	 G3 V& G2 V	\\
G061-014&6.832&0.348&0.170&0.378&2.621 & F8 V&  F8 V &	
G067-049&8.559&0.385&0.186&0.336&2.601&	G5&G2 V	\\
HD 31460&8.047&0.376&0.156&0.354&2.588	& G0V& 	F8 V	&
HD 90197&7.055&0.408&0.173&0.313&2.575& G3/G5 V&	G2 V	\\
HD 120237&6.584&0.355&0.167&0.364&2.642	& G0 V&F8 V&
G026-020&8.910&0.392&0.173&0.278&2.571& G0 V&	G2 V	\\
G043-033&7.887&0.365&0.159&0.311&2.585& Gwl& 	F9 V &	
G102-042&8.180&0.407&0.177&0.329&2.585& G0&	G2 V	\\
HD 4156&9.347&0.381&0.157&0.312&2.585	& G3 V&	F9 V	&	
G091-020&7.301&0.393&0.190&0.369&2.603& G0&	G2 V	\\
HD 203448&7.821&0.375&0.175&0.392&2.606	&G1/2 V &	F9 V	&
HD 3628&7.347&0.406&0.182&0.350&2.592& G3 V &	G2 V	\\
G171-064&7.387&0.350&0.177&0.348&2.613& F8	&F9 V& 
G034-016&8.441&0.397&0.182&0.331&2.598& G5&	G2 V	\\
HD 32023&9.174&0.341&0.182&0.355&2.586	& F7 V&	F9 V	&	
HD 100504&8.560&0.389&0.185&0.329&2.592&G1 V&	G2 V	\\
HD 114729&6.687&0.391&0.163&0.344&0.000	& G0 V &	G0 V	&
HD 190649&8.837&0.415&0.173&0.318&2.566& G5 V&	G2 V	\\
HD 43745&6.062&0.355&0.192&0.418&2.614	& F8.5 V&	G0 V	&	G034-036&9.043&0.405&0.172&0.289&2.578&G5&	G2 V	\\
G099-040&9.202&0.376&0.162&0.298&2.582	& F8 &	G0 V	&
HD 178496&8.172&0.405&0.182&0.343&2.573& G3 V&	G2 V	\\
G105-038&7.615&0.387&0.167&0.347&2.595	& G0& 	G0 V	&
HD 28571&8.982&0.405&0.183&0.346&2.584& G5 V&	G2 V	\\
HD 30361&8.346&0.385&0.162&0.315&2.576	&G0&	G0 V	&	
HD 196800&7.210&0.388&0.196&0.380&2.611& G1-2&	G2 V	\\
G114-B8A&10.242&0.378&0.164&0.303&2.605	& dF8& G0 V & 
G031-003&8.380&0.397&0.188&0.354&2.593&G0&	G2 V	\\
CD-31 8004&9.812&0.383&0.177&0.380&2.642&G0	&	G0 V	&
HD 237822&9.988&0.397&0.177&0.294&2.568&G3 V&	G2 V	\\
G156-019&8.746&0.374&0.163&0.284&2.592	&G3 V&G0 V & 
G013-001&9.020&0.368&0.183&0.265&2.590&G3 V&	G2 V	\\
HD 83529&6.971&0.377&0.171&0.325&2.602& G3 V	&G0 V &
G095-004&8.621&0.410&0.176&0.313&2.592&	G0&G2 V	\\
G219-020&9.505&0.383&0.162&0.287&2.579& G2	&G0 V& 
G065-016&8.556&0.399&0.179&0.301&2.580&	 G2 &G2 V	\\
HD 157214&5.384&0.400&0.163&0.327&2.572	&G0 V&	G0 V	&
HD 25535&6.738&0.393&0.190&0.347&2.598&	 G1-2 V& G2 V	\\
HD 212231&7.874&0.398&0.165&0.333&2.579	& G2 V& 	G0 V	&	
HD 184700&8.837&0.416&0.182&0.351&2.581&G3 V&	G2 V	\\
G089-034&8.197&0.398&0.163&0.322&2.578	& G0 &	G0 V	&	
HD 10785&8.520&0.381&0.192&0.326&2.593& G1-2 V&	G2 V	\\
G103-031&10.253&0.384&0.162&0.287&2.580	& G0&	G0 V	&
HD 98966&9.356&0.421&0.180&0.342&2.635&	G3wF7V &G2 V\\
HD 132996&7.773&0.393&0.170&0.349&2.604	& G3 V& 	G0 V	&	
G066-035&8.725&0.402&0.189&0.349&2.594&G3 V&	G2 V	\\
G188-029&9.144&0.379&0.181&0.378&2.624	& G0 &	G0 V	 &
G130-029&7.841&0.407&0.198&0.408&2.608&	G0 &G2 V	\\
HD 102158&8.066&0.393&0.163&0.309&2.581& G2 &	G1 V	&
HD 185203&9.012&0.416&0.183&0.345&2.581&	G5 V& G2 V\\
HD 111564&7.615&0.381&0.181&0.381&2.620& G0 V& G1 V	&
HD 156802&7.939&0.416&0.180&0.328&2.574& G2 V&	G2 V	\\
G126-059&7.571&0.389&0.174&0.351&2.597&	G4 &G1 V	&
G063-051&6.925&0.411&0.193&0.387&2.588&G2 V&	G2 V	\\
G001-051&7.224&0.375&0.178&0.343&2.602&	G0&	G1 V	&
HD 115031&8.334&0.395&0.192&0.348&2.602& G2-3 V &	G2 V	\\
G162-016&9.799&0.396&0.167&0.323&2.586&	 G5 V&	G1 V	&
G029-065&7.690&0.400&0.193&0.358&2.598& G5 V&	G2 V	\\
HD 165401&6.804&0.392&0.163&0.293&2.580&G2  V& 	G1 V	&	
HD 165271&7.646&0.416&0.196&0.407&2.592& G5 V&	G2 V	\\
HD 149996&8.495&0.396&0.164&0.305&2.572&G1 V&		G1 V	&	G099-050&10.227&0.387&0.182&0.267&2.560&F2&	G2 V	\\
G097-025&7.008&0.375&0.176&0.324&2.595&	 G0 V&	G1 V	&
HD 88371&8.414&0.407&0.186&0.329&2.584& G2 V&	G2 V	\\
LFT1036&6.335&0.401&0.182&0.409&2.597&	G0 V&	G1 V	&	
G134-003&10.321&0.406&0.177&0.278&2.582&G6&	G2 V	\\
HD 62549&7.723&0.385&0.182&0.375&2.605&	G2-3 V&	G1 V	&	
HD 53705&5.557&0.383&0.194&0.319&2.595&	G1.5 V& G3 V	\\
HD 101614&6.861&0.377&0.168&0.282&2.602&G0 V&	G1 V	&
HD 56196&8.944&0.388&0.198&0.350&2.595&	G5 &G3 V	\\
G071-027&8.898&0.347&0.177&0.266&2.590&	G0 IV-V& G1 V	&
G071-040&7.426&0.400&0.189&0.326&2.582& G5 V&	G3 V	\\
G192-023&7.470&0.377&0.183&0.357&2.599&	G0 &G1 V	&
G005-044&9.170&0.403&0.190&0.337&2.594&G0 &	G3 V	\\
HD 158630&7.609&0.381&0.174&0.312&2.600& G0	 V&	G1 V	&	
HD 68978&6.719&0.373&0.203&0.342&0.000& G0.5 V&	G3 V	\\
HD 2070&6.818&0.375&0.185&0.358&2.602&	G0 V& 	G1 V	&	
G065-047&6.270&0.410&0.178&0.284&2.574&	G1.5 V&G3 V	\\
HD 98868&9.640&0.375&0.176&0.304&2.611	&G3 V&	G1 V	&
HD 199190&6.872&0.397&0.202&0.383&2.607&G1 IV-V&	G3 V	\\
G060-031&9.517&0.387&0.170&0.294&2.586&	 G3V&	G1 V	&
HD 110313&7.886&0.386&0.201&0.354&2.612& F8& 	G3 V	\\
HD 211476&7.046&0.389&0.175&0.325&2.599& G2 &G1 V	&	
G132-073&7.832&0.397&0.198&0.360&2.604&	G0 &G3 V	\\
G093-047&10.700&0.396&0.172&0.322&2.586&G5 &	G1 V	&
G054-021&7.583&0.386&0.194&0.313&2.602&G0 V&	G3 V	\\
G033-048&8.754&0.388&0.169&0.289&2.586&	G0&	G1 V	&
HD 18605&9.566&0.405&0.181&0.282&2.573&	G6-8 V& G3 V	\\
HD 167425&6.192&0.376&0.187&0.359&2.614& F9.5 V&	G1 V	&
HD 118475&6.976&0.386&0.206&0.372&2.630&F9VFe+0.3 &	G3 V	\\
G014-026&9.732&0.387&0.181&0.349&2.590&G1 V& 	G1 V	&
HD 3074&6.414&0.382&0.206&0.363&2.616&F8-G0V&	G3 V	\\
HD 198188&8.129&0.385&0.181&0.340&2.599&G3 V&	G1 V	&
G101-041&7.135&0.396&0.181&0.257&2.565&G1 V& 	G3 V	\\
G054-007&7.782&0.375&0.180&0.313&2.593&	G0 &	G1 V	&
G018-005&9.262&0.401&0.183&0.276&2.579&G3 V& 	G3 V	\\
HD 162756&7.642&0.392&0.182&0.358&2.587&F8-G0V &	G1 V	&	
HD 11828&9.280&0.406&0.188&0.310&2.575&G5 V& 	G3 V	\\
G103-058&10.018&0.395&0.170&0.299&2.594&sd?G2&	G1 V	&
G131-053&9.376&0.395&0.203&0.368&2.606&	G0& G3 V	\\
G018-035&8.448&0.388&0.173&0.298&2.587&	G0&G1 V	&
G059-025&8.736&0.388&0.196&0.315&2.593&	G3 V& G3 V\\
HD 78558&7.297&0.391&0.174&0.307&2.585&	G0.5 V&G1 V	&	
G066-033&9.528&0.399&0.190&0.304&2.584&F8&	G3 V	\\
G085-045&8.451&0.385&0.187&0.359&2.601&	G5&	G2 V	&
G073-051&8.837&0.398&0.188&0.290&2.583&G5 &G3 V	\\
\hline
\end{tabular}
    \begin{tablenotes}
      \small
      \item The identifier for each star is in column 1. \textit{V} magnitude and the Str\"omgren photometry is from \citet{schuster1988} and \citet{schuster1993, schuster2006}. PC is the published classification reported in SIMBAD and SC is the synthetic classification derived from Str\"omgren data in this study.
    \end{tablenotes}
    
  \end{threeparttable}
\end{table*}

\begin{table*}\centering
\begin{threeparttable}
  \caption{Individual Synthetic MK Classifications for the stars in the studies of  \citet{schuster1988} and \citet{schuster1993, schuster2006}, continuation. }   \label{onlinetabb}
\tiny
 \begin{tabular}{lcccccll lcccccll}
 \hline
 \hline
  \multicolumn{1}{l}{ID} &
  \multicolumn{1}{c}{V} &
  \multicolumn{1}{c}{$(b-y)_0$} &
  \multicolumn{1}{c}{$m_0$}& 
  \multicolumn{1}{c}{$c_0$}& 
  \multicolumn{1}{c}{H$\beta$}& 
  \multicolumn{1}{c}{PC}& 
    \multicolumn{1}{c}{SC}& 
      \multicolumn{1}{l}{ID} &
  \multicolumn{1}{c}{V} &
  \multicolumn{1}{c}{$(b-y)_0$} &
  \multicolumn{1}{c}{$m_0$}& 
  \multicolumn{1}{c}{$c_0$}& 
  \multicolumn{1}{c}{H$\beta$}& 
    \multicolumn{1}{c}{PC}& 
    \multicolumn{1}{c}{SC}\\
\hline
HD 60298&7.373&0.401&0.197&0.340&2.586&--&	G3 V	&
G087-002&8.816&0.417&0.208&0.298&2.584&G0&	G5 V	\\
HD 89832&9.036&0.410&0.186&0.296&2.576&	G3-5 V&G3 V	&
HD 22380&8.636&0.414&0.217&0.339&2.584&	G5 V&G5 V	\\
HD 61986&8.684&0.390&0.191&0.279&0.000&	 G5 V&G3 V	&
HD 108309&6.246&0.422&0.220&0.371&2.609& G2V &G5 V\\
HD 184704&9.337&0.410&0.204&0.391&2.599&G0-2 V& G3 V&
G062-023&8.423&0.420&0.213&0.328&2.583&	G9 &G5 V	\\
HD 97998&7.362&0.397&0.185&0.260&2.599&	G1 V& G3 V	&
G092-016&10.178&0.441&0.199&0.296&2.565&G0&	G5 V	\\
G018-060&8.188&0.415&0.192&0.334&2.579&G6V&	G3 V	&
G114-048&10.649&0.409&0.213&0.304&2.586&--&	G5 V	\\
HD 4308&6.552&0.408&0.190&0.307&2.578&G6 VFe-0.9&	G3 V	&	
HD 123682&8.312&0.416&0.212&0.313&0.000&G3-5 V&	G5 V	\\
G069-008&9.015&0.404&0.186&0.275&2.586&G0&	G3 V	&
G127-012&8.938&0.415&0.219&0.347&2.595&	G0 &G5 V	\\
HD 189567&6.080&0.407&0.190&0.302&2.583&G2 V&	G3 V	&
G030-034&9.168&0.423&0.201&0.265&2.551&G6& 	G5 V	\\
G131-059&7.573&0.408&0.198&0.345&2.594&F8 &	G3 V	&
G023-006&7.918&0.418&0.217&0.340&2.586&G5&	G5 V	\\
HD 223171&6.888&0.414&0.203&0.382&2.598&G2 V&	G3 V	&
HD 4096&9.242&0.418&0.215&0.326&2.582&G3 V&	G5 V	\\
HD 224383&7.896&0.403&0.200&0.341&2.593&G3 V&	G3 V	&	
G044-044&8.160&0.407&0.218&0.319&2.579&G2 &	G5 V	\\
HD 30562&5.770&0.395&0.216&0.409&2.615& G2 IV&	G3 V	&	
HD 184768&7.557&0.427&0.216&0.348&2.587&G5 V &	G5 V	\\
G061-030&8.493&0.388&0.203&0.323&2.596&G0 &	G3 V	&
G126-012&8.614&0.428&0.219&0.365&2.590&G5 &	G5 V	\\
HD 65243&8.013&0.388&0.202&0.317&0.000&	G3-5 V &G3 V	&	
G078-041&10.231&0.419&0.205&0.270&2.561&G0&	G5 V	\\
G082-012&7.880&0.408&0.197&0.329&2.581&	G3 V &G3 V	&
G028-035&9.244&0.413&0.209&0.278&2.574&	G6 V & G5 V	\\
HD 189931&6.911&0.395&0.202&0.328&2.596& G1 V&	G3 V	&
HD 110619&7.530&0.413&0.205&0.256&2.580&G5 V&	G5 V	\\
HD 174995&8.494&0.410&0.193&0.310&2.578&	G9 &G3 V	&
CD-26 3087&9.759&0.438&0.204&0.302&2.538&	G8& G5 V	\\
G058-030&7.933&0.388&0.211&0.356&2.610&G0&	G3 V	&
G013-054&7.841&0.424&0.226&0.391&2.592&G6-8 V&	G5 V	\\
G088-008&7.995&0.394&0.201&0.312&2.599&	G4 &G3 V	&
G034-003&7.619&0.424&0.211&0.309&2.575&G5 &	G5 V	\\
G092-015&9.171&0.418&0.189&0.296&2.578&G5 V& 	G3 V	&
G048-031&8.820&0.425&0.227&0.397&2.596&G0 &	G5 V	\\
HD 12387&7.376&0.410&0.196&0.316&2.582&	G3 V&G3 V	&	
CD-49 6451&10.938&0.422&0.205&0.271&2.573&--&	G5 V	\\
HD 195019&6.882&0.418&0.203&0.365&2.597&G1 V& 	G3 V	&
G061-011&9.047&0.407&0.226&0.351&2.598&F8 &	G5 V	\\
G074-033&9.659&0.427&0.202&0.376&2.595&	G5 &G3 V	&
HD 128674&7.391&0.422&0.206&0.272&2.551&G7 V&	G5 V	\\
HD 184392&9.025&0.423&0.189&0.297&2.570&G5 &	G3 V	&	
HD 219657&7.889&0.422&0.222&0.358&2.591&G6 V &	G5 V	\\
HD 210918&6.223&0.408&0.201&0.329&2.590&G2 V&	G4 V	&	
HD 103459&7.598&0.434&0.229&0.416&2.603&G5 V&	G5 V	\\
HD 189566&8.193&0.422&0.201&0.357&2.573&G3 IV-V&	G4 V	&	
HD 102365&4.890&0.409&0.213&0.277&2.588&G2 V&	G5 V	\\
HD 216777&8.014&0.411&0.191&0.279&2.574&G6 V&	G4 V	&	
HD 202457&6.595&0.433&0.221&0.370&2.589&G5 V& 	G5 V	\\
HD 130265&8.524&0.402&0.196&0.287&2.586&G3 V&	G4 V	&	
HD 121849&8.154&0.432&0.207&0.292&0.000&G5 V&	G5 V	\\
HD 41323&8.732&0.385&0.199&0.267&2.575&	G3-5 V &G4 V	&	
G133-035&10.170&0.432&0.203&0.270&2.563&G0 &G5 V	\\
HD 136352&5.654&0.401&0.198&0.291&2.590&G3-5 V&G4 V	&
HD 45289&6.674&0.403&0.229&0.346&0.000&	G2 V&G5 V	\\
G045-030&8.250&0.399&0.208&0.339&2.593&	G5 &G4 V	&
G131-029&7.936&0.416&0.225&0.350&2.585&	G3 V &G5 V	\\
G088-020&8.018&0.402&0.190&0.248&2.575&G2 V &  G4 V	&
G079-065&8.624&0.426&0.209&0.283&2.563&G5 &	G5 V	\\
G106-055&9.411&0.414&0.197&0.309&2.571&	G0 &G4 V	&
G093-028&8.554&0.417&0.215&0.290&2.575& G6 V &	G5 V	\\
G085-017&10.480&0.408&0.190&0.255&2.559&F8 &	G4 V	&
G006-040&7.837&0.430&0.223&0.358&2.582&	G0 & G5 V	\\
HD 117126&7.440&0.414&0.206&0.353&2.585&G3-5 V&	G4 V	&	
HD 1779&8.933&0.425&0.210&0.276&2.561&	G5 V&G5 V	\\
G102-051&8.478&0.415&0.200&0.321&2.586&	G6 &G4 V	&
G060-067&6.795&0.415&0.228&0.345&2.592&G5 IV&	G5 V	\\
G079-029&8.459&0.432&0.198&0.345&2.583&	G0&G4 V	&
HD 207700&7.437&0.434&0.225&0.368&2.579&G4 V&	G5 V	\\
G130-032&8.509&0.417&0.198&0.311&2.582&G0&	G4 V	&
G039-005&8.347&0.427&0.212&0.281&2.567&	F5&G5 V	\\
HD 18757&6.652&0.408&0.202&0.313&2.579&G1.5 V &	G4 V	&	
G057-011&7.615&0.411&0.229&0.334&2.593&G5 &	G5 V	\\
G007-006&7.497&0.416&0.215&0.396&2.607&	G1&G4 V	&
G005-042&8.081&0.434&0.231&0.392&2.589&G5 &	G5 V	\\
HD 204670&9.022&0.412&0.200&0.305&2.571&G3-5 V&	G4 V	&
G022-009&10.130&0.441&0.208&0.281&2.559&G6 V&	G5 V	\\
G025-005&10.130&0.419&0.196&0.293&2.574&G5 &G4 V	&	
G023-003&9.411&0.435&0.217&0.307&2.565&	G5 V&G5 V	\\
G066-015&9.560&0.403&0.207&0.318&2.582&	G &G4 V	&
HD 183877&7.151&0.424&0.222&0.310&2.575&G8 VFe-1CH&	G5 V	\\
G130-048&7.785&0.421&0.205&0.343&2.589&G0 &	G4 V	&	
G101-047&10.104&0.432&0.215&0.286&2.561&G4 &	G5 V	\\
HD 16141&6.832&0.421&0.213&0.381&2.604&G2 V&	G4 V	&	
HD 3795&6.145&0.450&0.213&0.304&2.565&G3-5 V&	G5 V	\\
G069-021&10.336&0.427&0.196&0.300&2.577&G3 &G4 V	&
G074-031&7.585&0.426&0.228&0.334&2.584&G0 &	G5 V	\\
G102-057&6.855&0.405&0.206&0.306&2.588&	G4 V & G4 V	&	
HD 215696&7.343&0.432&0.223&0.315&2.591&G1 V&	G5 V	\\
G012-024&7.580&0.421&0.206&0.338&2.578&	G3 V& G4 V	&	
G078-040&8.612&0.428&0.230&0.344&2.592&G9 &	G5 V	\\
G078-017&7.367&0.413&0.213&0.355&2.606&G0& 	G4 V	&	
G127-055&7.989&0.434&0.222&0.310&2.581&K0 &	G5 V	\\
HD 216436&8.607&0.429&0.200&0.316&2.574&G3-5 V&	G4 V	&	
G014-005&8.131&0.432&0.213&0.257&2.554&	G6-8 V&G5 V	\\
G053-030&10.242&0.424&0.204&0.322&2.566&--&	G4 V	&	
HD 39427&8.733&0.410&0.228&0.280&2.568&	G6 V&G5 V	\\
G097-045&8.634&0.417&0.203&0.297&2.570&	G4 &G4 V	&	
G095-031&8.747&0.431&0.231&0.340&2.581&	G5 &G5 V	\\
HD 191069&8.120&0.422&0.212&0.354&2.584&G5 V&	G4 V	&	
G068-031&7.637&0.435&0.229&0.333&2.579&	G6 &G5 V	\\
HD 190333&9.230&0.427&0.200&0.297&2.553&G3-5 V&	G4 V	&
CD-02 0181&8.947&0.420&0.223&0.267&2.574&G0 V& 	G5 V	\\
HD 101563&6.439&0.399&0.219&0.341&2.601& G 2 III-IV&	G4 V	&	
G041-004&8.615&0.424&0.243&0.382&2.594&	G0 &G5 V	\\
G069-021&10.339&0.423&0.203&0.301&2.580&G3 &	G4 V	&
HD 8638&8.296&0.430&0.223&0.285&2.561&G6 V(w)&	G5 V	\\
G075-062&8.078&0.426&0.199&0.283&2.567&G5 V& 	G4 V	&	
HD 108754&9.006&0.435&0.217&0.254&2.545&G6 V&	G5 V	\\
HD 106589&8.874&0.404&0.211&0.300&2.612&G5 V&G4 V	&	
G056-040&8.656&0.418&0.236&0.320&2.591&G4 &	G5 V	\\
HD 117939&7.283&0.415&0.207&0.301&2.594& G4 V& 	G4 V	&
HD 73393&8.002&0.418&0.237&0.323&2.588&	G3 V&G5 V	\\
HD 140690&8.089&0.409&0.216&0.335&2.595&G5 V&	G4 V	&	
G116-026&10.198&0.439&0.218&0.263&2.563&G8&	G5 V	\\
G100-027&7.655&0.421&0.202&0.280&2.567&G4 IV-V&	G4 V	&
G081-019&6.713&0.430&0.246&0.393&2.606&	G0&G5 V	\\
HD 33449&8.488&0.423&0.201&0.273&2.564&	G6 V&G5 V	&	
G127-034&11.001&0.440&0.221&0.272&2.559&--&G5 V	\\
\hline
\end{tabular}

    \begin{tablenotes}
      \small
      \item The identifier for each star is in column 1. \textit{V} magnitude and the Str\"omgren photometry is from \citet{schuster1988} and \citet{schuster1993, schuster2006}. PC is the published classification reported in SIMBAD and SC is the synthetic classification derived from Str\"omgren data in this study.
    \end{tablenotes}
      \end{threeparttable}

\end{table*}

\begin{table}\centering
  \caption{Summary of the  synthetic classifications for stars in the studies of   \citet{schuster1988} and \citet{schuster1993, schuster2006}.} 
  \label{newclasstab}
  \begin{tabular}{lccc}
 \hline
 \hline
  \multicolumn{1}{l}{Spectral} &
  \multicolumn{1}{l}{Total} &
  \multicolumn{1}{c}{Spectral} &
  \multicolumn{1}{l}{Total} \\
  type  & &type &  \\
\hline
 F5 V & 3 & G0 V & 16 \\
 F6 V & 3 & G1 V & 28   \\
F7 V & 2 & G2 V & 45  \\
 F8 V & 6 & G3 V &  45 \\
 F9 V & 5 & G4 V & 37  \\
&  & G5 V &  64 \\
\hline
\end{tabular}
\end{table}

\section{Conclusions}
\label{summ}
\begin{itemize}

\item We have shown the effectiveness of the Str\"omgren system in the investigation of the physical properties of stars, avoiding the degeneracies inherent to broad-band photometry.

\item We have derived the physical parameters using mean color and indices  using a sample of 7054 reddening-free stars. Figures \ref{f5}-\ref{fc1} show the distribution of the indices as a function of spectral type. We present the main results in  Table \ref{tab70} and Table \ref{tabtodas}. We have found a new calibration for $T_{eff}$ as a function of  $(b-y)_0$ and $(b-y)_0$ as a function of H$_\beta$.

\item We have provided an empiral definition for solar-twin candidates, see Table \ref{solartwin}. In this paper we found 122 new solar-twin candidates.

\item We have applied PCA to Str\"omgren photometry  using the mean properties of  A0-G2 dwarfs and MK standards for higher luminosity classes. This has allowed us to introduce  a new formalism based on metaindices, namely SM1, SM2 and SM3, which are defined Equations \ref{pcaeq}-6. Using this formalism we can introduce a new technique to generate synthetic spectral types. We are able to  automatically segregate dwarfs from supergiants and bright giants, as we can assign spectral types to main sequence stars. The supergiants define a sequence in the SM1 vs. SM2 diagram, but it doesn't show a simple correspondence with spectral type. 

\item We  have confirmed the results of \citet{arellano} and \citet{LC91}  and have provided new spectral classifications for 254  metal rich stars in \citet{schuster1988} and \citet{schuster1993, schuster2006} samples. Using this last sample we corroborated that we can provide synthetic spectral classifications for dwarf stars with a precision of about $\pm$ two spectral MK types and a 98 \% accuracy in luminosity  class. This is comparable to the results of \citet{SDB90} using the Vilnius photometric system. 

\item We also found that we can extend our approach by the generation a new set of metaindices to cover wider range of spectral types that could include  G6-K4 stars. Hence, we can extend the original range of applicability of the Str\"omgren photometric system without including new filters (e.g., the $hk$ filter). 

\item We found four stars whose MK spectral types with discrepant photometric indices. The high latitude star BD +45 1459 (F5 III) has indices corresponding to F2 V, while the high-velocity stars  G014-032 (G8 V), G080-016(K1 V), and G016-032 (K0 IV) have indices corresponds to late-type supergiants. Similar cases  have been reported in previous studies \citep[e.g.,][]{gray2,gray}.

\end{itemize}

\section*{Acknowledgements}

\small We  thank the anonymous referee for providing us constructive comments that helped us to improve the analysis in this paper. We also thank  Dra.  M\'onica Rodr\'iguez, Dr. Miguel Ch\'avez Dagostino and Prof. Richard. O. Gray for helpful comments. An special thank you to  Prof. Eric Mamajek for very enriching discussions and comments. 

This research has made use of HIPPARCOS database. We also used SIMBAD database, which is operated at CDS, Strasbourg, France. WJS and CC would like to thank the PAPIIT project IN103014 (UNAM-M\'exico) for financial support. 

 We wish to dedicate this work in memory of the late Professors  William W. Morgan (1906-1994) and Robert F. Garrison (1936-2017) for their outstanding contributions to the development of the MK System and the MK Process, and for actively supporting the formation of MK classifiers in Mexico. This paper is also dedicated to our colleague and collaborator Mr. Juan Gabriel Garc\'ia-Ru\'iz who suffered a terrible accident which took his life away,  while on  duty at the Observatorio Astorn\'omico Nacional de San Pedro M\'artir (OAN-SPM) in 2010. 








\bsp	
\label{lastpage}
\end{document}